\documentstyle[epsfig]{mn} 
\onecolumn

\newcommand{\subI}{_{\rm _I}}
\newcommand{\subR}{_{\rm _R}}
\newcommand{\subs}[1]{{\rm _{s#1}}}
\newcommand{\sube}[1]{{\rm _{e#1}}}
\newcommand{\subi}[1]{{\rm _{i#1}}}
\newcommand{\dR}{\delta\subR}
\newcommand{\dI}{\delta\subI}
\newcommand{\dIO}{\delta_{_ {\rm IO}} }
\newcommand{\dC}{\delta_{_ {\rm C}} }

\newcommand{\vff}{v_{\rm ff}}

\newcommand{\rhoa}{\rho_{\rm a}}

\newcommand{\vb}{{\bf v}}

\newcommand{\tG}{{\tilde \Gamma}}
\newcommand{\tL}{{\tilde \Lambda}}
\newcommand{\psic}{\psi_{\rm c}}
\newcommand{\psiei}{\psi_{\rm ei}}

\newcommand{\kb}{k_{_{\rm B}}}

\newcommand{\GS}{\ga}
\newcommand{\LS}{\la}

\newcommand{\etal}{et al.\ }
\newcommand{\eg}{e.g.\ }
\newcommand{\ie}{i.e.\ }

\title[Stability analyses of two-temperature radiative shocks]{
Stability analyses of two-temperature radiative shocks:
formulation, eigenfunctions, luminosity response and boundary conditions }
\author[C.~J.~Saxton \& K.~Wu]{
Curtis~J.~Saxton$^{1,2,3}$ \& Kinwah~Wu$^{1,4}$\\ 
$^1$ School of Physics, 
  University~of~Sydney, NSW 2006, Australia  \\ 
$^2$ 
  Research School of Astronomy \& Astrophysics,
  Australian National University, ACT 0200, Australia,
  {\tt saxton@mso.anu.edu.au} \\
$^3$ Department of Theoretical Physics, Faculty of Science,
  Australian National University, ACT 0200, Australia \\
$^4$ Mullard Space Science Laboratory, University College London,
  Holmbury St Mary, Dorking, Surrey RH5 6NT }

\date{Received: }

\begin{document}

\maketitle 

\begin{abstract}  
We present a general formulation for stability analyses 
of radiative shocks with multiple cooling processes, longitudinal and 
transverse perturbations, and unequal electron and ion temperatures. 
Using the accretion shocks of magnetic cataclysmic variables 
as an illustrative application, 
we investigate the shock instabilities 
by examining the eigenfunctions 
of the perturbed  hydrodynamic variables. 
We also investigate 
the effects of varying the condition at the lower boundary 
of the post-shock flow from a zero-velocity fixed wall 
to several alternative type of boundaries involving the perturbed 
hydrodynamic variables, and the variations 
of the emission from the post-shock flow  
under different modes of oscillations.  
We found that the stability properties
for flow with a stationary-wall lower boundary
are not significantly affected by perturbing
the lower boundary condition, and 
they are determined mainly by the energy-transport processes.
Moreover, there is no obvious correlation 
between the amplitude or phase
of the luminosity response and the stability properties of the system.
Stability of the system can, however, be modified
in the presence of transverse perturbation.
The luminosity responses are also altered by transverse perturbation.
\end{abstract}



\section{Introduction} 

The time-dependent properties of radiative shocks have been investigated 
by many researchers in different settings, 
for example the interactions between supernova shocks 
and the interstellar medium, and accretion flow onto compact objects 
(\eg Falle 1975, 1981; Langer, Chanmugam \& Shaviv 1981, 1982; 
Chevalier \&  Imamura 1982; Imamura, Wolff \& Durisen 1984; 
Chanmugam, Langer \& Shaviv 1985; Imamura 1985; Bertschinger 1986; 
Innes, Giddings \& Falle 1987a, b; Gaetz, Edgar \& Chevalier 1988; 
Wolff, Gardner \& Wood 1989; Imamura \& Wolff 1990; 
Houck \& Chevalier 1992; Wu, Chanmugam \& Shaviv 1992;  
T\'{o}th \& Draine 1993; Dgani \& Soker 1994; 
Strickland \& Blodin 1995; Wu \etal 1996; Imamura \etal 1996;
Saxton, Wu \& Pongracic 1997; Saxton \etal 1998; 
Hujeirat \& Papaloizou 1998; Saxton \& Wu 1999). 
Many of these shocks are found to be thermally unstable. 
For instance, a numerical study by Langer \etal\shortcite{langer81} showed 
that the post-shock accretion flow in magnetic cataclysmic variables (mCVs), 
binaries in which a magnetic white dwarf accretes material 
from a red-dwarf companion star, 
suffers thermal instabilities and 
hence fails to attain a steady-state. 
The accretion shock is driven to oscillate, 
giving rise to quasi-periodic oscillations in the optical luminosity. 
A similar conclusion was obtained 
by Chevalier \& Imamura \shortcite{chevalier}, 
using a linear perturbative analysis.       

The stability of the radiative shock depends 
on the energy transport processes. 
In the case of accretion shocks in mCVs, 
thermal bremsstrahlung and cyclotron radiation
are the most important cooling processes
(\eg Lamb \& Masters 1979; King \& Lasota 1979).  
Bremsstrahlung and cyclotron cooling have 
very different temperature dependences, 
and hence influence the stability properties 
of the accretion shock differently. 

In the stability analysis of Chevalier \& Imamura \shortcite{chevalier},
the total cooling effects
were approximated by a single radiative loss term
$\Lambda \propto \rho^a T^b$ depending on temperature $T$ and density $\rho$.
Various choices of the power-law indices
(\eg $a = 0.5$ and $b = 2$ for bremsstrahlung cooling) 
were investigated, 
and they have found that radiative shocks 
with larger $b$ (\ie stronger temperature dependence)    
are more stable against perturbations.  

The individual cooling processes were subsequently considered explicitly 
in the numerical study of accretion onto magnetic white dwarfs 
by Chanmugam \etal \shortcite{cls}.  
An effective cooling term was constructed 
to mimic the effects due to optically thick cyclotron cooling. 
Their study showed that efficient cyclotron cooling stabilises the shock. 
Wu \etal (1992, 1996) further investigated the same system and found 
that in spite of the suppression of shock oscillations 
in the presence of cyclotron cooling, 
the oscillation frequency appears to increase quadratically 
with the magnetic-field strength. 
Moreover, each of the cooling processes, 
bremsstrahlung and cyclotron,
dominates in about half of the phases of an oscillatory cycle, 
allowing the perpetuation of small-amplitude oscillations 
provided that the magnetic field is moderate or weak ($B \LS $10~MG).     

Linear perturbative analyses of accretion shocks with 
bremsstrahlung and cyclotron cooling 
were carried out by Saxton \etal \shortcite{saxton97}. 
A composite cooling function (following Wu, Chanmugam \& Shaviv 1994) 
was considered, 
comprising the sum of a term for bremsstrahlung cooling  
and an effective term for cyclotron cooling.  
The analysis was further extended by Saxton \etal \shortcite{saxton98}, 
in which the cooling function is a sum of terms for bremsstrahlung cooling 
($\Lambda_{\rm br}\propto\rho^2 T^{0.5}$) 
and a second power-law process with a destabilizing influence 
($\Lambda_{_2}\propto\rho^a T^b$ for $b\ge 1$).   
The cases considered included that of a cooling term
($\Lambda_{\rm cy}\propto\rho^{0.15} T^{2.5}$)
effectively approximating the energy loss 
due to cyclotron radiation
in the geometry and flow conditions of the post-shock regions of mCVs. 
It was found that a simple comparison of cooling and oscillation timescales 
is insufficient to understand the instabilities of the shock
under various modes.

When the radiative cooling is fast compared to the electron-ion energy exchange,
the electron and ion temperatures are generally unequal.
Imamura \etal \shortcite{aboasha}
considered bremsstrahlung and Compton cooling,
and a more general perturbation of the shock 
in both longitudinal and transverse directions. 
Their study showed that 
the electron-ion exchange process and
the presence of transverse perturbations 
can destabilize each mode
compared to the purely longitudinal and one-temperature cases.
Saxton \& Wu (1999)
generalised the works of Chevalier \& Imamura (1982), 
Imamura \etal \shortcite{aboasha}, 
Saxton \etal (1997) and Saxton \etal (1998) 
for radiative accretion shocks
by considering multiple cooling processes explicitly,
the two-temperature effects 
and transverse perturbations (as for a corrugated shock).
In the case of mCVs the introduction of two-temperature effects
complicated and broke down
the strictly monotonic stabilisation of modes
with increasing cyclotron efficiency
in one-temperature shocks with bremsstrahlung and cyclotron cooling, 
and the influence of transverse perturbation
was not always able to destabilize oscillatory modes in the presence of both 
bremsstrahlung and cyclotron cooling.
(For a review of stability of accretion shocks, see Wu 2000.)

The present paper expands upon the studies 
of the eigenvalue in Saxton \& Wu (1999)
to examine the amplitudes and phases
of the eigenfunctions 
which describe the response of the post-shock structure 
to perturbations of the shock position. 
As an illustrative case we consider 
the accretion shocks of magnetic white dwarfs. 
Knowing the response of the hydrodynamic-variable profiles
in turn provides information about other characteristics
of thermally unstable shocks, 
such as the responses of post-shock emissions 
due to the shock oscillations.

\section{Formulation}  

\subsection{Hydrodynamics}
\label{'formulation'}

In accretion onto white dwarfs, 
the supersonic flow meets a stand-off shock 
where the inwards falling matter is abruptly decelerated 
to attain a subsonic speed. 
The shock sits above the white-dwarf surface 
at a height $x\subs{}\approx{\frac14}\vff t_{\rm cool}$, 
where the free-fall velocity
is $\vff={(2GM_{\rm wd}/R_{\rm wd})}^{1/2}$, 
the cooling timescale is
$t_{\rm cool}\sim n\sube{}\kb T\subs{}/\Lambda$, 
and $\Lambda$ is a radiative cooling function.
($M_{\rm wd}$ and $R_{\rm wd}$ 
are the white-dwarf mass and radius respectively;
$\kb$ is the Boltzman constant; 
$n\sube{}$ is the electron number density;
and $T\subs{}$ is the shock temperature.) 

The time-dependent mass continuity, momentum and energy equations 
for the post-shock accretion flow are
\begin{equation}
\left({
{\partial\over{\partial t}} + \vb\cdot\nabla
}\right)
\rho
+\rho\left({\nabla\cdot\vb}\right)
= 0 \ ,
\end{equation}
\begin{equation}
\rho\left({
{\partial\over{\partial t}} + \vb\cdot\nabla
}\right)
\vb
+\nabla P
= 0 \ ,
\end{equation}
\begin{equation}
\left({
{\partial\over{\partial t}}
+\vb\cdot\nabla
}\right) P
-\gamma{P\over\rho}
\left({
{\partial\over{\partial t}}
+\vb\cdot\nabla
}\right) \rho
= -(\gamma-1)\Lambda \ ,
\end{equation}
\begin{equation}
\left({
{\partial\over{\partial t}}
+\vb\cdot\nabla
}\right) P\sube{}
-\gamma{{P\sube{}}\over\rho}
\left({
{\partial\over{\partial t}}
+\vb\cdot\nabla
}\right) \rho
= (\gamma-1)(\Gamma-\Lambda) \ ,
\end{equation}
where the total radiative cooling function $\Lambda$ 
and the electron-ion energy exchange term $\Gamma$ 
are local functions of the density $\rho$, total pressure $P$, 
the electron partial pressure $P\sube{}$, and the flow velocity $\vb$. 
The explicit form of the exchange function is
\begin{equation}
\Gamma={{4\sqrt{2\pi}e^4n\sube{}n\subi{}\ln C}\over{m\sube{} c}}
\left[{
{\theta\subi{}-(m\sube{}/m\subi{})\theta\sube{}}
\over{(\theta\sube{}+\theta\subi{})^{3/2}}
}\right] \ , 
\end{equation}
where $n_{\rm i,e}$ are the ion and electron number density, 
and $\theta_{\rm i,e}=\kb T_{\rm i,e}/m_{\rm i,e}c^2$ 
with $T_{\rm i,e}$ being the corresponding temperatures. 
The constants $m_{\rm i,e}$ are the ion and electron masses, 
$e$ is the electron charge, 
$c$ is the speed of light
and $\ln C$ is the Coulomb logarithm (as in \eg Melrose 1986). 
An adiabatic index $\gamma={5/3}$ for an ideal gas is assumed, 
and the equation of state 
$P={{\rho\kb T}/{\mu m_{_{\rm H}}}}$ is considered,
where $m_{_{\rm H}}$ is the mass of hydrogen atom.

The total cooling function is written 
in a form of the bremsstrahlung-cooling term 
and a multiplicative term 
expressing the ratio of the losses 
due to the second cooling process and bremsstrahlung cooling. 
The second process is characterised by its power-law indices 
of density and electron pressure
($\alpha=b-1/2$ and $\beta=3/2-a+b$ 
for a general cooling term $\Lambda_2\propto\rho^a T^b$), 
and by the parameter $\epsilon\subs{}$,
which is the relative efficiency evaluated at the shock. 
(Larger $\epsilon\subs{}$ implies a more efficient second process.) 
\begin{equation}
\Lambda \equiv \Lambda_{_{\rm br}}+\Lambda_{_2}
=\Lambda_{\rm br}
\left[
1+\epsilon\subs{}f(\tau_0,\pi\sube{})
\right] \ , 
\label{'eq.lambda.total'}
\end{equation}
and a function is defined 
to relate the primary and secondary cooling processes:
\begin{equation}
f(\tau_0,\pi\sube{})\equiv{{4^{\alpha+\beta}}\over{3^\alpha}}
\left({
{1+\sigma\subs{}}\over{\sigma\subs{}}
}\right)^{\alpha}
\pi\sube{}^\alpha \tau_0^\beta
=\left({P\sube{} \over{P\sube{,s}}}\right)^\alpha
\left({\rho\over{\rho_{\rm s}}}\right)^{-\beta}
\ , 
\label{'eq.define.f'}
\end{equation}
where $P\sube{,s}$ and $\rho\subs{}$
are the shock values of electron pressure and density.
The dimensionless parameter
$\sigma\subs{}\equiv (P\sube{}/P\subi{})\subs{}$
is the ratio of electron and ion pressures at the shock.
The bremsstrahlung cooling term
$\Lambda_{\rm br}={\mathcal C}\rho^2\left({P\sube{}/\rho}\right)^{1/2}$,
where the constant is 
${\mathcal C} = (2\pi k_{\rm _B}/3 m_{\rm e})^{1/2} 
 (2^5\pi e^6/3 h m_{\rm e} c^3)
 (\mu/ k_{\rm _B} m_{\rm p}^3)^{1/2} g_{\rm _B}$,
with $m_{\rm p}$ the proton mass, 
$h$ the Planck constant, 
$\mu$ the mean molecular weight 
and $g_{\rm _B}\approx 1$ the Gaunt factor 
(see Rybicki \& Lightman 1979).
For completely ionised hydrogen plasma $\mu=0.5$ 
and the constant has a value
${\mathcal C}\approx3.9\times10^{16}$ in c.g.s.\ units.   

\subsection{Perturbation }
\label{'perturbation'}

A first-order perturbation is considered 
for the shock position $x\subs{}$ and velocity $v\subs{}$: 
\begin{equation}
v\subs{}=v\subs{1} e^{iky+\omega t} \ , 
\end{equation}
\begin{equation}
x\subs{}=x\subs{0}+x\subs{1} e^{iky+\omega t}\ , 
\end{equation}
where $\omega$ is the frequency, 
and $k$ is the transverse wavenumber of perturbation 
in the $y$ (transverse) direction. 
The shock is at rest in the stationary solution, 
$v\subs{0}=0$, and the perturbed motion of the shock has 
$v\subs{1}=x\subs{1}\omega$. 
The dimensionless frequency and transverse wavenumber are 
\begin{eqnarray}
\kappa&=&x\subs{0} k\ ,
\\
\delta&=&{{x\subs{0}}\over{\vff}}\omega \ . 
\end{eqnarray}
The eigenfrequencies are complex, $\delta=\dR+i\dI$, 
with $\dI$ being the dimensionless frequencies of the oscillations, 
and $\dR$ the stability term. 
When $\dR$ is positive the perturbation grows; 
when $\dR$ is negative the perturbation is damped.

The post-shock position coordinate is labelled 
by $\xi\equiv x/x\subs{}$, 
which is $\xi=1$ at the shock and $\xi=0$ at the white-dwarf surface. 
The size of the perturbation is parameterised by
$\varepsilon\equiv v\subs{1}/\vff=\delta x\subs{1}/x\subs{0}$.
The other scales to be eliminated are $x\subs{0}$ 
the stationary-state shock height 
and $\rhoa$ the mass density of the pre-shock accretion flow.
The hydrodynamic variables are expressed as
\begin{eqnarray}
\rho(\xi,y,t)&=&\rhoa\cdot
\zeta_0(\xi) \left({1+\varepsilon\lambda_\zeta(\xi) e^{iky+\omega t}
   }\right)\ , 
\\
\vb(\xi,y,t)&=&-\vff\cdot
\tau_0(\xi)
\left({
\left({1+\varepsilon\lambda_\tau(\xi) e^{iky+\omega t}}\right),
\varepsilon\lambda_y(\xi) e^{iky+\omega t}
}\right) \ , 
\\
P(\xi,y,t)&=&\rhoa\vff^2\cdot
\pi_0(\xi) \left({1+\varepsilon\lambda_\pi(\xi) e^{iky+\omega t}}\right) \ , 
\\
P\sube{}(\xi,y,t)&=&\rhoa\vff^2\cdot
\pi\sube{}(\xi) \left({1+\varepsilon\lambda\sube{}(\xi) 
   e^{iky+\omega t}}\right) \ .
\end{eqnarray}
where $\zeta_0$, $\tau_0$, $\pi_0$ and $\pi\sube{}$ 
are dimensionless density, velocity, total pressure and electron pressure 
in the stationary solution; 
and $\lambda_\zeta$, $\lambda_\tau$, $\lambda_y$, $\lambda_\pi$ 
and $\lambda\sube{}$ are complex functions 
representing the response of the downstream structure 
to perturbation of the shock height. 
These five functions describe perturbations 
of the density, longitudinal velocity, transverse velocity 
the total pressure and the electron pressure respectively.  

Separating the time-independent terms from the hydrodynamic 
equations yields two algebraic equations and two differential equations 
for the stationary case:
\begin{equation}
\zeta_0=1/\tau_0 \ , 
\label{'eq.unitless.mass'}
\end{equation} 
\begin{equation}
\pi_0=1-\tau_0 \ , 
\label{'eq.unitless.pressure'}
\end{equation}
\begin{equation}
{{d\xi}\over{d\tau_0}}
={{\gamma\pi_0-\tau_0}\over{\tL}} \ , 
\label{'eq.stationary.velocity'}
\end{equation}
\begin{equation}
{{d\pi\sube{}}\over{d\tau_0}}
={1\over{\tau_0}}
\left[{
\left({
1-{\tG\over\tL}
}\right)
(\gamma\pi_0-\tau_0)-\gamma\pi\sube{}
}\right]\ .
\label{'eq.stationary.pressure'}
\end{equation}
The electron-ion energy exchange and cooling processes are described by
appropriate dimensionless forms:
\begin{equation}
\tG=(\gamma-1)(\rhoa\vff^3/x\subs{0})^{-1}\Gamma
=(\gamma-1)\psic\psiei
{ {1-\tau_0-2\pi\sube{}} \over { \sqrt{\tau_0^5\pi\sube{}^3} } }\ ,
\label{'eq.bigG'}
\end{equation}
\begin{equation}
\tL=(\gamma-1) 
(\rhoa\vff^3/x\subs{0})^{-1}
\Lambda
=(\gamma-1)\psic\sqrt{{\pi\sube{}}\over{\tau_0^3}}
\left[{
1+\epsilon\subs{}f\left({\tau_0,\pi\sube{}}\right)
}\right] \  ,
\label{'eq.bigL'}  
\end{equation}
where the constant $\psi_{\rm c}$ is determined by normalisation of the 
integrated stationary solution and the parameter $\psiei$ is a ratio 
between the electron-ion energy exchange and radiative cooling timescales,
as described in Imamura \etal (1996) and Saxton \& Wu (1999). The physical 
values of $\psi_{\rm c}$ and $\psiei$ are given in Saxton \& Wu (1999) and 
Saxton (1999).

The first-order perturbation is determined by the matrix differential equation:
\begin{equation}
{d\over{d\tau_0}}
{
\left[
\begin{array}{c}
\lambda_\zeta\\
\lambda_\tau\\
\lambda_y\\
\lambda_\pi\\
\lambda\sube{}
\end{array}
\right]
}
=
{1\over\tL}
{
\left[
\begin{array}{ccccc}
1&-1&0&{1/{\tau_0}}&0\\
{-{\gamma\pi_0}/{\tau_0}}&1&0&-{1/{\tau_0}}&0\\
0&0&-{{(\gamma\pi_0-\tau_0)}/{\tau_0}}&0&0\\
\gamma&-\gamma&0&{1/{\pi_0}}&0\\
\gamma&-\gamma&0&{\gamma/{\tau_0}}&
 -{{(\gamma\pi_0-\tau_0)}/{\tau_0\pi\sube{}}} 
\end{array}
\right]
}
{
\left[
\begin{array}{c}
F_1\\
F_2\\
F_3\\
F_4\\
F_5\\
\end{array}
\right] \ , 
}
\label{'eq.2d2t.matrix.2'}
\end{equation}
where the $F$ functions,
which are comprised of terms that do not include derivatives of the $\lambda$ 
variables,
are given by
\begin{eqnarray}
F_1(\tau_0,\pi\sube{},\xi)
&=&
-\xi(\ln \tau_0)'
-\delta\lambda_\zeta
+i\kappa \tau_0\lambda_y \ , 
\\
F_2(\tau_0,\pi\sube{},\xi)
&=&
-(\delta-\tau_0')\lambda_\tau
+\xi(\ln \tau_0)'
+\tau_0'\lambda_\zeta
-\tau_0'(\lambda_\pi-\lambda_\tau) \ , 
\\
F_3(\tau_0,\pi\sube{},\xi)
&=&
-(\delta-\tau_0')\lambda_y
+i\kappa(1-\tau_0)\lambda_\pi
+i\kappa \tau_0'\xi/\delta \ ,    
\\
F_4(\tau_0,\pi\sube{},\xi)
&=&
-\pi_0\delta(\lambda_\pi -\gamma\lambda_\zeta)
-\tL\left[{
 {3\over2}g_2(\tau_0,\pi\sube{})\lambda_\zeta
+{1\over2}g_1(\tau_0,\pi\sube{})\lambda\sube{}
-\lambda_\tau-\lambda_\pi+{1\over\delta}-{\xi\over{\tau_0}}
}\right] \ , 
\\
F_5(\tau_0,\pi\sube{},\xi)
&=&
-\pi\sube{}\delta(\lambda\sube{} -\gamma\lambda_\zeta)
-\tL\left[{
 {3\over2}g_2(\tau_0,\pi\sube{})\lambda_\zeta
+{1\over2}g_1(\tau_0,\pi\sube{})\lambda\sube{}
-\lambda_\tau-\lambda\sube{}+{1\over\delta}-{\xi\over{\tau_0}}
}\right]
\nonumber \\
& &+\tG \left[{
{5\over2}\lambda_\zeta -{3\over2}\lambda\sube{}
+{{\pi_0\lambda_\pi -2\pi\sube{}\lambda\sube{} }\over{\pi_0-2\pi\sube{}}}
-\lambda_\tau-\lambda\sube{}+{1\over\delta}-{\xi\over{\tau_0}}
}\right] \ , 
\end{eqnarray}
where the primed quantities are derivatives in terms of $\xi$.
The functions $g_1(\tau_0,\pi\sube{})$ and $g_2(\tau_0,\pi\sube{})$
are defined as
\begin{equation}
g_1(\tau_0,\pi\sube{}) = 1
+{{2{\epsilon_{\rm s}}\alpha f(\tau_0,\pi\sube{})}
\over{1+\epsilon_{\rm s}f(\tau_0,\pi\sube{})}} \ ,
\end{equation}
\begin{equation}
g_2(\tau_0,\pi\sube{}) = 1-{2\over 3} \left[{
{{{\epsilon_{\rm s}}\beta f(\tau_0,\pi\sube{})}
\over{1+\epsilon_{\rm s}f(\tau_0,\pi\sube{})}}
 }\right]  \ .
\end{equation}
The complex matrix differential equation (\ref{'eq.2d2t.matrix.2'})
can be decomposed
into ten first-order real differential equations in terms of
the functions of the stationary solution $\xi(\tau_0)$ and 
$\pi\sube{}(\tau_0)$, and the real and imaginary parts of each of the 
perturbed variables.
It can be shown that equation (\ref{'eq.2d2t.matrix.2'})
is the general description and that the more restricted formulations
in Chevalier \& Imamura \shortcite{chevalier},
Saxton \etal (1997, 1998)
can be recovered from it under specific assumptions
(Appendix~\ref{'appendix.reductions'}).

\subsection{Cooling functions}  

Optically-thick cyclotron cooling 
depends not only on the local properties of the hydrodynamic variables 
but also on the radiation field.
A self-consistent description of cyclotron loss
usually requires solving the equations for
radiative transfer and the hydrodynamics simultaneously. 
The particular geometry and physical conditions
of the magnetically channelled accretion flow in mCVs, however, 
permit a simplification (see \eg Cropper \etal 1999).
A functional fit involving the density and temperature dependences
of the cut-off frequency yields an approximate cooling term
$\Lambda_{\rm cy}\propto\rho^{0.15}T\sube{}^{2.5}$
(see Langer \etal 1982; Wu \etal 1994).  

In our analysis the effective cyclotron cooling term
has power-law indices $(\alpha,\beta)=(2.0,3.85)$, and 
the efficiency parameter $\epsilon\subs{}$ (Wu 1994) 
depends upon the temperature, density, magnetic field, and  
geometry of the emission region.
In the limit of a one-temperature accretion flow
the efficiency of electron-ion energy exchange $\psiei$ is large,
and the ratio of pressures $\sigma\subs{}$ tends to unity.
For two-temperature shocks, $\sigma\subs{}$
is determined by the electrons carrying energy
into the region above the ion shock,
which is a complication beyond the scope of this paper. 
Following Imamura \etal (1996) and Saxton \& Wu (1999),
we treat $\sigma\subs{}$ as a parameter. 

\subsection{Stationary structure} 

\begin{figure}
\begin{center}
\epsfxsize=17cm
\epsfbox{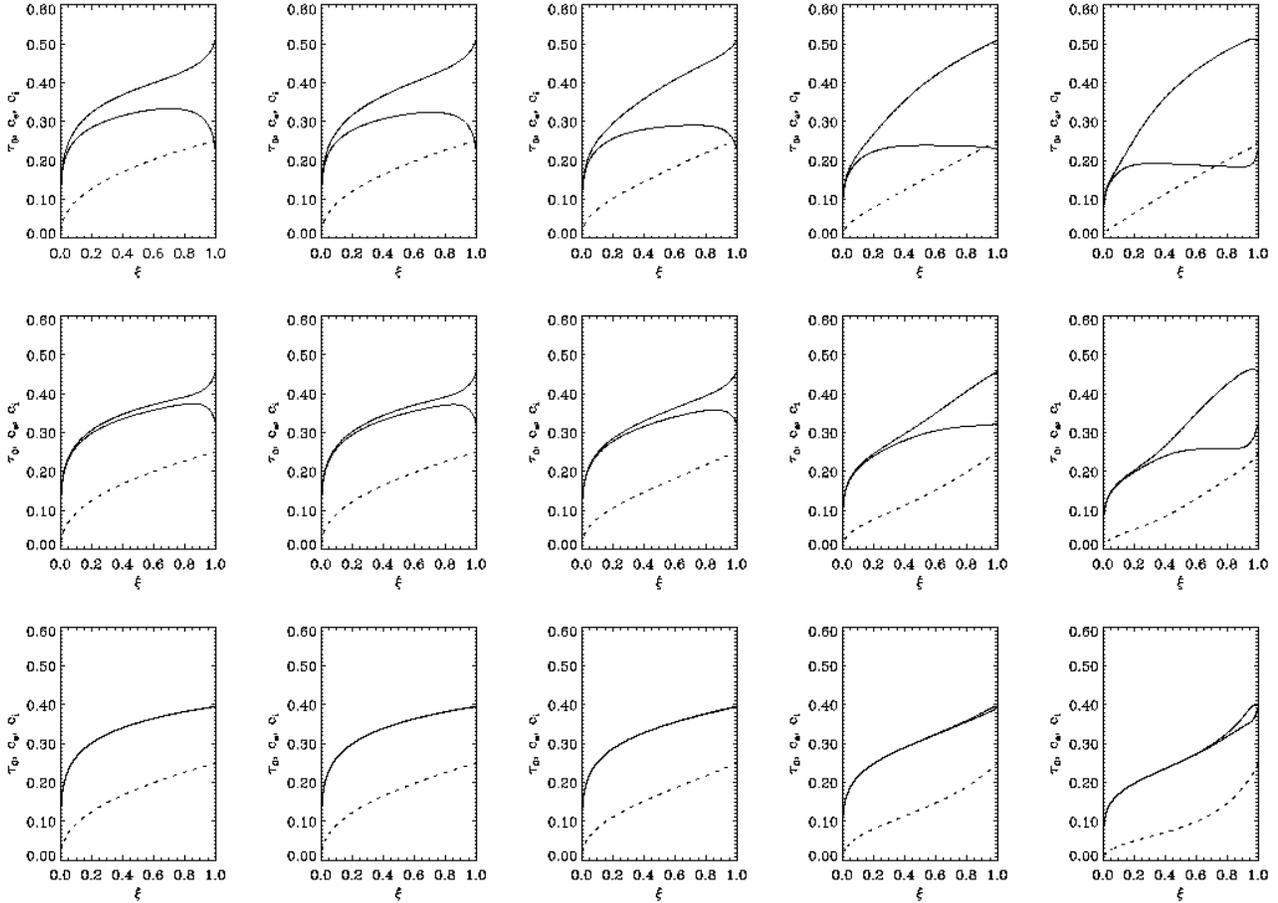}
\end{center}
\scriptsize
\caption{
Statinary structures of the post-shock flow,
with $\tau_0$ the flow velocity normalised to $\vff$
(dashed lines) as functions of the normalised position $\xi$.
The electron and ion sound speeds are represented by
the upper and lower solid curves respectively.
The upper, middle and lower rows represent parameter choices
$(\sigma\subs{},\psiei)=(0.2,0.1)$,
$(\sigma\subs{},\psiei)=(0.5,0.5)$ and,
$(\sigma\subs{},\psiei)=(1,10)$ respectively.
The columns represent values
$\epsilon\subs{}=0,0.1,1,10,100$ from left to right.
}
\label{'ssa1b2c4.ps'}
\end{figure}

We assume the stationary-wall condition, 
which requires a zero terminal velocity at the lower boundary. 
The stationary solution can be obtained by direct integration, 
after substituting equations (16) and (17) into equations (18) and (19). 
In Figure ~\ref{'ssa1b2c4.ps'} we show the stationary velocity structures
($\tau_0$)
of the post-shock region for various choices of the system parameters.
The electron and ion sound speeds ($c\sube{}$, $c\subi{}$)
are also plotted on the same scale.
The density is related to the velocity by
$\rho_0 = \rhoa/\tau_0$
and the respective temperatures are proportional to the squared sound speeds.
Two-temperature effects are more significant in cases
when cyclotron cooling is efficient
(\eg $\epsilon\subs{}=100$),
and when the electron-ion exchange
is inefficient (\ie small $\psiei$).
When two-temperature effects are unimportant,
we recover the velocity, density and temperature structure
of the one-temperature calculations (Wu 1994; Chevalier \& Imamura 1982).
A detailed discussion on the two-temperature stationary structures
of the post-shock flows 
and their emission will be presented elsewhere
(Saxton, Wu \& Cropper, in preparation).

\subsection{Time-dependent solutions}  

For the time-dependent solution, 
the set of differential equations for the perturbed variables
are integrated numerically from the shock down to the lower boundary 
for trial values of the complex eigenfrequency $\delta$. 
The variables at the shock are determined by the shock-jump conditions:  
$\lambda_\zeta=0$, $\lambda_\tau=-3$, $\lambda_y=3i\kappa/\delta$, and 
$\lambda_\pi=\lambda\sube{}=2$. 
The condition at the lower boundary is less well defined.  
The stationary-wall condition only requires a zero terminal velocity 
at the bottom for the stationary solution.  
A lower boundary condition for the perturbed variables, 
such as the requirement that the flow should stagnate ($\lambda_\tau=0$), 
the total pressure be constant ($\lambda_\pi=0$), 
or some other condition relating density, pressure and longitudinal 
velocity (\eg $\lambda_\zeta+\lambda_\tau=0$), may be chosen.

The time-dependent solution
can be expressed as a linear combination of oscillations of 
different eigenmodes. For a hydrodynamic variable $X$, we have  
\begin{equation}
X(\xi,t)=X_0(\xi)\sum_n
\left[{
a_n \lambda_{x,n}(\xi)\exp\left({ {{\vff}\over{x\subs{0}}}\delta_n t}\right)
}\right] \ ,
\end{equation} 
where $X_0(\xi)$ is the stationary solution, $a_n$ is the relative 
strength of the mode $n$, $\delta_n$ the eigenfrequency and $\lambda_{x,n}$ the 
eigenfunction.
While the complex-$\delta$ eigenvalues provide information about 
the global stability properties and oscillating frequencies of the modes, 
the $\lambda$-eigenfunctions describe the more local dynamical properties.
The absolute value of the $\lambda$-function determines the local 
oscillation amplitude of the mode, and the phase indicates whether the 
oscillation of the hydrodynamic variables lags or leads in one region with 
respect to another. 

\section{Eigenvalues}
\label{'eigenvalues'}

The eigenvalues for two-temperature oscillating shocks 
with a ``perfect'' stationary-wall lower boundary condition, 
\ie $\tau = \lambda_\tau = 0$ (see Saxton 2001),  
have been discussed
in our previous paper (Saxton \& Wu 1999, see also Imamura \etal 1996).
More restricted studies on the one-temperature radiative shocks 
were presented in Saxton \etal (1997) and Saxton \& Wu (1998)
(see also Chevalier \& Imamura 1982).
Particular results of these studies include  
\begin{enumerate} 
\item 
In the one-temperature case, the frequencies are quantised like modes 
of a pipe open at one end, $\dI\approx\dIO(n-1/2)+\dC$ with a small 
correction $\dC$. When two-temperature effects are strong 
the ``stationary-wall'' condition at the lower boundary loses importance, 
and the frequency quantisation becomes more like that of a 
doubly-open pipe, i.e.\ $\dI\approx\dIO n$. 
\item 
The frequency spacing $\dIO$ decreases as the efficiency 
of cyclotron cooling ($\epsilon\subs{}$)  
increases, but tends to increase when the efficiency of  
electron-ion energy exchange ($\psiei$) decreases.  
\item
Increasing $\epsilon\subs{}$ generally stabilises 
the modes, but when two-temperature effects are extreme there are 
situations in which an increase of $\epsilon\subs{}$ destablises 
some modes. 
\item 
In presence of transverse perturbation, there are maxima of instability 
at certain values of transverse wavenumber $\kappa$ for each longitudinal 
mode. However, there are values of  
($\sigma_{\rm s}, \psiei, \epsilon_{\rm s}, \alpha, \beta$)
for which some longitudinal modes are stable at all $\kappa$ ranges. 
\item 
For a given longitudinal mode, the shocks are generally stable 
against transverse perturbation of sufficiently large $\kappa$.  
\item
Two-temperature effects affect the stability of cyclotron-cooling dominated 
shocks ($\epsilon\subs{} \gg 1$), but have less influence when 
bremsstrahlung cooling dominates ($\epsilon\subs{} \approx 0$).   
\end{enumerate}

\section{Eigenfunctions}
\label{'chapter.efunctions'}

\subsection{Amplitude-profile}  

In one-temperature accretion shocks, the non-trivial eigenfunctions are  
$\lambda_\tau$, $\lambda_\pi$, and $\lambda_\zeta$, 
corresponding to total pressure, longitudinal velocity and density. 
In the two-temperature shocks, 
the non-trivial eigenfunctions also include 
that of the electron pressure $\lambda_{\rm e}$
When there is a transverse perturbation,
a transverse perturbed velocity eigenfunction $\lambda_y$
becomes important also (see \S\ref{'transverse'}).

Our choice of boundary conditions requires that
the perturbed longitudinal velocity $\lambda_\tau$ 
has a fixed value of $-3$ at the shock, 
and a stagnant-flow at the white-dwarf surface 
implies a zero $\lambda_\tau$ at $\xi =0$.   
(Here and hereafter in this section, unless specified, 
the stagnant-flow condition is assumed.)  
The amplitude profiles generally have local minima and maxima 
(see Appendix B), 
which are analogous to the nodes and antinodes in the oscillations of a pipe. 
The number of nodes and antinodes depend on the harmonic number $n$, 
and their positions are determined 
by the radiative-transport processes and the system geometry.  
Unlike the nodes of an ideal pipe, 
the amplitude minima of oscillating shocks do not always go to zero, 
and the amplitude maxima do not have the same values.
Moreover, the nodal features of one hydrodynamic variable 
may not coincide with those of another hydrodynamic variable.   
  
\begin{figure}
\begin{center}
$
\epsfxsize=14cm
\epsfbox{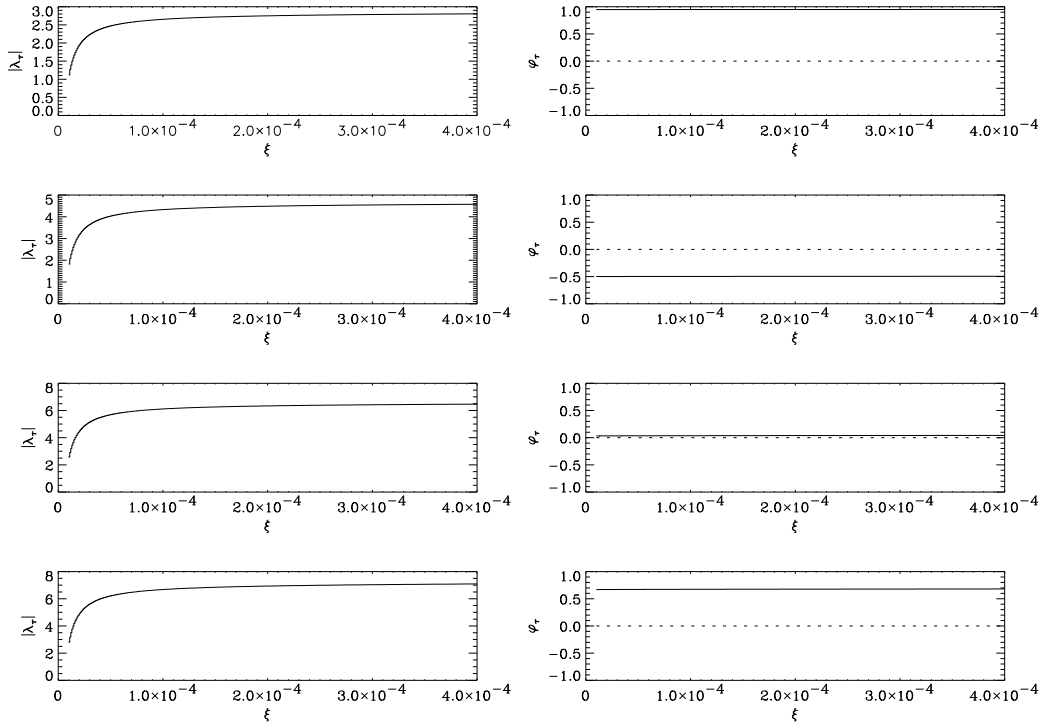}
$
\scriptsize
\caption{The amplified view $|\lambda_\tau|$ profiles (left column) and the 
 phases $\varphi_\tau$ (right column) near the lower boundary for the case 
 with $(\sigma\subs{},\psiei,\epsilon\subs{}) = (0.5,0.5,1.0)$ and 
 a stagnant-flow boundary condition ($\lambda_\tau =0$ at $\xi =0$). The 
 $|\lambda_\tau|$ profiles and the phases of the modes $n = 1$, 2, 3 and 4 
 are shown in rows from top to bottom.
 The $|\lambda_\tau|$ profiles are similar for the four modes,
 but the phases are very different. }
\label{'zoom'}
\end{center}
\end{figure}
 
\subsubsection{$|\lambda_\tau|$-profile} 

The $|\lambda_\tau|$ profiles 
do not show very distinctive nodes and antinodes. 
The amplitude
$|\lambda_\tau|$ increases very rapidly with $\xi$   
from a zero value at the lower boundary, 
reaches a plateau or a peak at larger $\xi$
and then declines with a gradual tail towards the shock boundary value.  
The rapid rise of $|\lambda_\tau|$ occurs only in a small region 
where $\xi\LS 10^{-4}$, 
and the size of the region seems
to be independent of the harmonic number $n$ 
(see Figure~\ref{'zoom'}).
The profiles for $\epsilon\subs{}$ = 0 and 1 are similar, 
with the plateau value of $|\lambda_\tau|$
of the fundamental mode $\approx 3$
(\eg Figure\ \ref{'b2ef1'}).   
For higher harmonics, the plateau values of $|\lambda_\tau|$ are larger. 
The maxima (antinodes) 
generally become more visible when $n$ increases 
(e.g.\ Figures~\ref{'b2ef2'},\ \ref{'b2ef3'},\ \ref{'b2ef4'}).  

For the cases with large $\epsilon\subs{}$, 
$|\lambda_\tau|$ has a strong peak near the lower boundary.   
The peak heights tends to increase with $n$. 
It is worth noting that 
in the extreme of small $\psiei$, large $\sigma\subs{}$ and 
large $\epsilon\subs{}$ 
(strong two-temperature effects 
and equal temperature for electrons and ions at the shock), 
a small peak in $|\lambda_\tau|$ is also present near the shock   
(see Appendix D.2.4 in Saxton 1999).  

\subsubsection{$|\lambda_\pi|$-profile}  

The boundary value of $|\lambda_\pi|$ is 2 at the shock, 
but it is not defined at the white-dwarf surface 
for the stagnant-flow condition. 
The $|\lambda_\pi|$ profiles   
do not show distinctive nodes and antinodes 
for $n =1$ and 2.  
The nodes and antinodes, however, become more visible 
when $n$ increases and $\psiei$ becomes small
(Figure~\ref{'b2ef3'}-\ref{'b2ef6'}).  

\subsubsection{$|\lambda\sube{}|$-profile} 

The electron pressure eigenfunctions $\lambda\sube{}$ show 
nodes and antinodes. 
The sharpness of the nodes 
depends on the strength of the electron-ion energy exchange, 
$\psiei$. 
When the electron-ion energy exchange is efficient, 
the nodes are weak.
When the electron-ion energy exchange is inefficient,
(\eg $\psiei=0.1$),
the amplitude minima are narrow in $\xi$
and deep in terms of amplitude.
(See \eg the $\epsilon\subs{}=100$ curves of 
Figures~\ref{'a1ef1'}-\ref{'a1ef4'}). 

For two-temperature shocks, it is the electron pressure $\pi\sube{}$
rather than the total pressure $\pi_0$
that appears in the cooling terms and electron-ion exchange term.
The electron pressure eigenfunction $\lambda\sube{}$
therefore adopts the role that is played by the total pressure eigenfunction 
$\lambda_\pi$ in the one-temperature case. 
When the electron-ion energy exchange is weak (\ie small $\psiei$),
the $\lambda_\pi$ and $\lambda\sube{}$ eigenfunctions
are very different.
The different modes have
sharp and strikingly different features in $|\lambda\sube{}|$,
whereas the total pressure eigenfunction
is relatively featureless and smoothly varying.
When the energy exchange between electrons and ions
is efficient (\ie large $\psiei$),
the disparties between their pressures becomes small
throughout most of the post-shock region.
When the system approaches the one-temperature condition,
and the total pressure follows the behaviour of the electron pressure,
and the variations of the electron pressure 
and total pressure eigenfunctions,
$\lambda\sube{}$ and $\lambda_\pi$,
are comparable in amplitude over all $\xi$. 

Because of the lower boundary conditions that we assume, 
the electron and ion temperatures are zero and the 
electron and ion pressures always become equal at $\xi=0$.  
The relative oscillation of the electron pressure
must equal the relative oscillation of the total pressure
at the lower boundary.
I.e.
$\lambda\sube{}(\xi)\rightarrow\lambda_\pi(\xi)$
as $\xi\rightarrow0$. However, there is no condition to determine the particular
value at which these quantities must meet at $\xi=0$,
for any given mode under particular conditions.

\subsubsection{$|\lambda_\zeta|$-profile} 
\label{'lambda.zeta.profile'}

The density eigenfunctions, $\lambda_\zeta$,
have strong and distinct nodes and antinodes.
For each mode the positions and magnitudes of the node-antinode-node sections
are affected by the efficiency of cyclotron cooling $\epsilon\subs{}$.
When $\epsilon\subs{}$ is small ($\LS 1$),
the nodes are nearly evenly spaced
and the antinodes near the shock tend to have lower amplitudes
than those nearer to the white-dwarf surface.
When $\epsilon\subs{}$ is larger,
the nodes near the shock are more widely spaced
and the antinodes near the shock tend to be higher in amplitude.
The innermost antinode in the region $\xi\LS0.1$
becomes higher as $\epsilon\subs{}$ increases
through the cases of $\epsilon\subs{}=0,1,100$ studied here.  

Because of the sharp node and antinode features appearing in the profiles of
$|\lambda_\zeta|$ (and $|\lambda\sube{}|$)
we suspect that the density and electron pressure may be the quantities
which determine the oscillatory properties of the shock.
These hydrodynamic variables appear explicitly
in the functions for the cooling and electron-ion energy exchange processes.
Other hydrodynamic variables
(\ie the flow velocities and the total pressure)
have amplitudes varying less rapidly in $\xi$,
probably indicating a less active involvement
in determining the oscillatory behaviour of the shock.

For the studied cases of $(\sigma\subs{},\psiei,\epsilon\subs{})$,
the modes divide into two classes based on the qualitative features
of their amplitude profiles.
Profiles of the fundamental and first overtone ($n=1,2$)
are more alike than the profiles of higher modes.
The $n>2$ modes have $n-2$ nodes
at intermediate $\xi$ positions,
whereas the $n=1,2$ profiles
are slowly-varying in $\xi$.
The actual number of density nodes in each profile is $n-1$,
including the shock (where $\lambda_\zeta=0$ as a boundary condition).
An additional shallow, and usually indistinct, node
occurs near the fixed wall boundary (small $\xi$)
for some cases of large $\epsilon\subs{}$.
When two-temperature effects are strong,
$|\lambda\sube{}|$ also has sharp nodes for $n>2$ modes.

\subsubsection{nodes, antinodes and sound speeds}
 
The dimensionless electron and ion sound speeds are given by
\begin{equation}
c\sube{}=\sqrt{\gamma\tau_0\pi\sube{}}\ ,
\end{equation}
and
\begin{equation}
c\subi{}=\sqrt{\gamma\tau_0(1-\tau_0-\pi\sube{})}
\end{equation}
respectively, and the mean sound speed is
\begin{equation}
c\subs{}=\sqrt{\gamma\tau_0(1-\tau_0)}\ .
\end{equation}
The sound speeds determine the local dynamical time and length scales,
which influence the mode spacing.
In regions of the post-shock flow where the sound propagation is fast,
nodes of the eigenfunctions tend to be further apart.

The sound speeds generally decrease
from the shock down to the lower boundary
(see Figure~\ref{'ssa1b2c4.ps'}).
When the gradients of the sound speed have less dramatic change
(\eg the case $\epsilon\subs{}=0$
as in leftmost column of Figure~\ref{'ssa1b2c4.ps'})
the nodes of the eigenfunctions are more evenly spaced.
When $\epsilon\subs{}$ is large the sound speeds
are proportionately larger in the upper region of the flow,
resulting in wider node spacing in the high-$\xi$ region
(contrast the different $\epsilon\subs{}$ curves of $|\lambda_\zeta|$
in Figure~\ref{'a1ef4'}).

The local amplitudes of antinodes of the density eigenfunctions
are greatest in regions near the shock for large $\epsilon\subs{}$.
This is counterintuitive because previous studies
(\eg numerical simulations by Chanmugam \etal 1985, and Wu \etal 1992, 1996)
show that oscillations are {\em globally} suppressed by cyclotron cooling.
This global result would lead us to expect lower local amplitudes
in the cyclotron-dominated region,
and reduced amplitudes in systems with greater $\epsilon\subs{}$.
We believe that there is a connection between
the increased antinode amplitudes
and the increased node spacing in the cyclotron-dominated region
when $\epsilon\subs{}$ is great.
The amplitude enhancement
is not related to two-temperature effects,
because it occurs in the one-temperature extreme
(large $\psiei$ and $\sigma\subs{}\rightarrow 1$)
as well as the general two-temperature cases.
The behaviours of the nodes and antinodes (of the density eigenfunction)
probably depend more sensitively on the overall form of the stationary solution
than on the energy exchange processes present in each region of the flow.

\subsection{Phase}

The phases of the various perturbed hydrodynamic variables
are given the labels
$\varphi_\zeta$, $\varphi_\tau$, $\varphi_y$, $\varphi_\pi$ and
$\varphi\sube{}$
for density, longitudinal velocity, transverse velocity,
total pressure and electron pressure respectively.
The zero values correspond to oscillations in phase
with that of the shock height.

\subsubsection{Boundary conditions and phase}

At the shock ($\xi=1$)
there are boundary conditions on all of the $\lambda$ variables.
The only phase which is not explicitly determined is
$\varphi_\zeta$.
Our calculations, however, show that the density oscillation
is approximately in quarter-phase
ahead of or behind the shock height oscillation,
except when $\epsilon\subs{}\sim 100$
(see Figures~\ref{'a1ef2'}, \ref{'b2ef2'}). 

There are fewer boundary conditions on the hydrodynamic variables
at the bottom of the post-shock region.
For the stationary-wall condition,
only the longitudinal velocity variable is constrained,
$|\lambda_\tau|\rightarrow0$ as $\xi\rightarrow0$.
The phase $\varphi_\tau$ is not determined.
The other perturbed variables have no definite lower boundary values,
and their phases depend completely on
the energy transport processes.
 
The phases $\varphi\sube{}$ and $\varphi_\pi$ are not completely independent.
Since the electrons and ions have reached the same temperature
at the lower boundary,
the oscillations of electron pressure
and the oscillations of total pressure
are identical at $\xi=0$.
The variables $\lambda_\pi$ and $\lambda\sube{}$
have the same values at the shock and the base
but different values in between.
 
\subsubsection{Interpretation of $d\varphi/d\xi$}
\label{'subsection.winding'}

If the phase increases with $\xi$,
we consider the oscillation to be
propagating downwards from the shock towards the lower boundary.
(This is because the oscillations in the high-$\xi$ region lead
the oscillation in the lower-$\xi$ regions.)
If the phase decreases with $\xi$
then the oscillation is considered to be propagating upwards.
The upward and downward propagation
are labelled positive $[+]$ and negative $[-]$ respectively.
If a complex $\lambda$-function is viewed along the $\xi$ axis,
with the real and imaginary $\lambda$ parts horizontal and vertical,
then the complex function appears to wind about the origin
in either a clockwise manner for positive propagation,
or an anticlockwise manner for negative propagation.

The phase functions can be regarded as having
an overall winding between $\xi=1$ and $\xi=0$,
with local phase glitches.
For given $(\sigma\subs{},\psiei,\epsilon\subs{})$,
the total winding of each perturbed variable across the interval $0<\xi<1$
is a function of harmonic number $n$.
The number of winding cycles is not necessarily an integer.
Generally the number of cycles increases with $n$.
 
\subsubsection{Illustrative cases}

For the case of $(\sigma\subs{},\psiei)=(0.5,0.5)$
(see Figures~\ref{'b2ef1'}-\ref{'b2ef6'}),
the pressure phases $\varphi_\pi$ and $\varphi\sube{}$
wind positively for the first six modes when $\epsilon\subs{}=0,1$.
When $\epsilon\subs{}=100$,
$\varphi\sube{}$ may wind negatively in some regions.
For $n=1$ the winding is negative throughout the entire post-shock region.
For $n>1$ the winding is negative near the lower boundary ($\xi\LS0.05$),
but it can be positive or negative elsewhere.

The phase $\varphi_\zeta$
winds in a monotonically negative sense below $\xi\LS0.98$
for the fundamental and first overtone.
For the higher harmonics, the winding of $\varphi_\zeta$ undergoes
one or more abrupt jumps or reversals in narrow ranges of $\xi$.
There are $n-2$ phase jumps in $\varphi_\zeta$,
and their positions correspond to the distinct density-amplitude nodes.
For the harmonics with $n>2$,
$\varphi_\zeta$ begins with a negative winding near the shock ($\xi=1$)
and remains negatively winding throughout most of the flow,
except at the jumps.
The jumps can be either positive or negative.
Descending from the shock, the first jump is positive,
and for many choices of $(\sigma\subs,\psiei,\epsilon\subs{})$ and $n$
the second jump is negative.
The signs of the further jumps depend on the mode and
$(\sigma\subs,\psiei,\epsilon\subs{})$.
For $(\sigma\subs{},\psiei,\epsilon\subs{})=(0.5,0.5,1)$
the sequence of jumps is $[+,-,-]$ when $n=5$
($\epsilon\subs{}=0,1$ in Figure~\ref{'b2ef5'})
and $[+,-,+,-]$ when $n=6$
($\epsilon\subs{}=0,1$ in Figure~\ref{'b2ef6'}).
In cases where the cyclotron cooling dominates
the positive jump in $\varphi_\zeta$ occurs very close to the shock
and the subsequent jumps are all negative and indistinct.

\subsubsection{Phase ``discontinuities''}

\begin{figure}
\begin{center}
$
\epsfxsize=14cm
\epsfbox{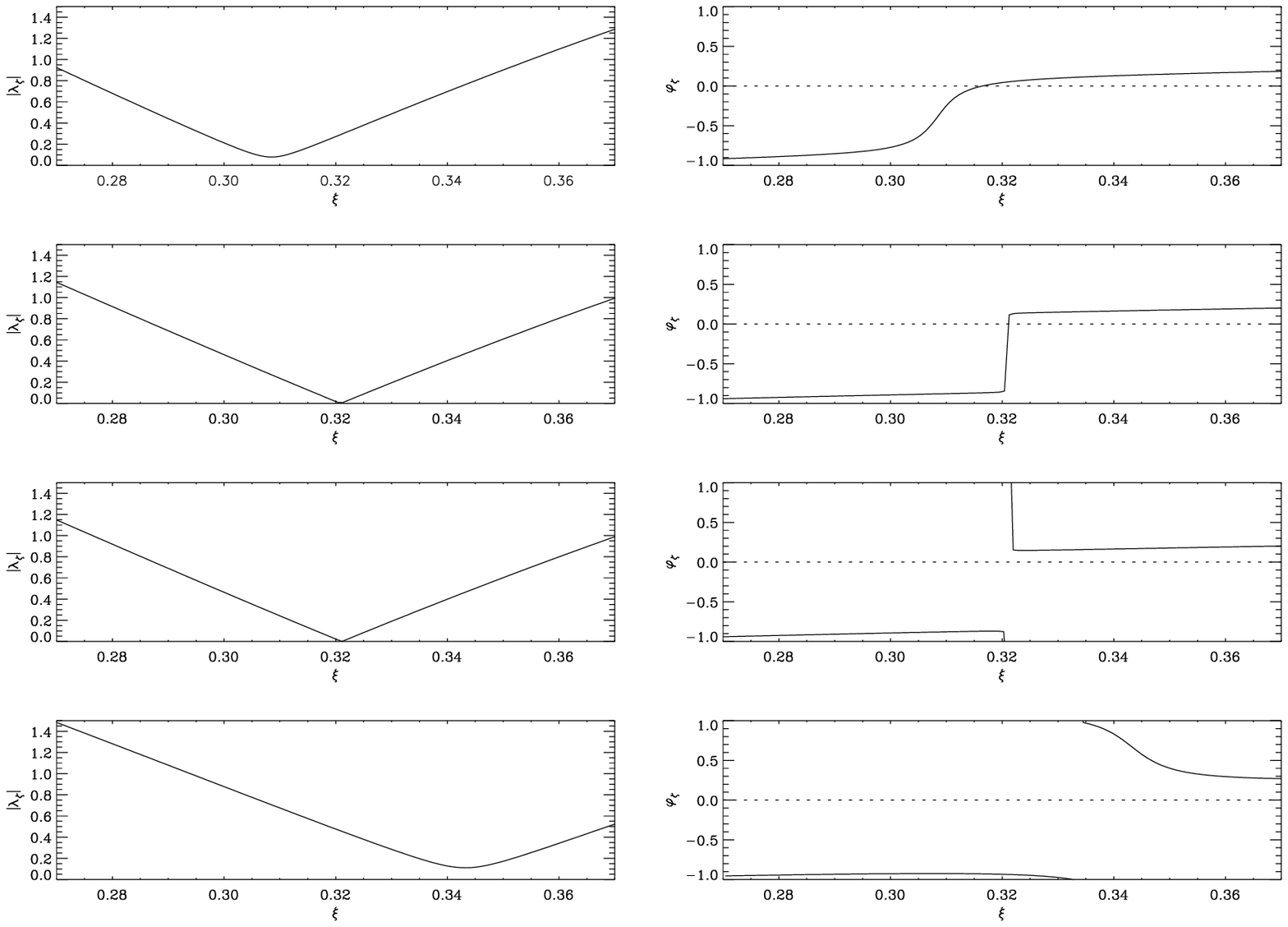}
$
\scriptsize
\caption{
Density eigenfunction of the $n=3$ eigenmode of a system with
$(\sigma\subs{},\psiei)=(0.2,0.1)$,
with varying $\epsilon\subs{}$.
Far from the critical value for these parameters
($\epsilon\subs{}\approx6.527$),
the phase jumps are broad:
first row ($\epsilon\subs{}=10$);
fourth row ($\epsilon\subs{}=3$).
Near the transition, the phase jump is abrupt
and the amplitude approaches zero at the node:
second row ($\epsilon\subs{}=6.55$);
third row ($\epsilon\subs{}=6.5$).
}
\label{'jump'}
\end{center}
\end{figure}

The sense of a phase jump, \eg either a modest negative jump in phase
or a positive jump by more than half a cycle,
is essentially distinct
because the phase is seen to either increase or decrease asymptotically
on either side of the node discontinuity, \ie it is
determined by the sign of $-d\varphi/d\xi$ in the neighbourhood.
In some situations
there are critical values of $\epsilon\subs{}$ at which
the sense of a phase jump changes from positive to negative.
For $\epsilon\subs{}$ far from the critical values,
phase jumps are gradual, being spread relatively broadly in $\xi$.
At the critical $\epsilon\subs{}$
the phase jump is a discontinuity without a well-defined sign.
Figure~\ref{'jump'} shows an example of the changes in structures
of a density phase jump
with $\epsilon\subs{}$ shown near and far from its transition value.
 
\subsubsection{Phases at $\xi=0$}

As not all the phases at the lower boundary
are directly constrained by boundary conditions,
they can only be obtained by integrating the hydrodynamic equations.

\begin{figure}   
\begin{center}
$  
\begin{array}{ccc}
{\underline{\epsilon\subs{}=0.1}}&{\underline{\epsilon\subs{}=1}}&{\underline{\epsilon\subs{}=10}}\\
&&\\
&&\\
\epsfxsize=4.5cm
\epsfbox{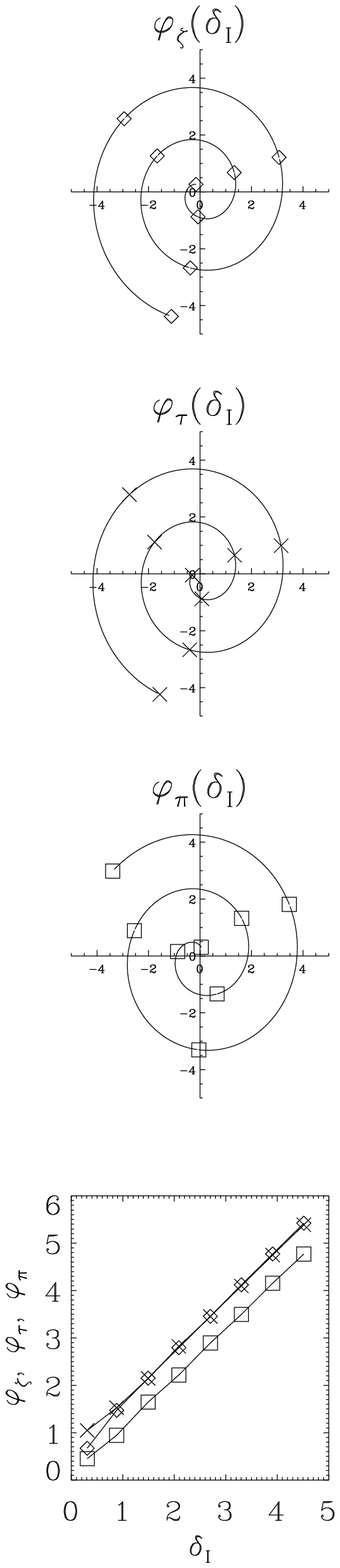}&
\epsfxsize=4.5cm
\epsfbox{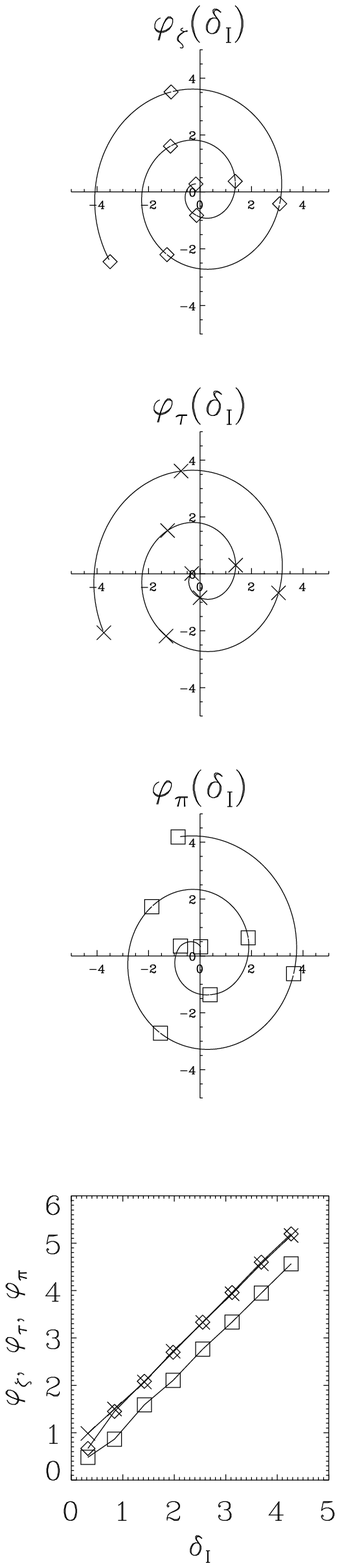}&
\epsfxsize=4.5cm
\epsfbox{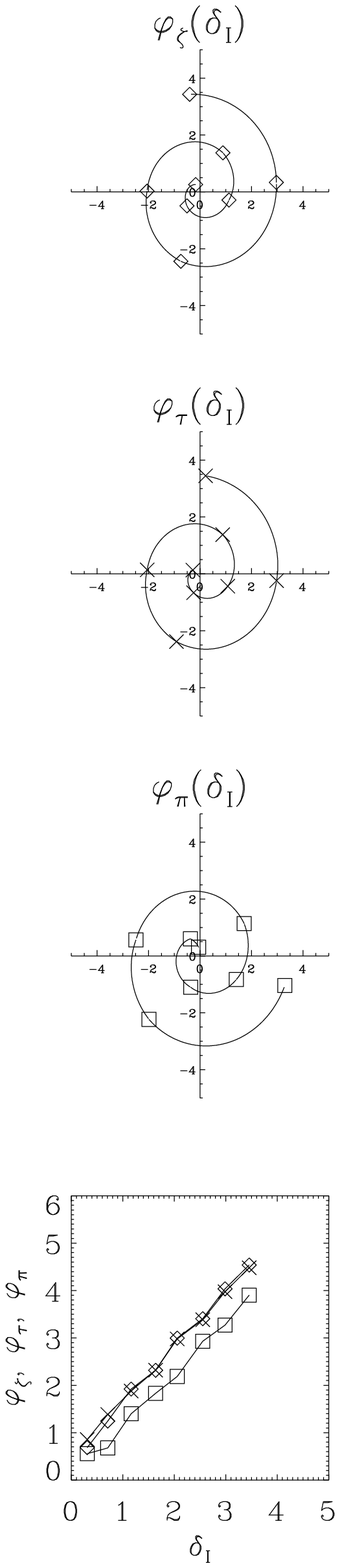}\\
\end{array}
$
\scriptsize
\caption{ 
Phase of perturbed variables evaluated at the lower boundary ($\xi=0$)
for the $n=1,\ldots,8$ modes of one-temperature systems.
Phases are defined in relation to the oscillation of shock position.
The left, centre and right columns show cases of the efficiency parameter
$\epsilon\subs{}=0.1, 1, 10$ respectively.
The polar plots show
density, longitudinal velocity and pressure phases
as diamonds, crosses ans squares.
The radius is the oscillatory part of the eigenvalue, $\dI$.
The bottom row shows equivalent linear plots of phase vs $\dI$
for the same modes, with the same symbols as the polar plots.
The phase variables are defined in terms of $\pi$,
so that a phase difference $\Delta\varphi=2$ means a full cycle.
}
\label{'figure.phase.1d'}
\end{center}   
\end{figure}

The lower-boundary phase of each perturbed variable increases
between a mode $n$ and the consecutive mode $n+1$,
when $(\sigma\subs{},\psiei,\epsilon\subs{})$ is fixed.
This inter-mode increment differs slightly
between the phases of different hydrodynamic variables.
It also depends weakly on harmonic number $n$.
(See Figure~\ref{'figure.phase.1d'}
for examples of lower-boundary phases
for one-temperature systems
with different values of the parameter $\epsilon\subs{}$.)
 
For a particular mode,
there are regular phase relationships between different perturbed variables
evaluated at the lower boundary.
The general trends are as follows:
(1)
For both one-temperature and two-temperature accretion flows,
the density and longitudinal velocity profiles
are approximately in phase
(\ie $\varphi_\zeta|_{\xi=0}\approx\varphi_\tau|_{\xi=0}$).
(2)
The phase $\varphi_\pi$
is approximately quarter-phase
behind the density and longitudinal velocity phases
($\varphi_\pi|_{\xi=0}
\approx\varphi_\zeta|_{\xi=0}-0.5\approx\varphi_\tau|_{\xi=0}-0.5$).
The phasings $\varphi_a-\varphi_b$
for two hydrodynamic variables $a$ and $b$,
are approximately constant for $n>2$.
For $n=1,$ and $2$ the phasings
conform less to the general trend.
 
\section{Boundary conditions}

\begin{figure}
\begin{center}
$
\begin{array}{cccc}
&\epsilon\subs{}=0&\epsilon\subs{}=1&\epsilon\subs{}=100\\
{\lambda_\tau=0}&
\epsfxsize=4.9cm
\epsfbox{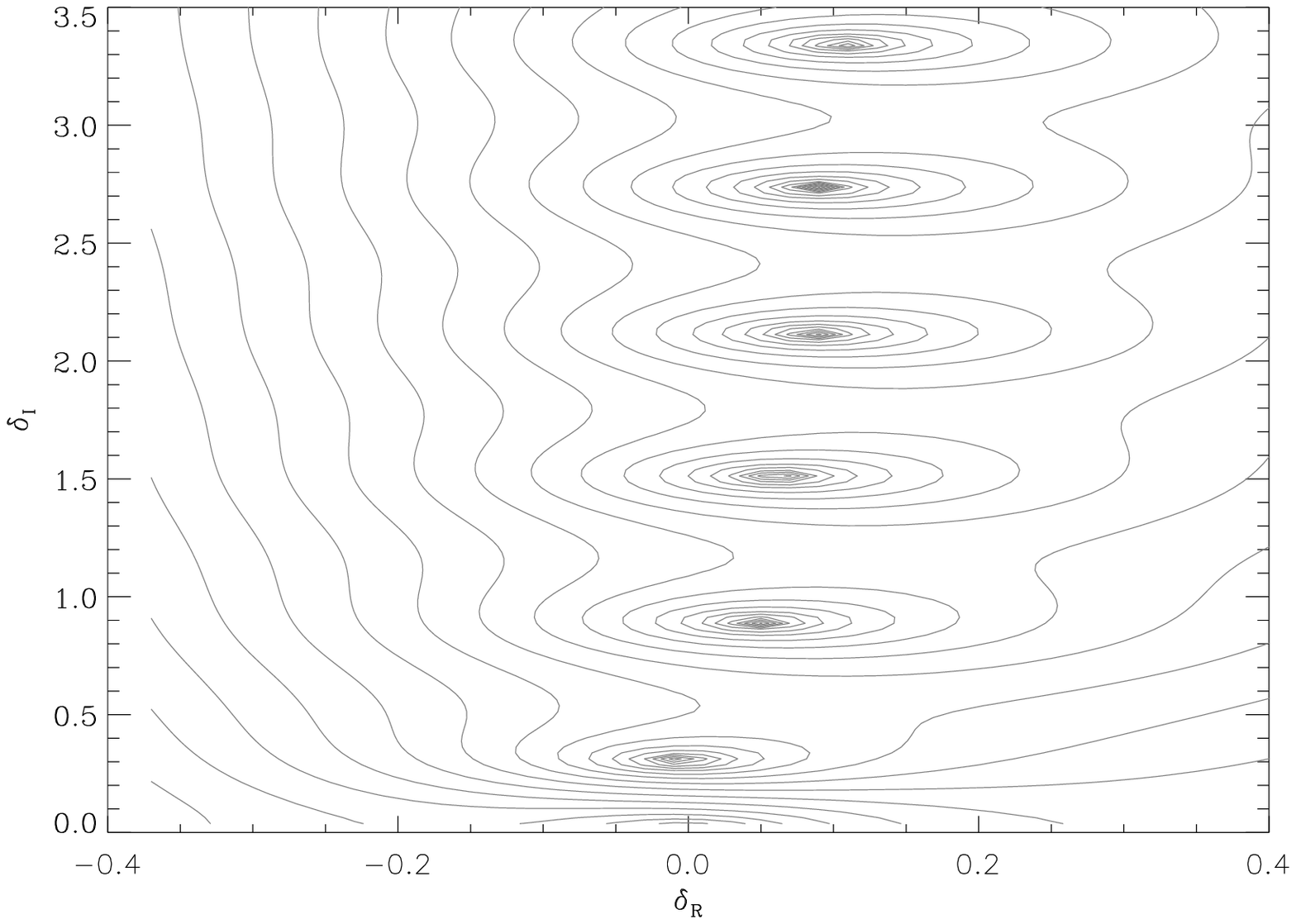}&
\epsfxsize=4.9cm
\epsfbox{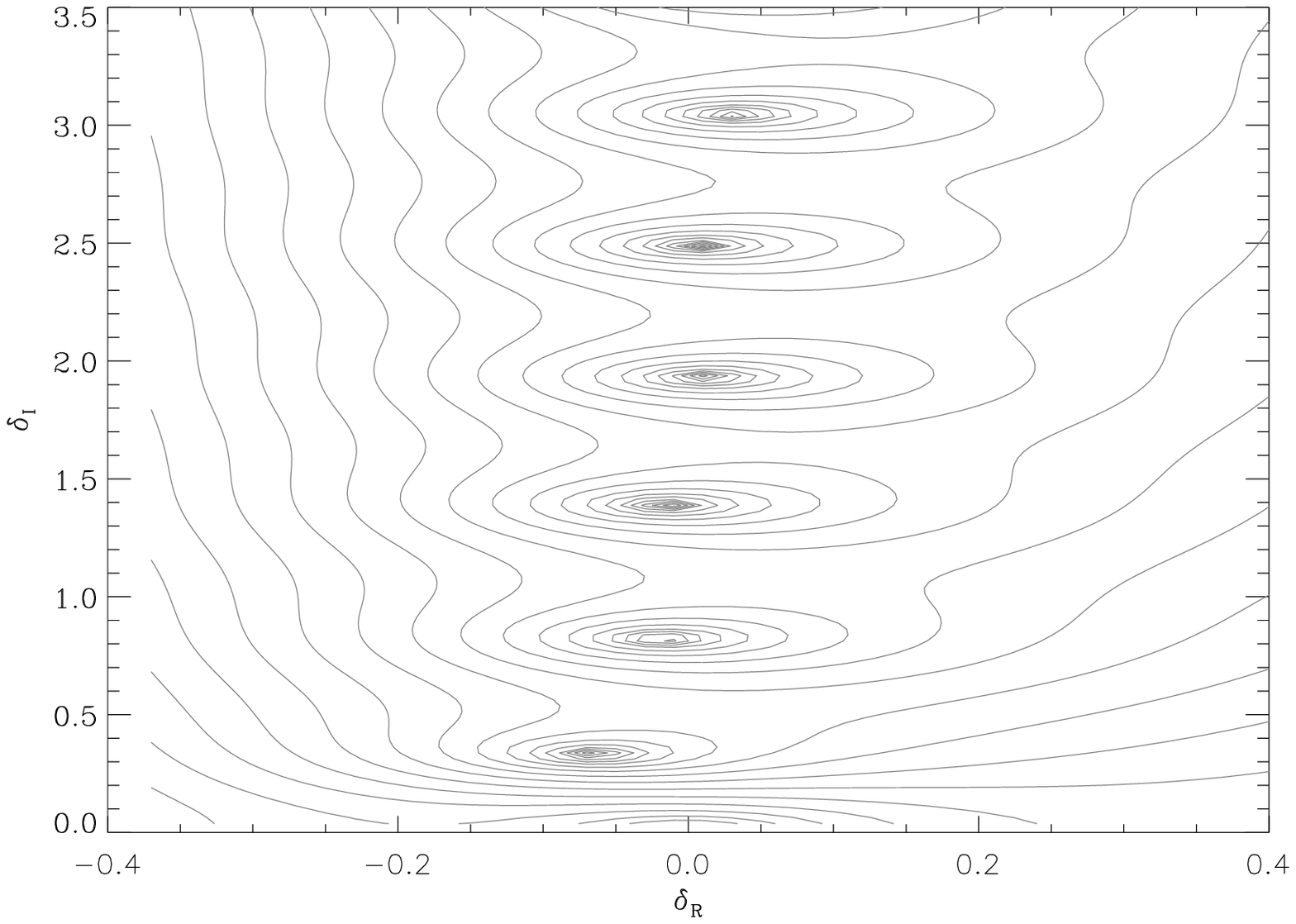}&
\epsfxsize=4.9cm
\epsfbox{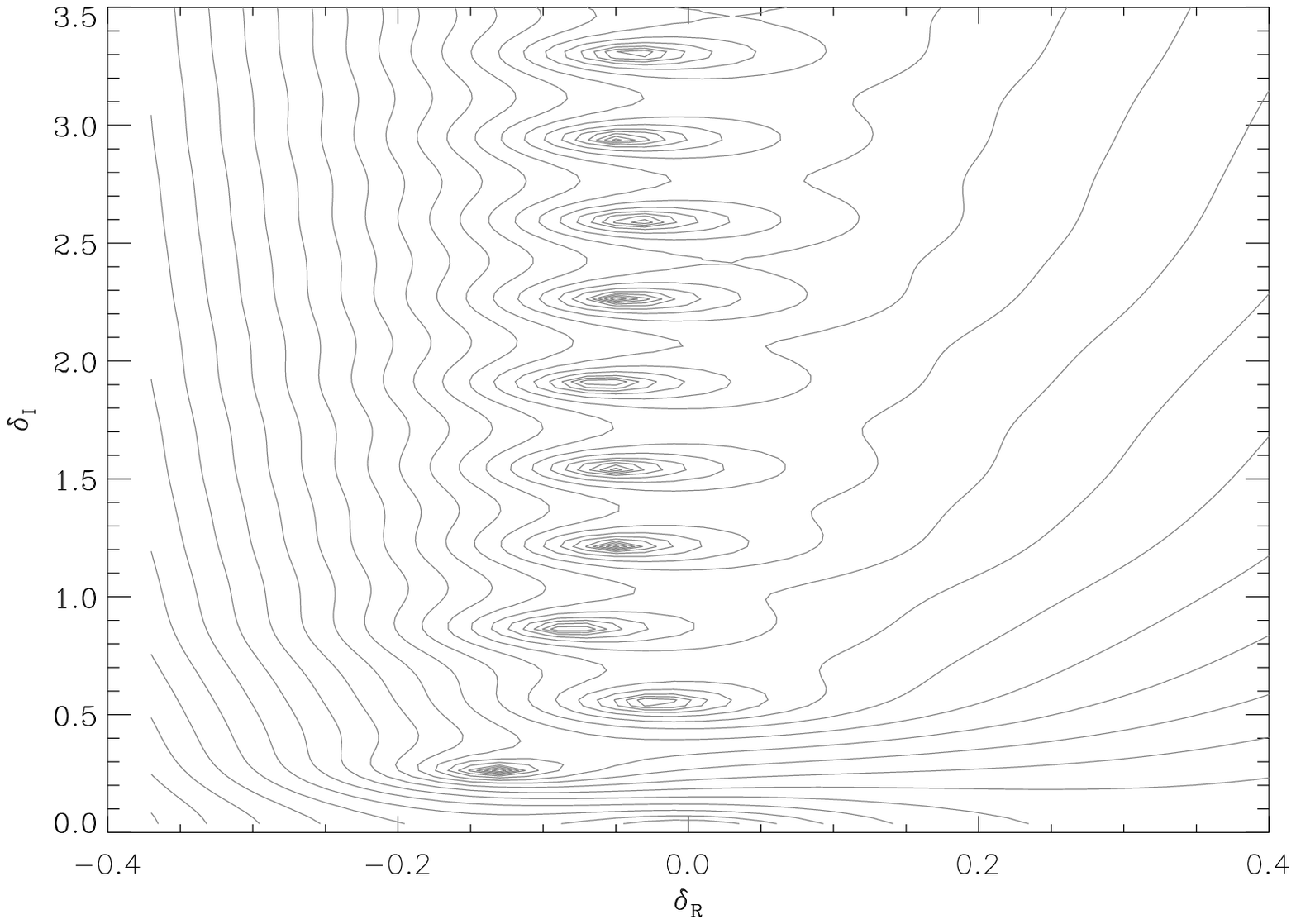}\\
{\lambda_\pi=0}&
\epsfxsize=4.9cm
\epsfbox{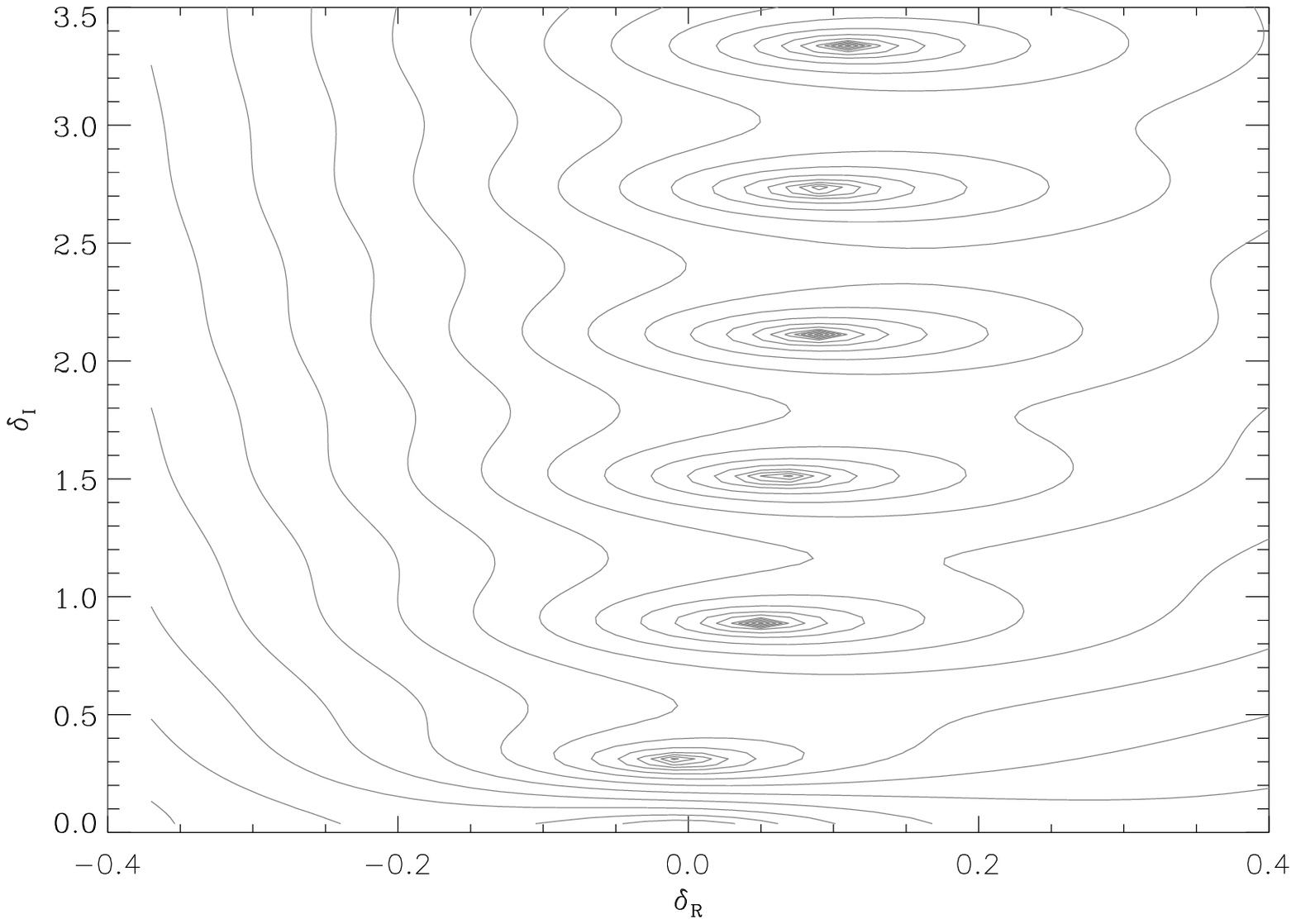}&
\epsfxsize=4.9cm
\epsfbox{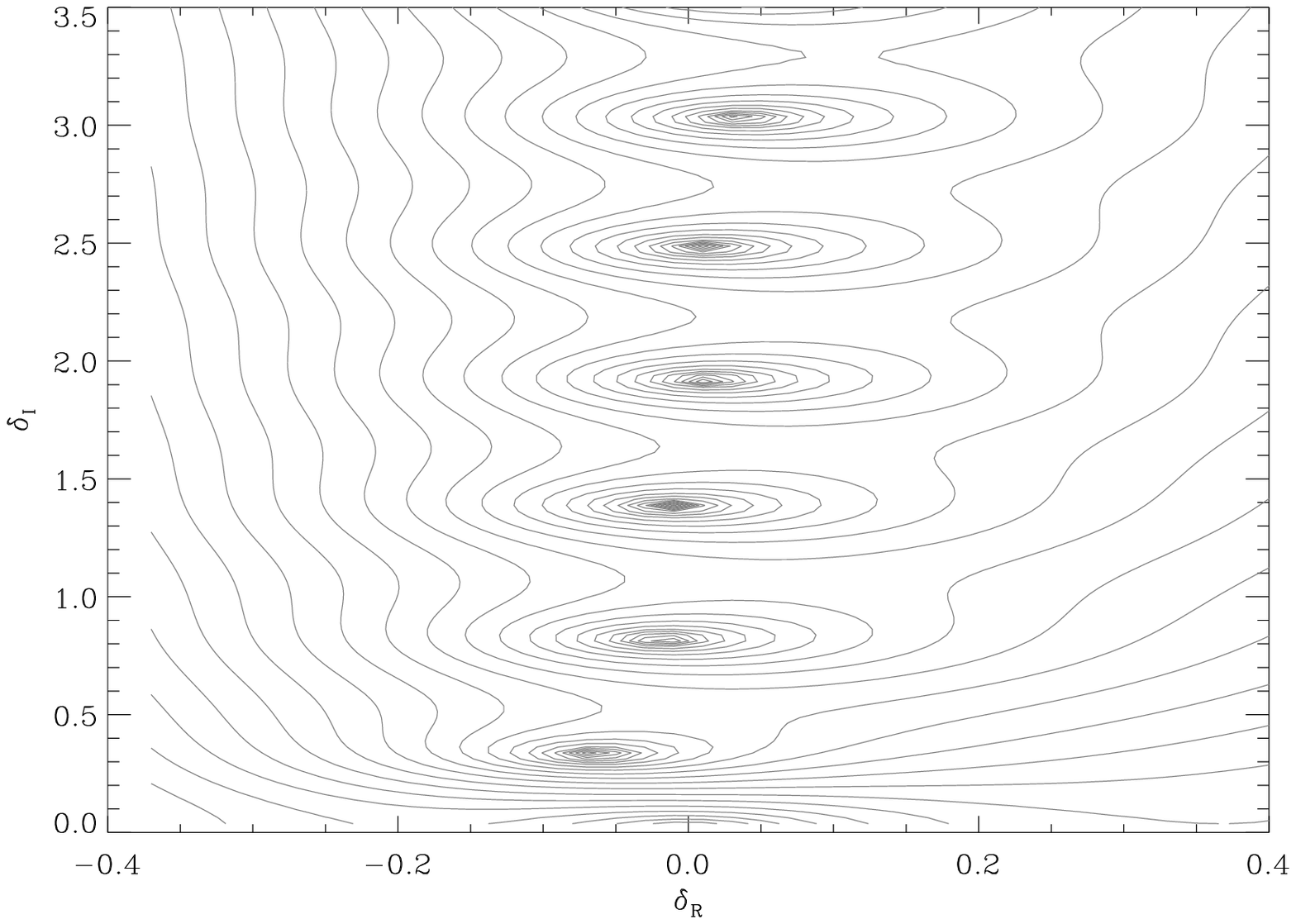}&
\epsfxsize=4.9cm
\epsfbox{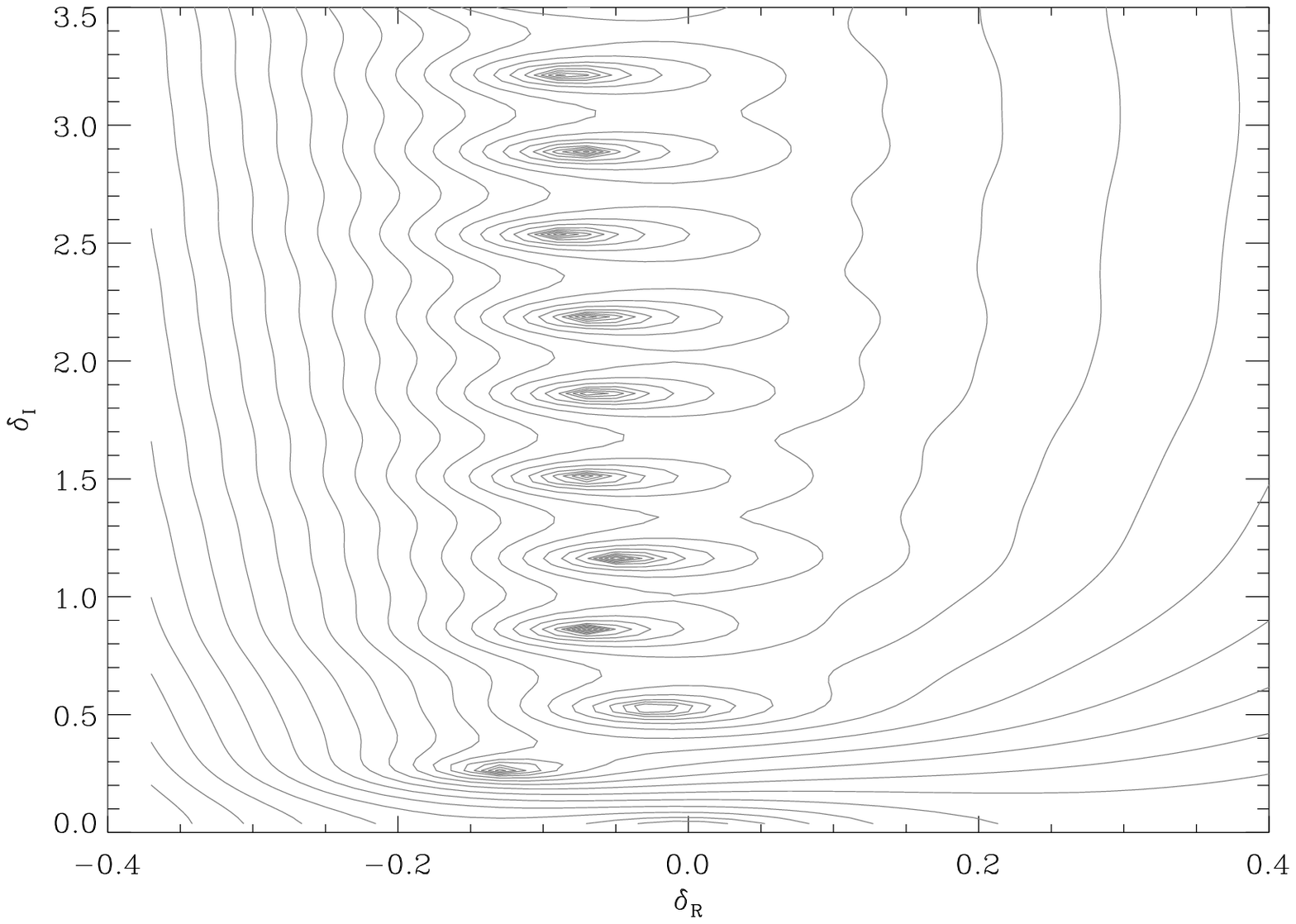}\\
{\lambda_\zeta+\lambda_\tau=0}&
\epsfxsize=4.9cm
\epsfbox{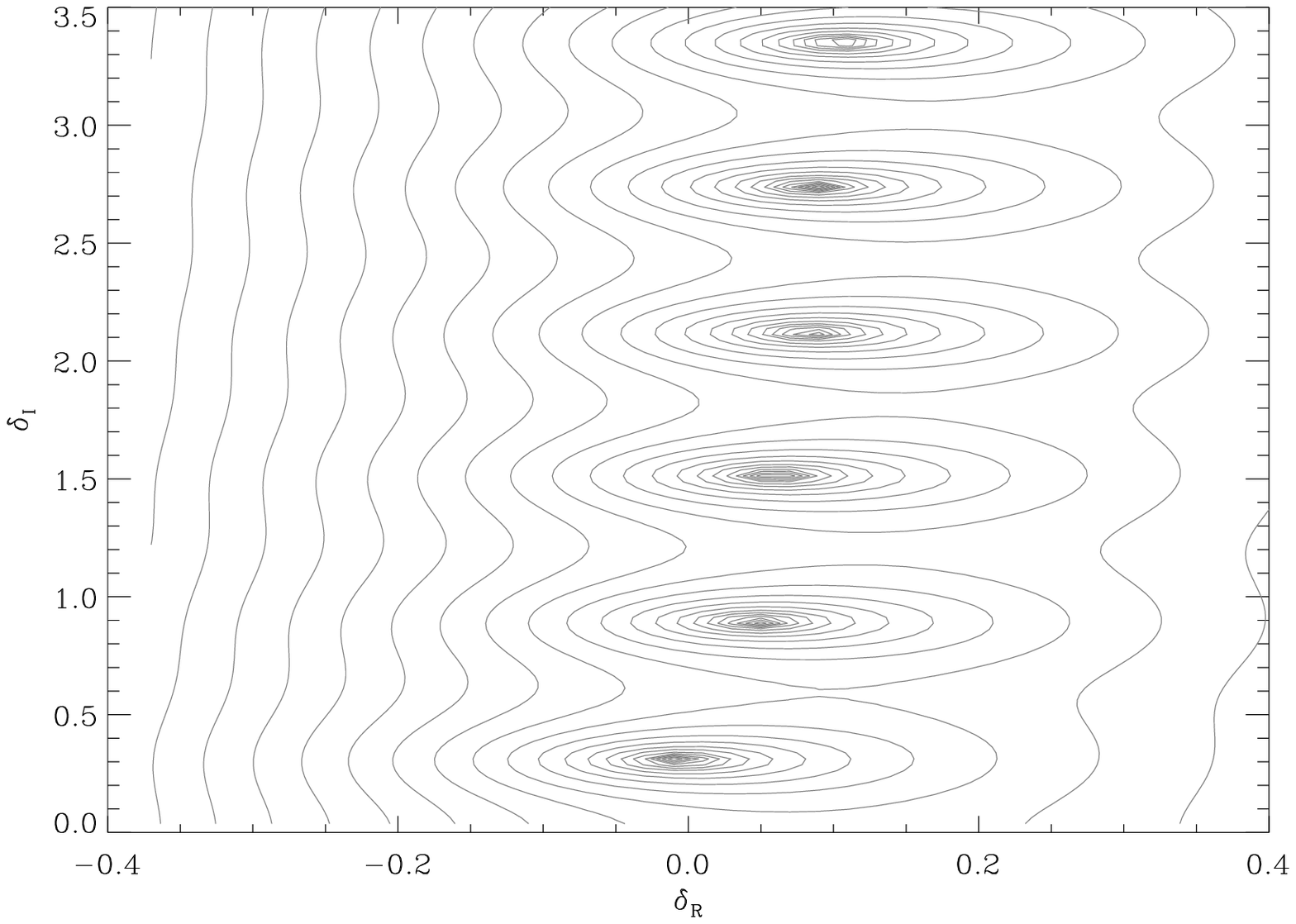}&
\epsfxsize=4.9cm
\epsfbox{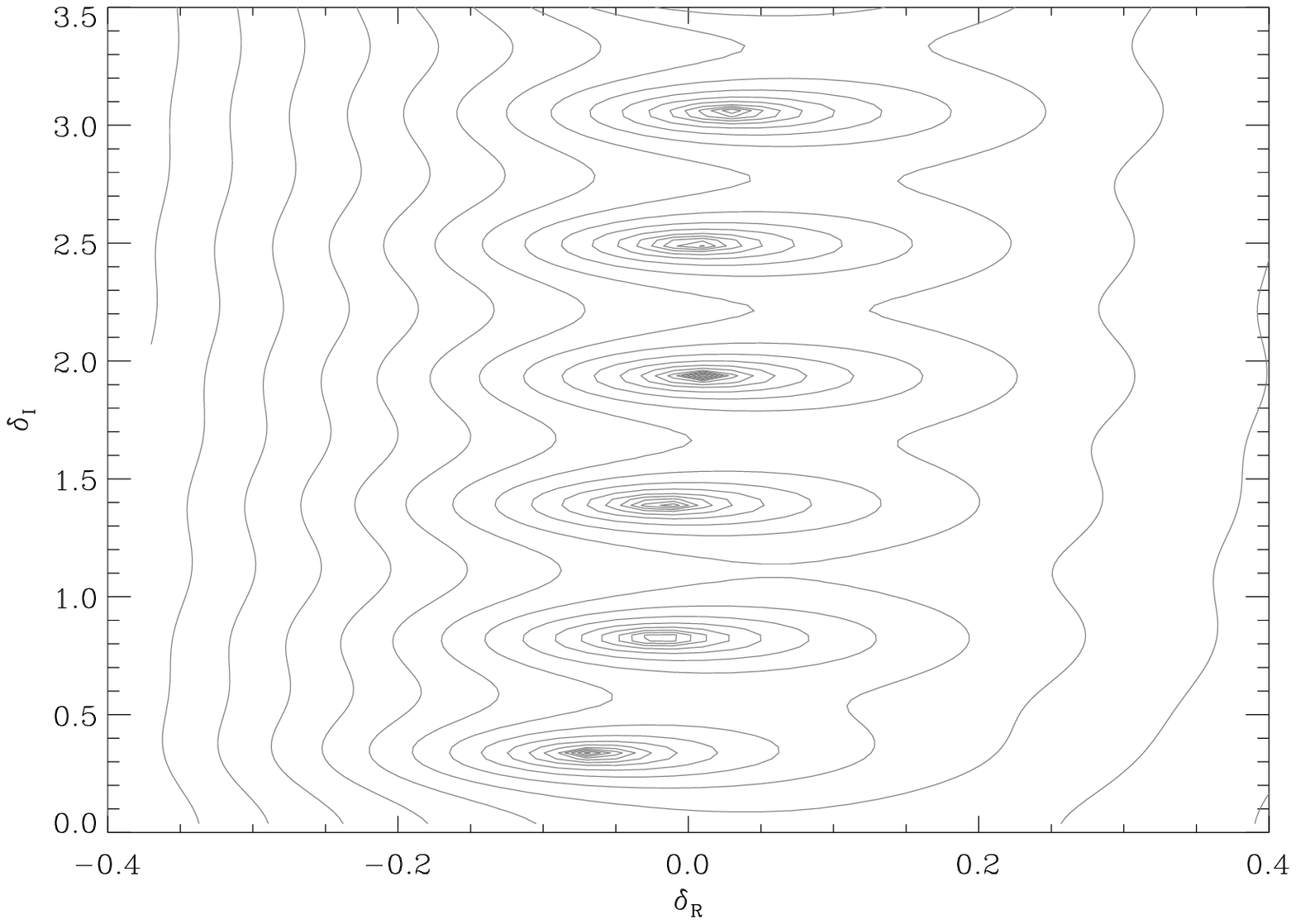}&
\epsfxsize=4.9cm
\epsfbox{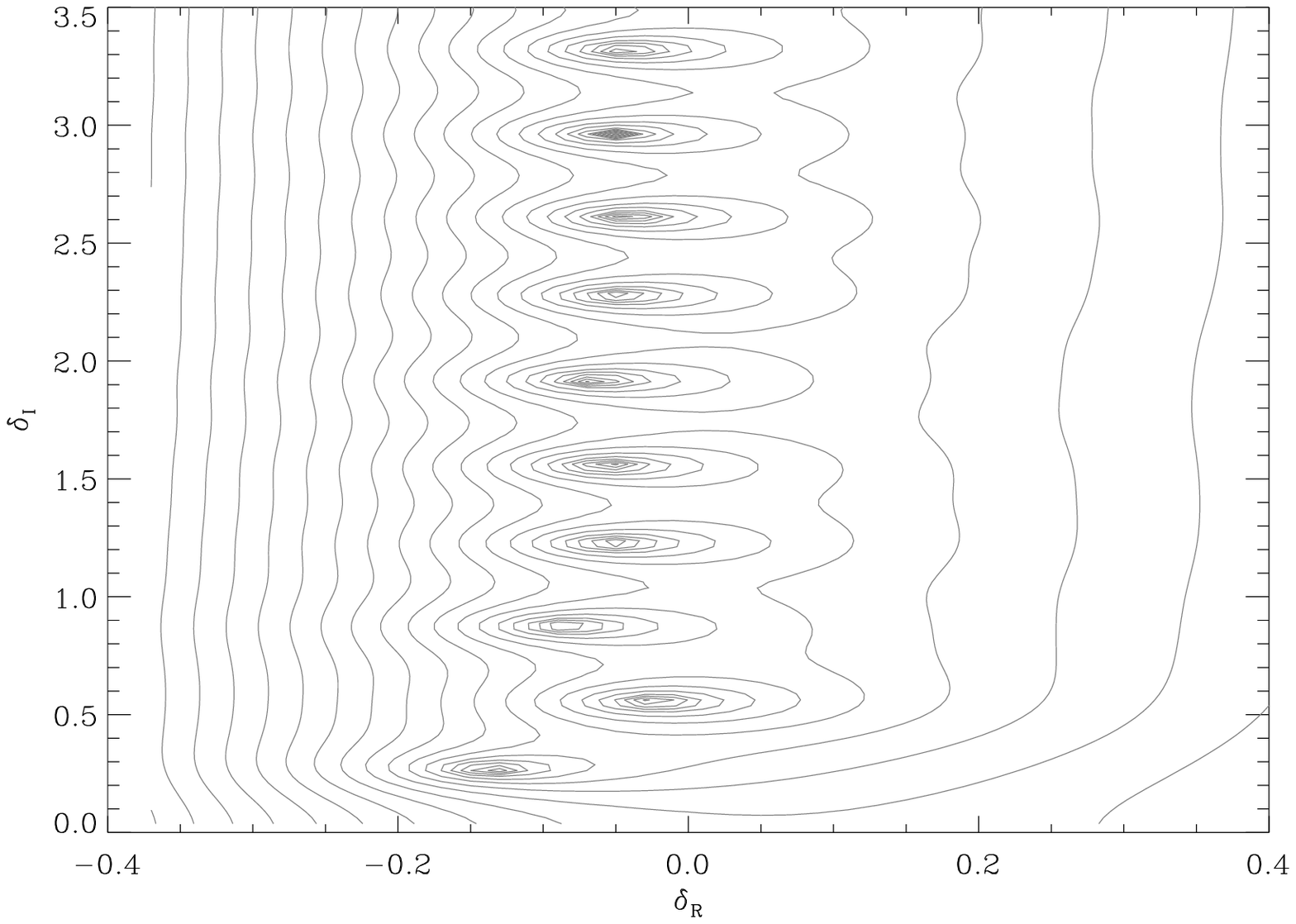}\\
{\lambda_\zeta=0}&
\epsfxsize=4.9cm
\epsfbox{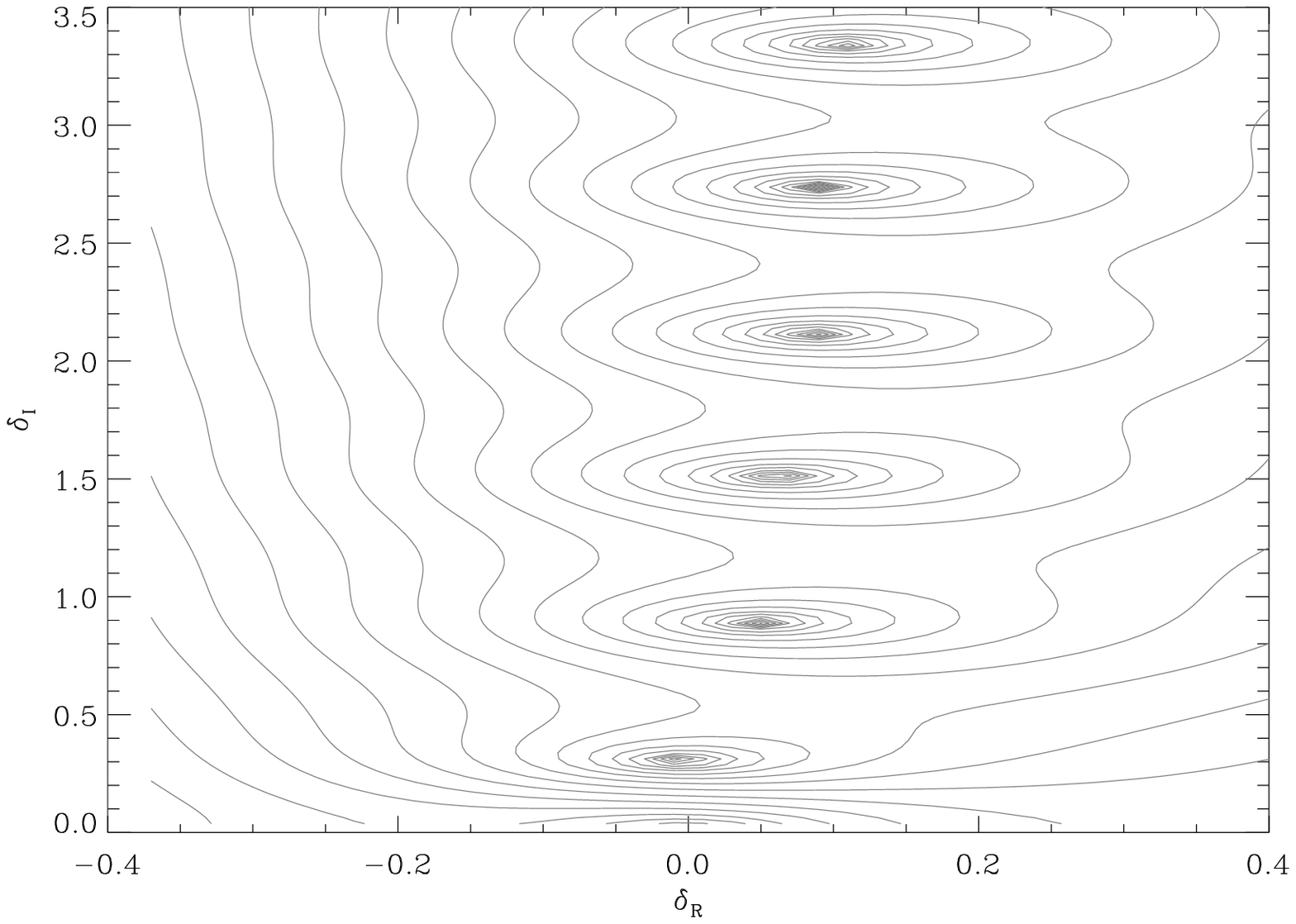}&
\epsfxsize=4.9cm
\epsfbox{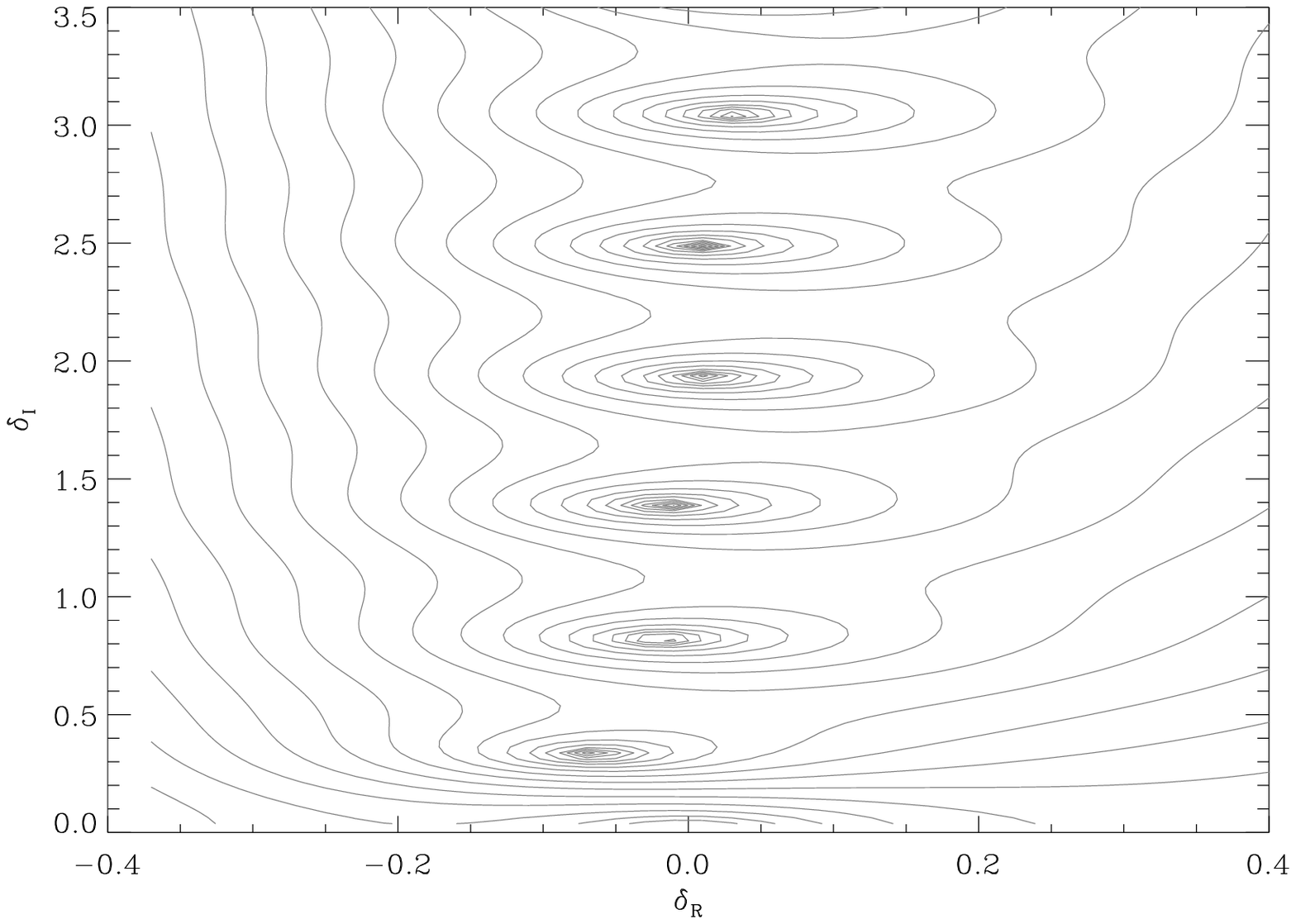}&
\epsfxsize=4.9cm
\epsfbox{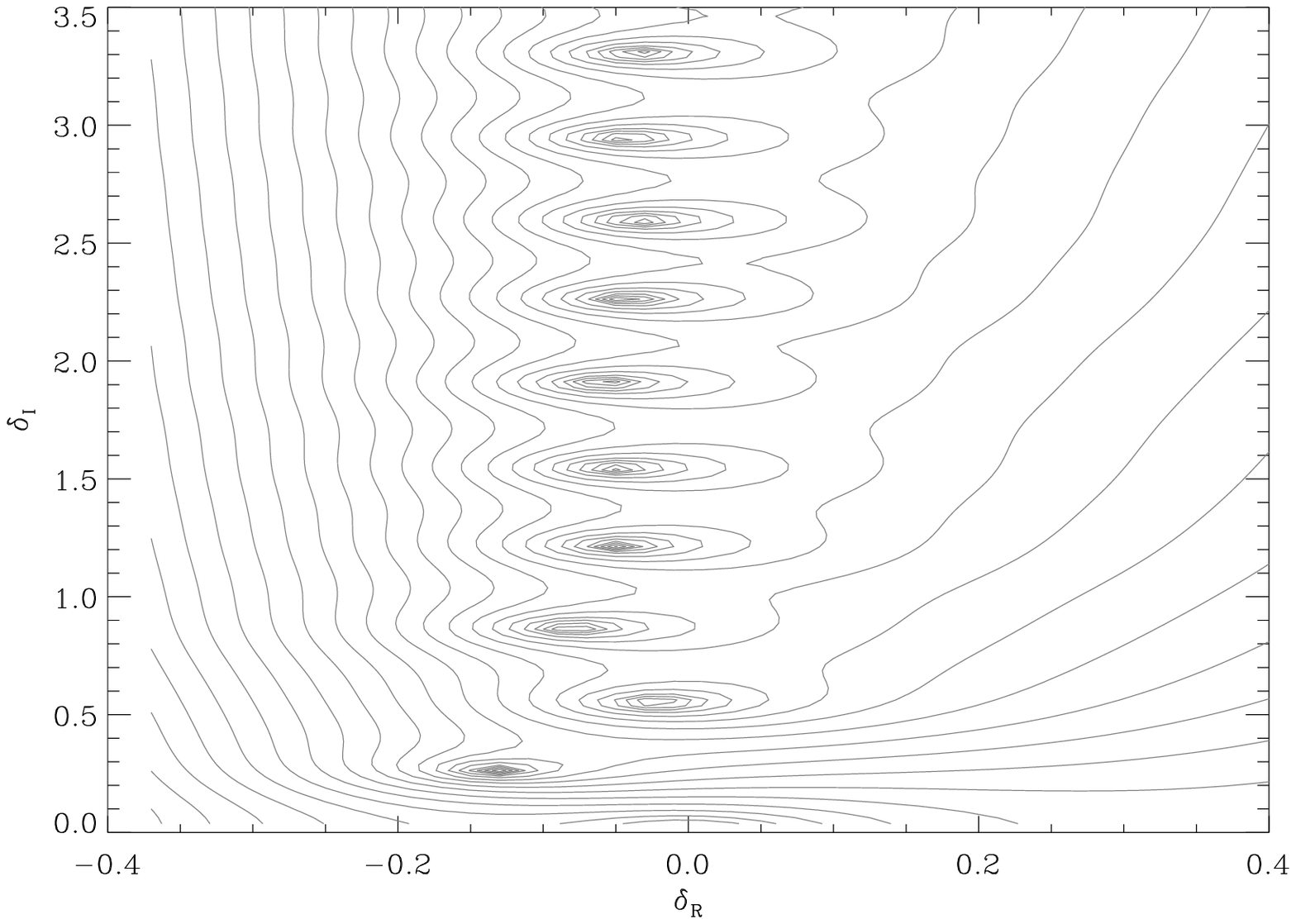}\\
\end{array}
$
\end{center}
\scriptsize
\caption{
Eigenplanes for $(\sigma\subs{},\psiei)=(0.5,0.5)$
and $\epsilon\subs{}=0,1,100$ with alternative lower boundary conditions.
Contours indicate the agreement between
integrated values of the perturbed variables at given complex $\delta$
and the boundary condition.
The top row is for the conventional zero-velocity boundary condition
$\lambda_\tau=0$.
The upper-middle row is for $\lambda_\pi=0$,
meaning constant pressure at the lower boundary;
the lower-middle row is for $\lambda_\zeta+\lambda_\tau=0$,
which relates to the strict relation $\rho\propto1/v$;
and the bottom row is for $\lambda_\zeta=0$,
which is a condition of constant density at the boundary.
}
\label{'figure.alt.bc'}
\end{figure}

So far we considered a special case of ``stationary wall''
in which both $\tau$ and $\lambda_\tau$ are zero at $\xi=0$.
We now consider modifications in which
the zero-velocity boundary condition for the stationary solution
($\tau_0=0$ at $\xi=0$) is retained
but the boundary value of $\lambda_\tau$ is not necessarily zero.
For example, the condition $\lambda_\pi=0$
implies constant (unoscillating) pressure at the lower boundary,
whilst allowing oscillations of the velocity at the base of the settling flow.
Other interesting conditions studied include
$\lambda_\zeta+\lambda_\tau=0$ (opposite density and velocity oscillations)
and $\lambda_\zeta=0$ (constant density at lower boundary),
(see Figure~\ref{'figure.alt.bc'}).
However it is found that the change of boundary conditions
has little effect on the frequency sequence $\dI$.
The mode stability $\dR$ only changes appreciably
in the regime of large $\epsilon\subs{}$ and high harmonic number $n$.
Some choices with reasonable physical interpretations
In general most of the choices that we consider
yield eigenvalues that are very close to those of the conventional
case ($\lambda_\tau=0$) at $\xi=0$.
Detailed discussions of the effects of boundary conditions
on the eigenfunctions can be found in Saxton (2001).

Alternative lower boundary conditions on the stationary variables
may be considered in conjunction with alternative conditions
on the perturbed variables.
However these choices would describe systems that are
physically very different from the white-dwarf accretion problem
illustrated in this paper.
An exploration of the properties of such systems
is beyond the scope of this paper,
and will be investigated in the future.
 
\section{Luminosity response} 

The total power (normalised to $\rhoa\vff^2$) is given by integrating
the cooling function over the whole post-shock structure
in the stationary solution:
\begin{equation}
L=\int_0^1{{\tL}\over{\gamma-1}} d\xi
=\int_0^{1/4} {{[\gamma(1-\tau_0)-\tau_0]}\over{\gamma-1}} d\tau_0
={{7\gamma-1}\over{32(\gamma-1)}} \ .
\end{equation}
For the adiabatic index $\gamma=5/3$,
the total power radiated via all processes is $L=1/2$,
consistent with energy-conservation.
The contribution of bremsstrahlung cooling is
\begin{equation}
L_{\rm br,0}
=\int_0^{1/4} {{[\gamma(1-\tau_0)-\tau_0]}\over{\gamma-1}}
{1\over{1+\tL_{\rm cy}/\tL_{\rm br} }} d\tau_0 \ . 
\end{equation}
The second cooling process (which is assumed to be cyclotron cooling)
contributes the difference between this value and the total,
$L_{\rm cy,0}=L-L_{\rm br,0}$.

It can be shown that the local luminosity responses
are described by complex $\lambda$-functions
that are analogous to the eigenfunctions of the hydrodynamic variables.
The bremsstrahlung and cyclotron luminosity response functions,
normalised to $\varepsilon$, are
\begin{equation}
\lambda_{\rm br}={\frac32}\lambda_\zeta+{\frac12}\lambda\sube{} 
\end{equation}
and
\begin{equation}
\lambda_{\rm cy}=
\left({0.15-2.5}\right)\lambda_\zeta+{2.5}\lambda\sube{}
\end{equation}
for the conditions of accreting magnetic white dwarfs.
For a particular mode the $\lambda_{\rm br}$ and $\lambda_{\rm cy}$ profiles
describe what could be regarded as eigenfunctions of
the effect of the oscillations on the cooling emission.
 
\begin{figure}
\begin{center}
\epsfxsize=17.2cm
\epsfbox{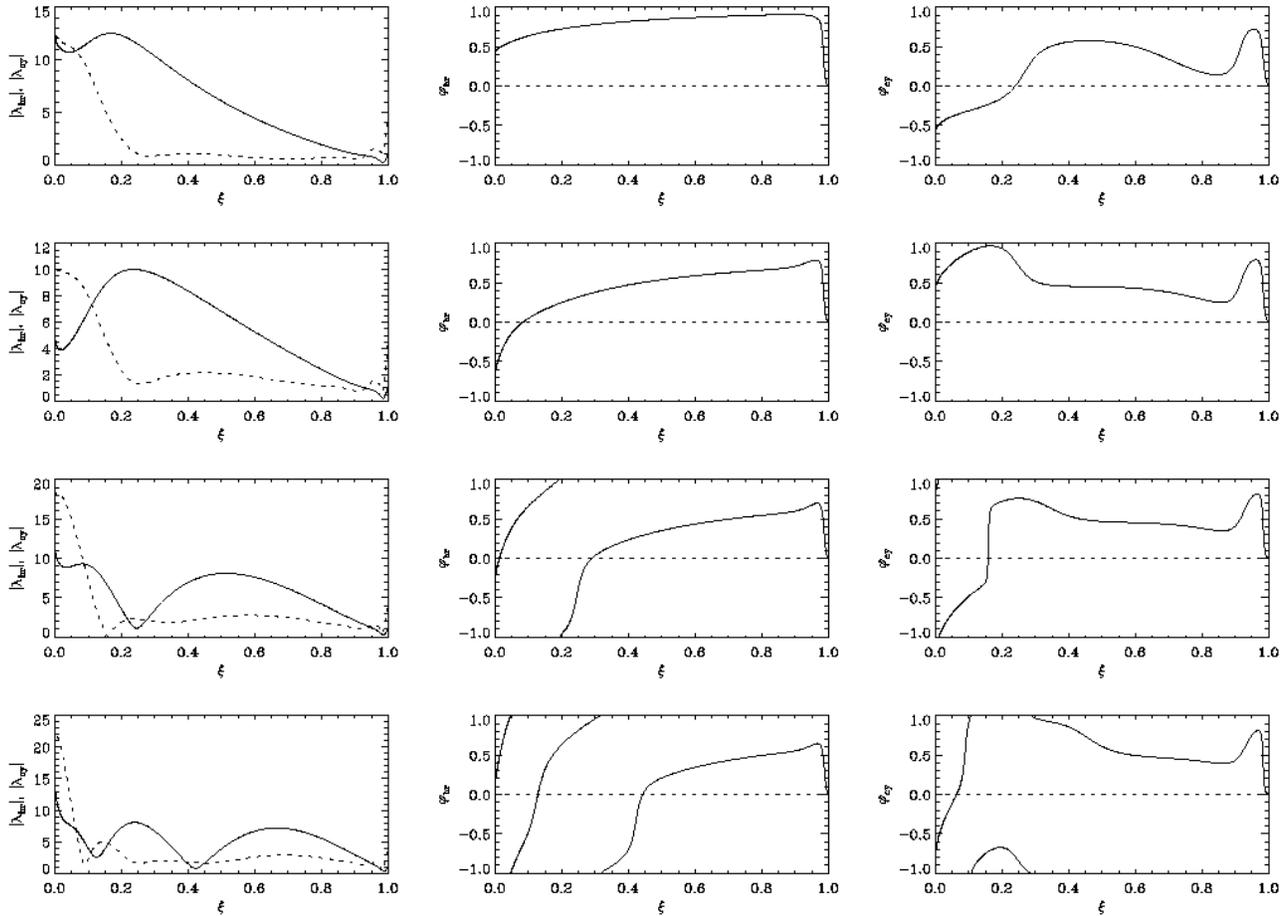}
\end{center}
\caption{
Profiles of luminosity perturbed variables,
$\lambda_{\rm br}$ and $\lambda_{\rm cy}$,
in modes $n=1, 2, 3, 4$ from top to bottom.
This choice of system parameters,
$(\sigma\subs{},\psiei,\epsilon\subs{})=(0.2,0.1,100)$,
gives strong two-temperature effects
and cyclotron cooling dominates the cooling.
The left column shows
the amplitudes for bremsstrahlung luminosity $|\lambda_{\rm br}|$ 
(solid line)
and cyclotron $|\lambda_{\rm cy}|$ (dotted line).
Central and right columns are the phase profiles,
$\varphi_{\rm br}$ and $\varphi_{\rm cy}$,
with the dotted straight line being a zero-phase reference.
}
\label{'a1c4l'}
\end{figure}

\begin{figure}
\begin{center}
\epsfxsize=17.2cm
\epsfbox{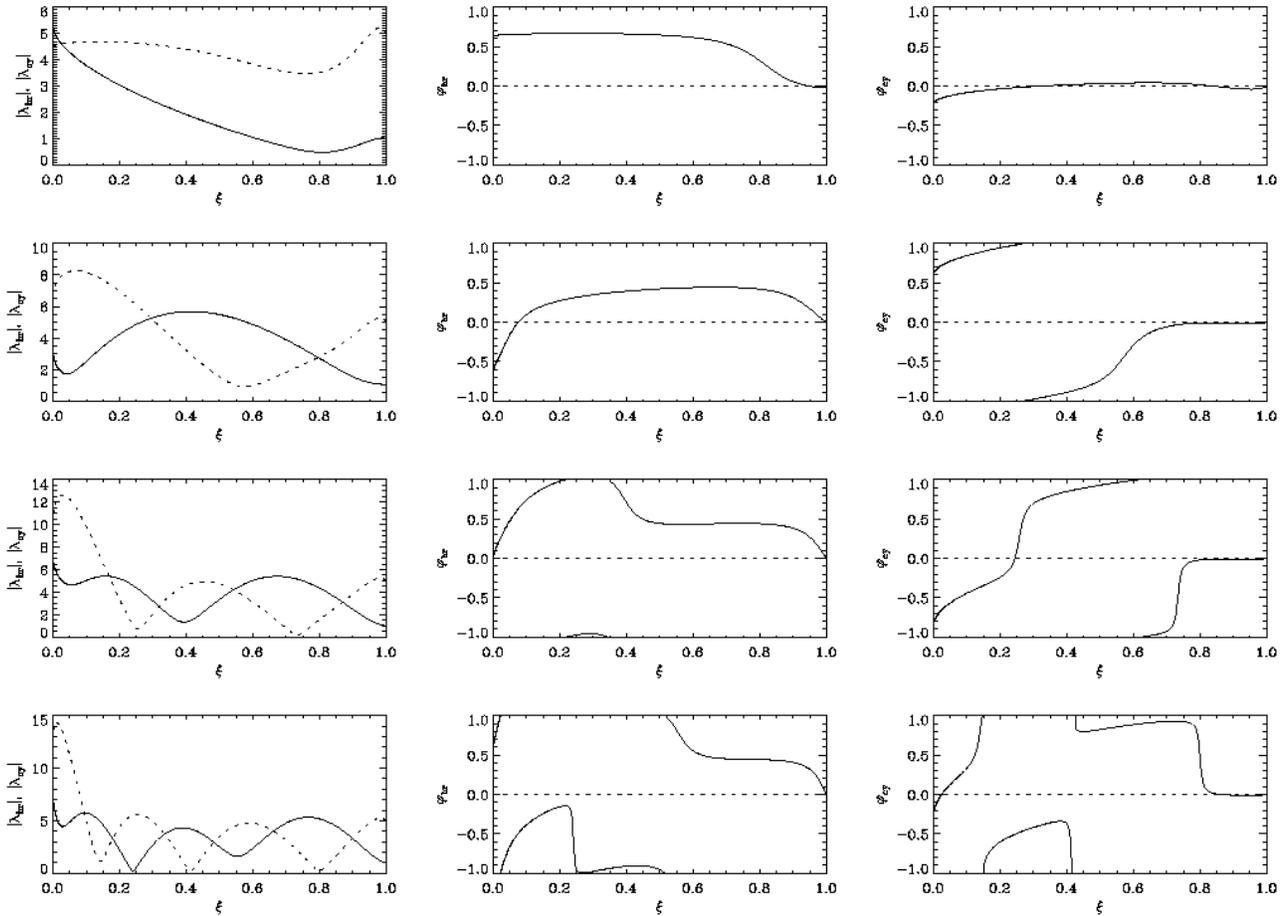}
\end{center}
\caption{
Same as Figure~\ref{'a1c4l'} but for
$(\sigma\subs{},\psiei,\epsilon\subs{})=(0.5,0.5,1)$.
This choice of system parameters gives modest two-temperature effects
and equal bremsstrahlung- and cyclotron-cooling efficiencies.
}
\label{'b2s4l'}
\end{figure}

The amplitude and phase profiles of the modes reveal several regularities.
The phases $\varphi_{\rm br}$ and $\varphi_{\rm cy}$
are both zero at the shock, $\xi=1$
(see Figures~\ref{'a1c4l'}-\ref{'b2s4l'}),
meaning that the emission due to both processes locally near the shock
oscillates in phase with the oscillation of the shock height.
The function $|\lambda_{\rm br}|$ has a minimum in amplitude
at or near the shock,
and the cyclotron function $|\lambda_{\rm cy}|$
has a maximum in the same vicinity.
In between the upper and lower boundaries there is no simple relationship
between the phases of the two functions.
Near the lower boundary, both eigenfunctions reach their maximum amplitudes
for any given mode,
and these maxima are higher for modes with higher harmonic number $n$.
At the lower boundary, $\xi=0$,
the oscillations of the two cooling processes are in antiphase.
However, in almost all cases
the antinodes of $\lambda_{\rm br}$ occur at the nodes of $\lambda_{\rm cy}$
and
the nodes of $\lambda_{\rm br}$ occur at the antinodes of $\lambda_{\rm cy}$.
There is no obvious relation between
the luminosity responses $\lambda_{\rm br}$, $\lambda_{\rm cy}$
and the eigenvalues $\delta$.

Multiplying $\lambda_{\rm br}$ and $\lambda_{\rm cy}$
by the respective cooling functions
and integrating over the entire post-shock region
yields the luminosity responses
for a small shock-height perturbation $\varepsilon$
(see subsection~\ref{'perturbation'}):
\begin{equation}
{L_{\rm br,1} }=\varepsilon\int^{1/4}_0
{{\tL_{\rm br}}\over{\gamma-1}}
\lambda_{\rm br}
{{d\xi}\over{d\tau_0}}
d\tau_0
=
\varepsilon\int^{1/4}_0
{{\left[{\gamma(1-\tau_0)-\tau_0}\right]}\over{\gamma-1}}
{{\tL_{\rm br}}\over{\tL_{\rm br}+\tL_{\rm cy}}}
\lambda_{\rm br}
d\tau_0 \ , 
\end{equation}
and
\begin{equation}
{L_{\rm cy,1} }=\varepsilon\int^{1/4}_0
{{\tL_{\rm cy}}\over{\gamma-1}}
\lambda_{\rm cy}
{{d\xi}\over{d\tau_0}}
d\tau_0
=
\varepsilon\int^{1/4}_0
{{\left[{\gamma(1-\tau_0)-\tau_0}\right]}\over{\gamma-1}}
{{\tL_{\rm cy}}\over{\tL_{\rm br}+\tL_{\rm cy}}}
\lambda_{\rm cy}
d\tau_0 \ . 
\end{equation}
$L_{\rm br,1}$ is different for different modes,
and so is $L_{\rm cy,1}$.
Moreover,
the modes which have high-amplitude $L_{\rm cy,1}$ oscillations
do not necessarily have strong oscillations in $L_{\rm br,1}$.
 
In Table~\ref{'table.luminosity.relamp'}
we show the integrated luminosity responses,
scaled according to the amplitude of shock-height oscillation $\varepsilon$.
These quantities are proportional to the relative variations
in bremsstrahlung (cyclotron) emission.

$L_{\rm br,1}$ and $L_{\rm cy,1}$ do not show obvious dependence on $\dR$,
however they seem to be dependent on the system parameters
$(\sigma\subs{},\psiei,\epsilon\subs{})$.
Whether or not $|L_{\rm br,1}|/L_{\rm br,0}>|L_{\rm cy,1}|/L_{\rm cy,0}$
depends strongly on $\epsilon\subs{}$,
but is only weakly dependent on $\sigma\subs{}$ and $\psiei$.
 
The complex phases of the values $L_{\rm br,1}$ and $L_{\rm cy,1}$,
are $\Phi_{\rm br}(n)$ and $\Phi_{\rm cy}(n)$ respectively
for each mode $n$,
are listed in Table~\ref{'table.luminosity.phasediff'}.
Depending on the difference between the phases,
the waxing and waning of the emission due to one cooling process
may follow or lead the other process,
or else they may be in phase or antiphase.
 
For small $\epsilon\subs{}$ and
for a given mode $n$,
the phases $\Phi_{\rm br}(n)$ and $\Phi_{\rm cy}(n)$
are nearly constant throughout the $(\sigma\subs{},\psiei)$ parameter space.
The two-temperature parameters $(\sigma\subs{},\psiei)$
are almost ineffectual in the small-$\epsilon\subs{}$ regime.
For constant $\sigma\subs{}$ and $\epsilon\subs{}$,
decreasing $\psiei$ causes the phase difference
$\Phi_{\rm cy}(n)-\Phi_{\rm br}(n)$
to decrease.
This means that if cyclotron luminosity lags then its lag increases;
else if cyclotron luminosity leads bremsstrahlung luminosity 
then cyclotron's lead decreases.
 
In general,
$\Phi_{\rm br}(1)\approx0.35\pi$ to $0.5\pi$
and $\Phi_{\rm cy}(1)\approx0.9\pi$
for the fundamental mode in a wide range of system parameters,
\ie cyclotron emission oscillation
almost always lags bremsstrahlung emission
by $\approx0.6\pi$.
In the extreme cases in which two-temperature effects
are so strong that the fundamental mode becomes unstable, \eg when
$(\sigma\subs{},\psiei,\epsilon\subs{})=(1.0,0.1,100)$,
this relation breaks down.
The phase properties are more complicated for the overtones
because of more complicated winding and nodes in the $\lambda$-functions
(see \ref{'subsection.winding'}).
Therefore no obvious relationships are found between
the stability of a mode, the phases
$\Phi_{\rm br}$, $\Phi_{\rm cy}$ and their differences.

\begin{table}
\scriptsize
\caption{
Amplitudes of the oscillations of the total emission
in bremsstrahlung and cyclotron radiation,
relative to these processes total luminosities in the stationary solution.
For each set of the system parameters,
the left column indicates whether the mode is stable ($-$) or unstable ($+$)
in the small-amplitude analysis
(see complete eigenvalue results in Saxton \& Wu 1999).
The middle and right columns are
the $\varepsilon$-normalised relative amplitudes
$\left|{L_{\rm br,1}}\right|/{\varepsilon}L_{\rm br,0}$
and $\left|{L_{\rm cy,1}}\right|/{\varepsilon}L_{\rm cy,0}$ respectively.
}
\begin{center}
$
\begin{array}{rrccc}
\hline
\sigma\subs{}&\psiei&\epsilon\subs{}=0&\epsilon\subs{}=1&\epsilon\subs{}=100\\
\begin{array}{r}~\\~\\~
\\0.2\\~\\~\\~\\~\\~\\
\\0.2\\~\\~\\~\\~\\~\\
\\0.2\\~\\~\\~\\~\\~\\
\\0.5\\~\\~\\~\\~\\~\\
\\0.5\\~\\~\\~\\~\\~\\
\\0.5\\~\\~\\~\\~\\~\\
\\1.0\\~\\~\\~\\~\\~\\
\\1.0\\~\\~\\~\\~\\~\\
\\1.0\\~\\~\\~\\~\\~\\
\\
\end{array}&
\begin{array}{r}~\\~\\~
\\0.1\\~\\~\\~\\~\\~\\
\\0.5\\~\\~\\~\\~\\~\\
\\1.0\\~\\~\\~\\~\\~\\
\\0.1\\~\\~\\~\\~\\~\\
\\0.5\\~\\~\\~\\~\\~\\
\\1.0\\~\\~\\~\\~\\~\\
\\0.1\\~\\~\\~\\~\\~\\
\\0.5\\~\\~\\~\\~\\~\\
\\1.0\\~\\~\\~\\~\\~\\
\\
\end{array}&
\begin{array}{crr}
{\dR}?&{|{L_{\rm br,1}}|}\over{{\varepsilon}L_{\rm br,0}}&{|{L_{\rm cy,1}}|}\over{{\varepsilon}L_{\rm cy,0}}\\
\hline\\
-&2.749&*\\
+&1.650&*\\
+&2.501&*\\
+&1.298&*\\
+&1.685&*\\
+&2.020&*\\
\\
-&2.732&*\\
+&1.626&*\\
+&2.493&*\\
+&1.294&*\\
+&1.651&*\\
+&1.977&*\\
\\
-&2.732&*\\
+&1.621&*\\
+&2.494&*\\
+&1.291&*\\
+&1.654&*\\
+&1.974&*\\
\\
+&2.800&*\\
+&1.628&*\\
+&2.530&*\\
+&1.335&*\\
+&1.652&*\\
+&2.049&*\\
\\
-&2.745&*\\
+&1.618&*\\
+&2.497&*\\
+&1.306&*\\
+&1.641&*\\
+&1.983&*\\
\\
-&2.739&*\\
+&1.617&*\\
+&2.496&*\\
+&1.297&*\\
+&1.650&*\\
+&1.976&*\\
\\
+&2.893&*\\
+&1.601&*\\
+&2.596&*\\
+&1.396&*\\
+&1.591&*\\
+&2.094&*\\
\\
-&2.769&*\\
+&1.604&*\\
+&2.505&*\\
+&1.329&*\\
+&1.624&*\\
+&1.998&*\\
\\
-&2.752&*\\
+&1.609&*\\
+&2.499&*\\
+&1.309&*\\
+&1.640&*\\
+&1.982&*\\
\\
\end{array}
&
\begin{array}{crr}
{\dR}?&{|{L_{\rm br,1}}|}\over{{\varepsilon}L_{\rm br,0}}&{|{L_{\rm cy,1}}|}\over{{\varepsilon}L_{\rm cy,0}}\\
\hline\\
-&0.924&2.615\\
-&0.333&0.736\\
-&0.945&0.588\\
+&0.859&0.473\\
-&1.085&0.375\\
+&0.978&0.352\\
\\
-&1.136&3.755\\
-&0.213&0.659\\
-&1.001&0.202\\
-&0.819&0.364\\
-&1.107&0.354\\
-&0.987&0.329\\
\\
-&1.219&4.227\\
-&0.168&1.193\\
-&1.054&0.260\\
-&0.788&0.099\\
-&1.099&0.214\\
-&0.998&0.248\\
\\
-&1.277&2.857\\
+&0.435&0.727\\
+&1.218&0.693\\
+&1.096&0.633\\
+&1.243&0.557\\
+&1.209&0.528\\
\\
-&1.318&4.079\\
-&0.345&0.513\\
-&1.197&0.190\\
+&1.015&0.344\\
+&1.228&0.339\\
+&1.109&0.331\\
\\
-&1.342&4.453\\
-&0.304&0.950\\
-&1.215&0.237\\
+&0.987&0.199\\
-&1.228&0.231\\
+&1.091&0.232\\
\\
-&1.635&2.419\\
+&0.516&1.024\\
+&1.438&1.070\\
+&1.278&1.043\\
+&1.319&0.975\\
+&1.388&0.917\\
\\
-&1.551&4.088\\
+&0.415&0.419\\
+&1.373&0.284\\
+&1.141&0.402\\
+&1.303&0.395\\
+&1.270&0.402\\
\\
-&1.544&4.516\\
+&0.384&0.780\\
+&1.374&0.246\\
+&1.108&0.251\\
+&1.299&0.246\\
+&1.245&0.240\\
\\
\end{array}
&
\begin{array}{crr}
{\dR}?&{|{L_{\rm br,1}}|}\over{{\varepsilon}L_{\rm br,0}}&{|{L_{\rm cy,1}}|}\over{{\varepsilon}L_{\rm cy,0}}\\
\hline\\
-&0.534&0.796\\
-&0.203&0.405\\
-&0.471&1.156\\
-&0.566&0.927\\
-&0.544&0.728\\
-&0.741&0.594\\
\\
-&1.128&4.057\\
-&0.145&0.825\\
-&0.471&0.964\\
-&0.525&1.093\\
-&0.484&0.923\\
-&0.688&0.795\\
\\
-&1.244&5.198\\
-&0.246&1.072\\
-&0.352&0.400\\
-&0.550&0.806\\
-&0.416&0.924\\
-&0.641&0.863\\
\\
-&0.661&0.695\\
-&0.279&1.751\\
-&0.612&1.494\\
+&0.693&1.429\\
+&0.860&1.315\\
+&0.923&1.215\\
\\
-&1.254&3.600\\
-&0.183&0.429\\
-&0.715&0.950\\
-&0.652&1.173\\
-&0.804&1.143\\
-&1.102&1.080\\
\\
-&1.425&4.649\\
-&0.294&0.898\\
-&0.677&0.171\\
-&0.666&0.758\\
-&0.643&0.929\\
-&1.109&0.959\\
\\
+&0.738&0.580\\
+&0.350&2.706\\
+&0.654&2.588\\
+&0.814&2.533\\
+&1.192&2.430\\
+&1.013&2.311\\
\\
-&1.277&3.048\\
+&0.211&0.512\\
-&0.806&1.494\\
-&0.662&1.737\\
-&1.128&1.778\\
-&1.202&1.739\\
\\
-&1.441&4.218\\
-&0.311&0.462\\
-&0.822&0.615\\
-&0.672&1.115\\
-&0.905&1.319\\
-&1.254&1.390\\
\\
\end{array}
\\
\hline
\end{array}
$
\end{center}
\label{'table.luminosity.relamp'}
\end{table}

\begin{table}
\scriptsize
\tiny
\caption{
Total bremsstrahlung and cyclotron emission phases,
$\Phi_{\rm br}$ and $\Phi_{\rm cy}$,
expressed as multiples of $\pi$,
for the modes $n=1\ldots6$ from top to bottom,
under given $(\sigma\subs{},\psiei,\epsilon\subs{})$ conditions.
In this convention the phase of the shock-position oscillation is defined as
$\Phi\subs{}=0$.
In each set, the third column described the difference between the phases of
integrated cyclotron- and bremsstrahlung-luminosity oscillations,
$\Phi_{\rm cy}(n)-\Phi_{\rm br}(n)$.
Negative values indicate the cyclotron luminosity oscillation
following the bremsstrahlung luminosity oscillation;
and positive values indicate cyclotron luminosity
leading bremsstrahlung luminosity.
}
\begin{center}
$
\begin{array}{rrccc}
\hline
\sigma\subs{}&\psiei&\epsilon\subs{}=0&\epsilon\subs{}=1&\epsilon\subs{}=100\\
\begin{array}{r}~\\~
\\0.2\\~\\~\\~\\~\\~\\
\\0.2\\~\\~\\~\\~\\~\\
\\0.2\\~\\~\\~\\~\\~\\
\\0.5\\~\\~\\~\\~\\~\\
\\0.5\\~\\~\\~\\~\\~\\
\\0.5\\~\\~\\~\\~\\~\\
\\1.0\\~\\~\\~\\~\\~\\
\\1.0\\~\\~\\~\\~\\~\\
\\1.0\\~\\~\\~\\~\\~\\
\\
\end{array}&
\begin{array}{r}~\\~
\\0.1\\~\\~\\~\\~\\~\\
\\0.5\\~\\~\\~\\~\\~\\
\\1.0\\~\\~\\~\\~\\~\\
\\0.1\\~\\~\\~\\~\\~\\
\\0.5\\~\\~\\~\\~\\~\\
\\1.0\\~\\~\\~\\~\\~\\
\\0.1\\~\\~\\~\\~\\~\\
\\0.5\\~\\~\\~\\~\\~\\
\\1.0\\~\\~\\~\\~\\~\\
\\
\end{array}&
\begin{array}{rrr}
\Phi_{\rm br}-\Phi\subs{}&\Phi_{\rm cy}-\Phi\subs{}&\Phi_{\rm cy}-\Phi_{\rm br}\\
\hline\\
-0.366&*&*\\
-0.763&*&*\\
-0.450&*&*\\
 0.141&*&*\\
 0.936&*&*\\
-0.454&*&*\\
\\
-0.374&*&*\\
-0.766&*&*\\
-0.458&*&*\\
 0.116&*&*\\
 0.913&*&*\\
-0.487&*&*\\
\\
-0.376&*&*\\
-0.767&*&*\\
-0.461&*&*\\
 0.111&*&*\\
 0.909&*&*\\
-0.494&*&*\\
\\
-0.361&*&*\\
-0.762&*&*\\
-0.448&*&*\\
 0.123&*&*\\
 0.931&*&*\\
-0.466&*&*\\
\\
-0.373&*&*\\
-0.766&*&*\\
-0.458&*&*\\
 0.112&*&*\\
 0.911&*&*\\
-0.490&*&*\\
\\
-0.375&*&*\\
-0.768&*&*\\
-0.461&*&*\\
 0.123&*&*\\
 0.907&*&*\\
-0.496&*&*\\
\\
-0.352&*&*\\
-0.757&*&*\\
-0.444&*&*\\
 0.091&*&*\\
 0.924&*&*\\
-0.487&*&*\\
\\
-0.371&*&*\\
-0.766&*&*\\
-0.458&*&*\\
 0.104&*&*\\
 0.906&*&*\\
-0.494&*&*\\
\\
-0.374&*&*\\
-0.768&*&*\\
-0.461&*&*\\
 0.105&*&*\\
 0.904&*&*\\
-0.499&*&*\\
\\
\end{array}
&
\begin{array}{rrr}
\Phi_{\rm br}-\Phi\subs{}&\Phi_{\rm cy}-\Phi\subs{}&\Phi_{\rm cy}-\Phi_{\rm br}\\
\hline\\
-0.353&-0.986&-0.663\\
 0.573&-0.336&-0.909\\
-0.891&-0.319& 0.572\\
-0.333&-0.263& 0.070\\
 0.305&-0.280&-0.585\\
 0.892&-0.302&-0.194\\
\\
-0.347&-0.989&-0.642\\
 0.517&-0.989& 0.494\\
-0.957&-0.334& 0.623\\
-0.441&-0.207& 0.234\\
 0.175&-0.168&-0.343\\
 0.674&-0.170&-0.844\\
\\
-0.339&-0.981&-0.642\\
 0.396& 0.993& 0.597\\
-0.973& 0.934&-0.093\\
-0.487&-0.107& 0.380\\
 0.147&-0.050&-0.197\\
 0.612&-0.071&-0.683\\
\\
-0.348&-0.976&-0.628\\
 0.635&-0.157&-0.792\\
-0.771&-0.192& 0.579\\
-0.863&-0.284& 0.579\\
 0.484&-0.252&-0.736\\
-0.863&-0.284& 0.579\\
\\
-0.349&-0.995&-0.646\\
 0.615& 0.825& 0.210\\
-0.820& 0.077& 0.897\\
-0.240&-0.024& 0.216\\
 0.406&-0.054&-0.460\\
-0.974&-0.105& 0.869\\
\\
-0.347&-0.992&-0.645\\
 0.593& 0.884& 0.291\\
-0.833& 0.660&-0.507\\
-0.260& 0.235& 0.495\\
 0.388& 0.137&-0.251\\
 0.996& 0.047&-0.949\\
\\
-0.342&-0.889&-0.547\\
 0.700&-0.134&-0.834\\
-0.690&-0.195& 0.495\\
-0.109&-0.243&-0.134\\
 0.573&-0.292&-0.865\\
-0.772&-0.325& 0.447\\
\\
-0.352&-0.986&-0.634\\
 0.684& 0.689& 0.005\\
-0.734& 0.113& 0.847\\
-0.141& 0.008& 0.149\\
 0.533&-0.047&-0.580\\
-0.817&-0.111& 0.706\\
\\
-0.353&-0.991&-0.638\\
 0.680& 0.807& 0.127\\
-0.743& 0.525&-0.732\\
-0.151& 0.241& 0.392\\
 0.526& 0.160&-0.366\\
-0.830& 0.010& 0.840\\
\\
\end{array}
&
\begin{array}{rrr}
\Phi_{\rm br}-\Phi\subs{}&\Phi_{\rm cy}-\Phi\subs{}&\Phi_{\rm cy}-\Phi_{\rm br}\\
\hline\\
-0.542& 0.928&-0.530\\
 0.388&-0.385&-0.773\\
 0.770&-0.429& 0.801\\
-0.855&-0.295& 0.560\\
-0.325&-0.271& 0.054\\
 0.144&-0.281&-0.425\\
\\
-0.410& 0.954&-0.636\\
-0.163&-0.645&-0.482\\
 0.725&-0.367& 0.908\\
 0.955&-0.373& 0.672\\
-0.600&-0.324& 0.276\\
-0.137&-0.265&-0.128\\
\\
-0.340&-0.984&-0.644\\
-0.284&-0.827&-0.543\\
 0.810&-0.607& 0.583\\
 0.879&-0.335& 0.786\\
-0.868&-0.324& 0.544\\
-0.244&-0.299&-0.055\\
\\
-0.486&-0.981&-0.495\\
 0.402&-0.231&-0.633\\
 0.775&-0.251& 0.974\\
-0.738&-0.192& 0.546\\
-0.241&-0.182& 0.059\\
 0.219&-0.195&-0.414\\
\\
-0.415& 0.976&-0.609\\
-0.047&-0.585&-0.538\\
 0.743&-0.202&-0.945\\
-0.923& 0.270& 0.807\\
-0.399&-0.211& 0.188\\
-0.021&-0.189&-0.168\\
\\
-0.359&-0.980&-0.621\\
-0.160&-0.878&-0.718\\
 0.782& 0.171&-0.611\\
 0.986&-0.187& 0.827\\
-0.458& 0.193& 0.651\\
-0.083&-0.187&-0.104\\
\\
-0.474&-0.299& 0.175\\
 0.418&-0.104&-0.522\\
 0.762&-0.146&-0.908\\
-0.639& 0.497&-0.864\\
-0.227&-0.159& 0.068\\
 0.251&-0.181&-0.432\\
\\
-0.428& 0.994&-0.578\\
 0.039&-0.067&-0.106\\
 0.731&-0.070&-0.801\\
-0.814&-0.121& 0.693\\
-0.306&-0.137& 0.169\\
 0.039&-0.149&-0.188\\
\\
-0.378&-0.975&-0.597\\
-0.088&-0.950&-0.862\\
 0.755& 0.056&-0.699\\
-0.928&-0.063& 0.865\\
-0.328&-0.093& 0.235\\
-0.008&-0.115&-0.107\\
\\
\end{array}
\\
\hline
\end{array}
$
\end{center}
\label{'table.luminosity.phases'}
\label{'table.luminosity.phasediff'}
\end{table}

\section{Transverse perturbation}
\label{'transverse'}

\subsection{eigen-function profiles}

\begin{figure}
\begin{center}
\epsfxsize=17cm
\epsfbox{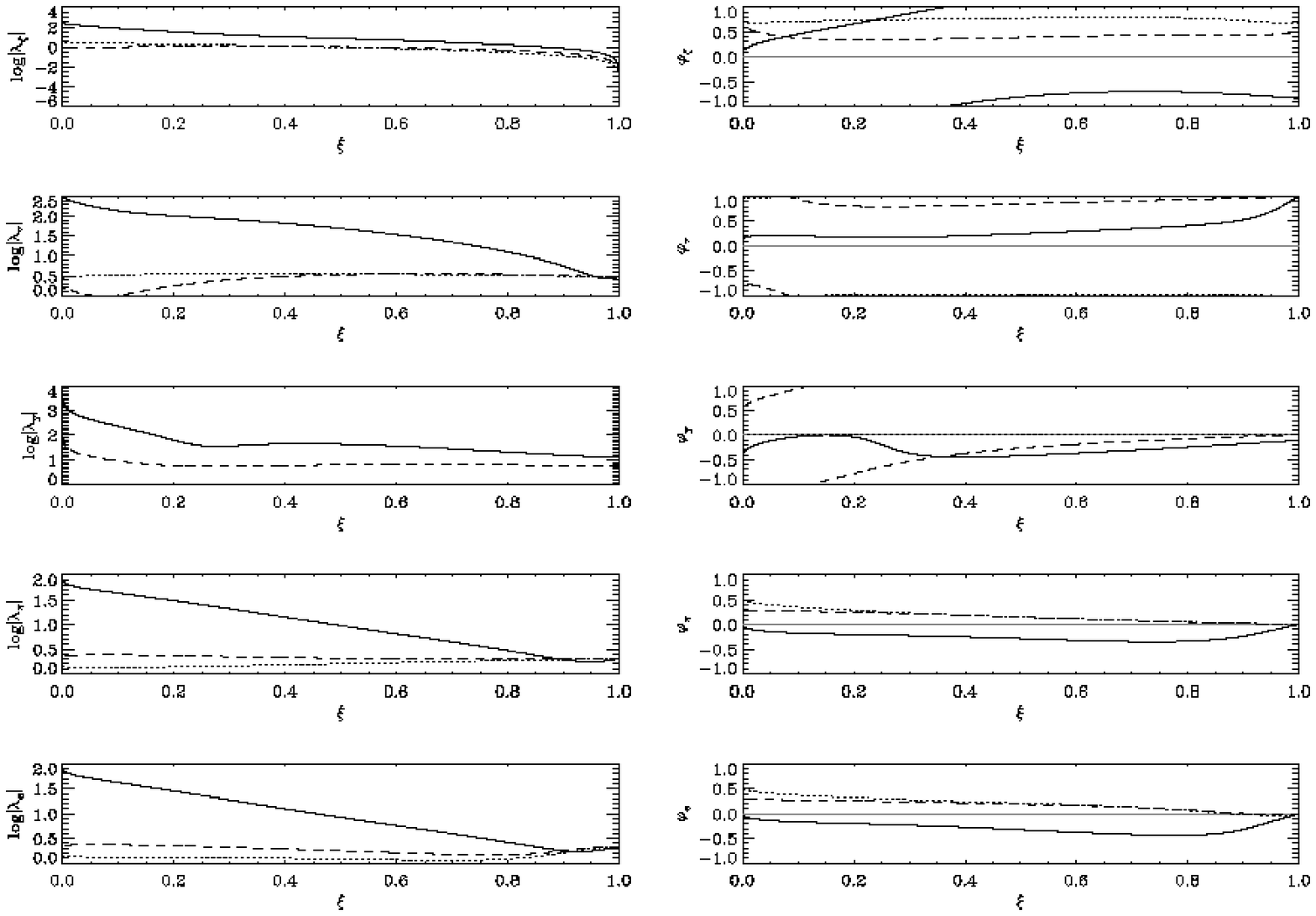}
\end{center}
\caption{
Eigenfunction profiles in the presence of transverse perturbation,
for density, longitudinal velocity, transverse velocity,
total pressure and electron pressure (from top to bottom),
for the $n=1$ mode with
$(\sigma\subs,\psiei,\epsilon\subs{})=(0.5,0.5,1)$.
Profiles for wavenumbers $\kappa=0,1,4$
are plotted in dotted, dashed and solid curves respectively.
The left column shows logarithms of the amplitudes,
and the right column shows the phases.
Phases are multiples of $\pi$.
The dotted line in the phase plots gives reference for zero phase,
which is defined to be the phase of the oscillation of shock height
(marked in grey).
}
\label{'b2befk1'}
\end{figure}

\begin{figure}
\begin{center}
\epsfxsize=17cm
\epsfbox{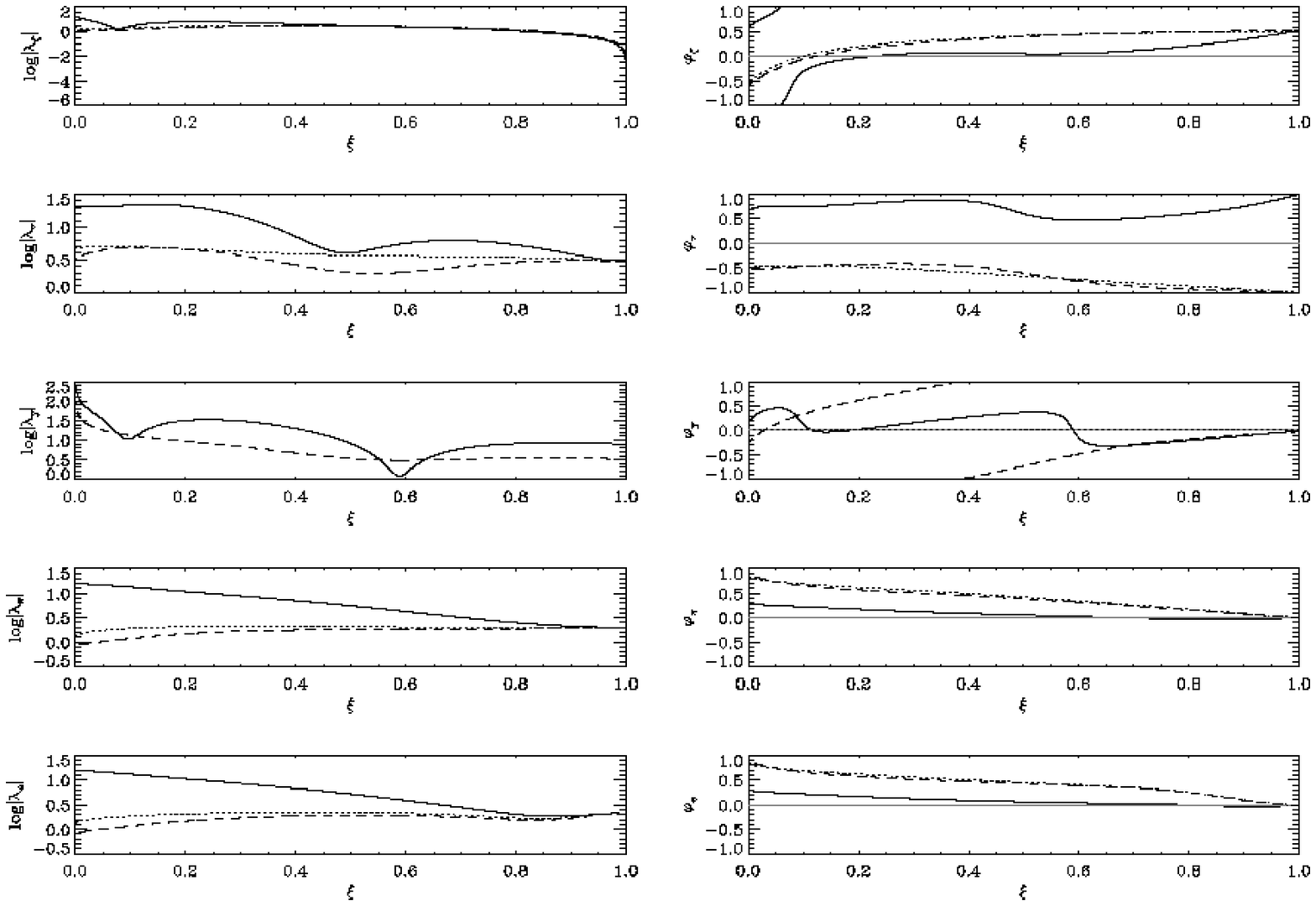}
\end{center}
\caption{
Same as Figure~\ref{'b2befk1'} but for
$n=2$.
}
\label{'b2sefk2'}
\end{figure}

In the absence of transverse perturbations ($\kappa=0$),
the transverse velocity eigenfunction $\lambda_y$
is zero everywhere.
When $\kappa>0$, 
the profiles of the other eigenfunctions are modified
(see Figures~\ref{'b2befk1'}, \ref{'b2sefk2'}).
The amplitude $|\lambda_y|$ generally has its maximum value
near the lower boundary,
and from that value it declines steeply in $\xi$
to the value fixed by the boundary condition at the shock.
Increasing $\kappa$ causes the amplitude $|\lambda_y|$
to increase in the region near the lower boundary,
and the slope $d|\lambda_y|/d\xi$ steepens throughout the profile.
Greater $\epsilon\subs{}$ makes the slope steeper,
when $\kappa$ is fixed.
For some $\kappa$ there are one or more local minima
in the $|\lambda_y|$ profile,
with the number of minima depending on the harmonic number $n$.

The transverse velocity phase $\varphi_y$
generally winds in a negative sense
from the shock to the lower boundary
(see \ref{'subsection.winding'}).
The total number of turns of $\varphi_y$ increases
with $\kappa$ until a threshold is reached.
Beyond the threshold there is no winding in $\varphi_y$
(see Figures~\ref{'b2befk1'}-\ref{'b2sefk2'}, third panel, right column).
The lower boundary values of
$|\lambda_\zeta|$, $|\lambda_\tau|$, $|\lambda_\pi|$
and $|\lambda\sube{}|$ all increase with $\kappa$
beyond the $\kappa$-threshold.

The $|\lambda_\zeta|$ and $|\lambda_\tau|$ profiles
have the most distinctive features.
However the $\lambda_\zeta$ and $\lambda_\tau$ features disappear
when $\kappa$ is sufficiently large.
These $\kappa$-dependent properties
depend on the harmonic number $n$ and the system parameters
(\eg $\kappa\GS 4$ for $(\sigma\subs{},\psiei,\epsilon\subs{})=(0.5,0.5,100)$
and $n=1,2$,
or $\kappa\GS 8$ for $(\sigma\subs{},\psiei,\epsilon\subs{})=(0.5,0.5,0)$ 
and $n=1,2$).

The amplitude and phase profiles of the total pressure $\lambda_\pi$
and electron pressure $\lambda\sube{}$
are less dependent upon $\kappa$
than the density and velocity eigenfunctions are.
In the case of $(\sigma\subs{},\psiei)=(0.5,0.5)$
the eigenfunctions for total pressure and electron pressure
are not greatly affected by the introduction of a transverse perturbation
when $\kappa$ is small.
The total pressure and electron pressure eigenfunctions are much alike because
the amplitude and phase profiles match
at both the upper and lower boundaries.
Increasing $\kappa$ causes decrease of
the lower boundary pressure phase,
$\varphi\sube{}|_{\xi=0}=\varphi_\pi|_{\xi=0}$.
The phases $\varphi_\pi$ and $\varphi\sube{}$
have similar profiles for small-$\epsilon\subs{}$ flows,
but the electron pressure develops
phase jumps and associated node-like amplitude features
when $\epsilon\subs{}$ is large.
(For a more detailed discussion of the transverse perturbation,
see Saxton 1999.)
 

\subsection{Luminosity response} 
 
\begin{figure}
\begin{center}
\begin{tabular}{ccc}
\epsfxsize=5cm
\epsfbox{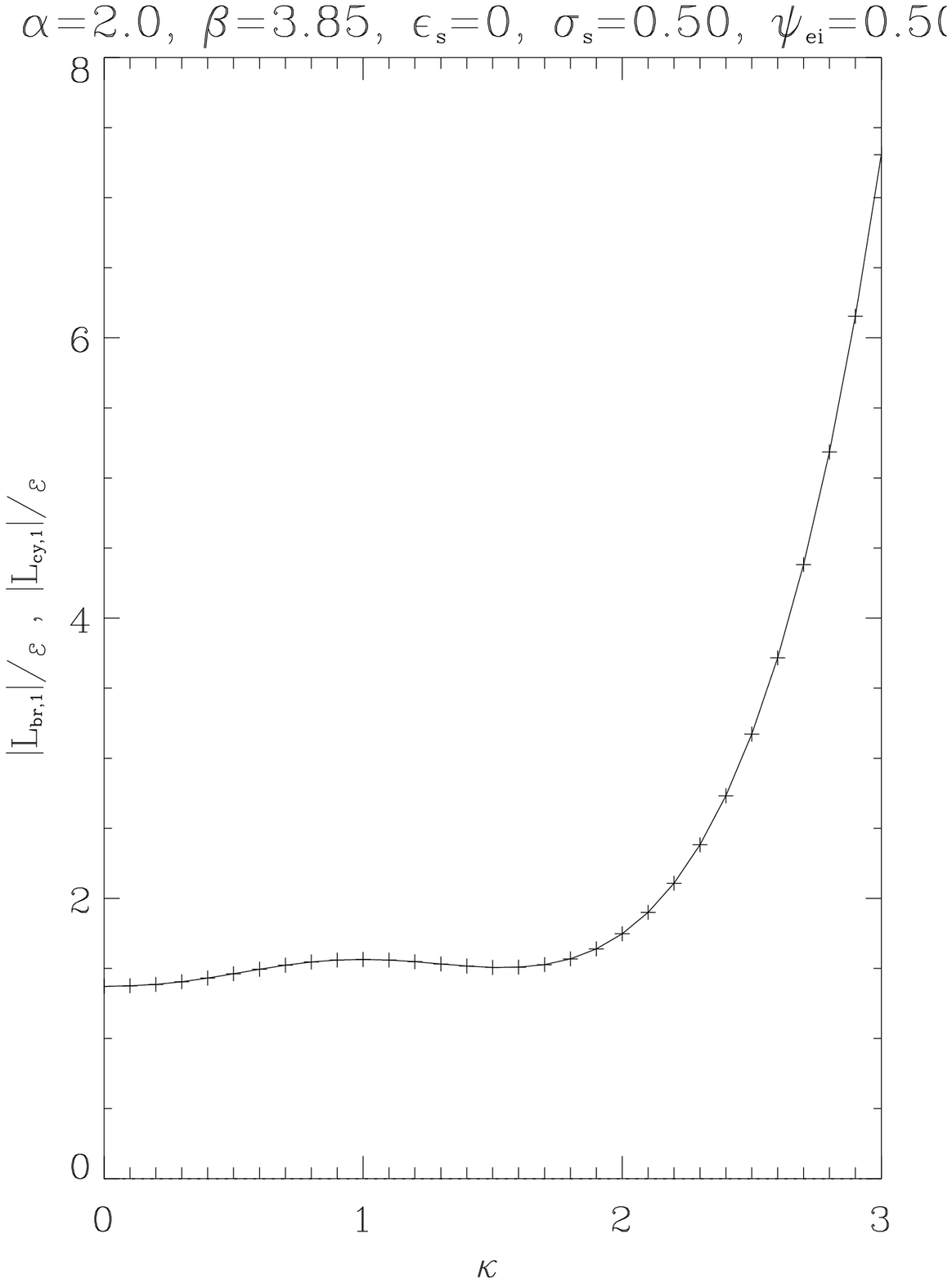}
&
\epsfxsize=5cm
\epsfbox{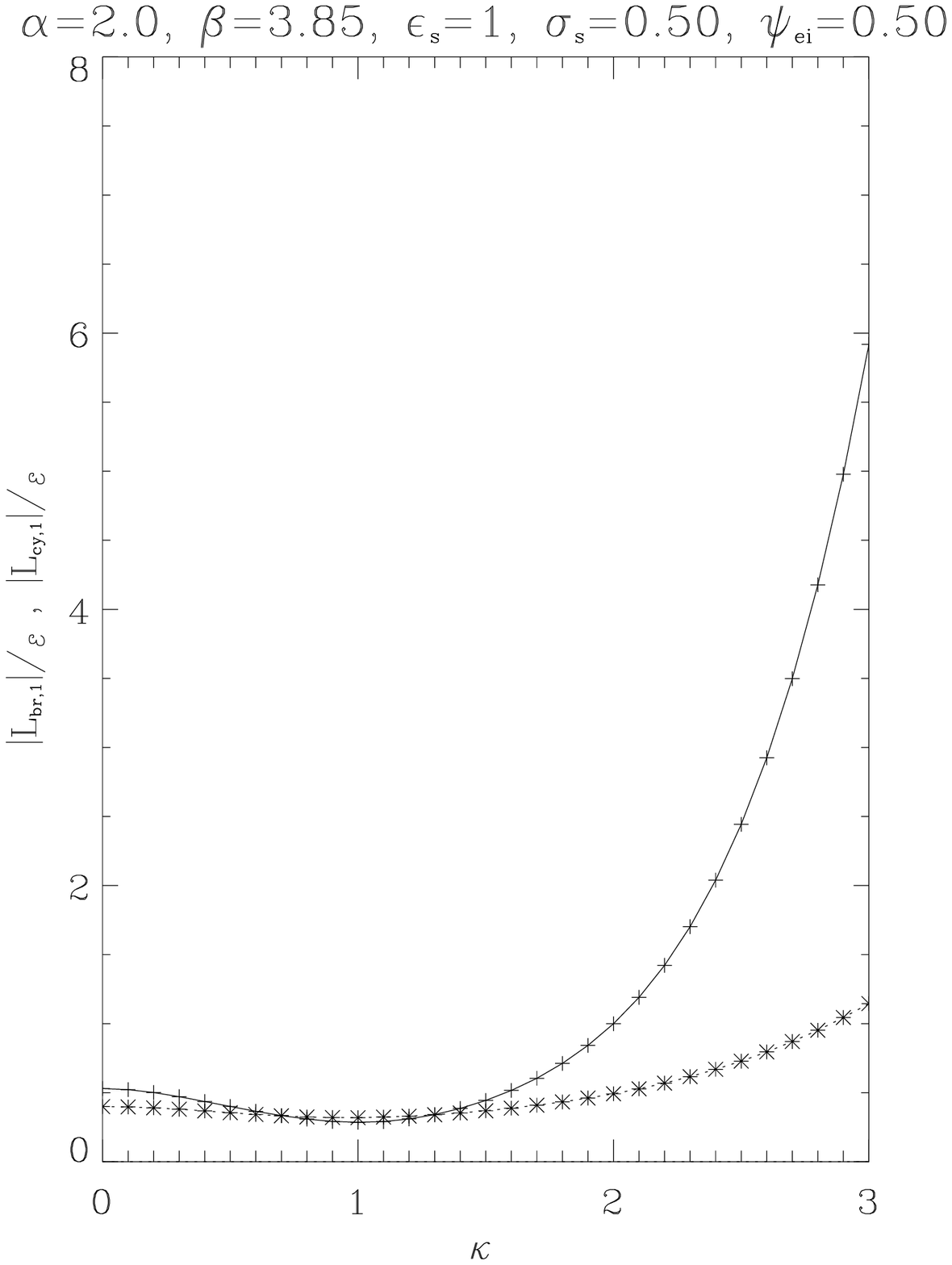}
&
\epsfxsize=5cm
\epsfbox{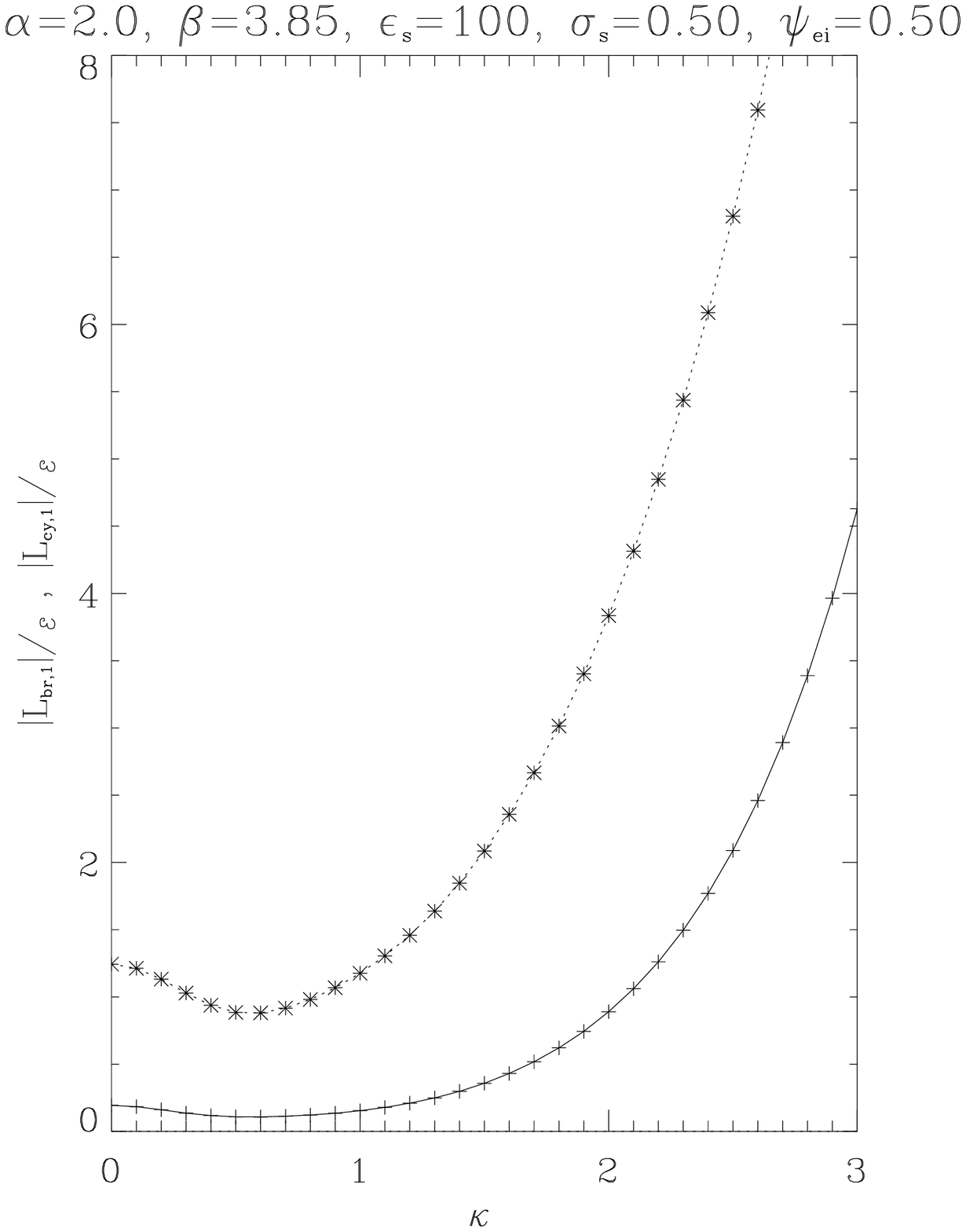}
\\
\epsfxsize=5cm
\epsfbox{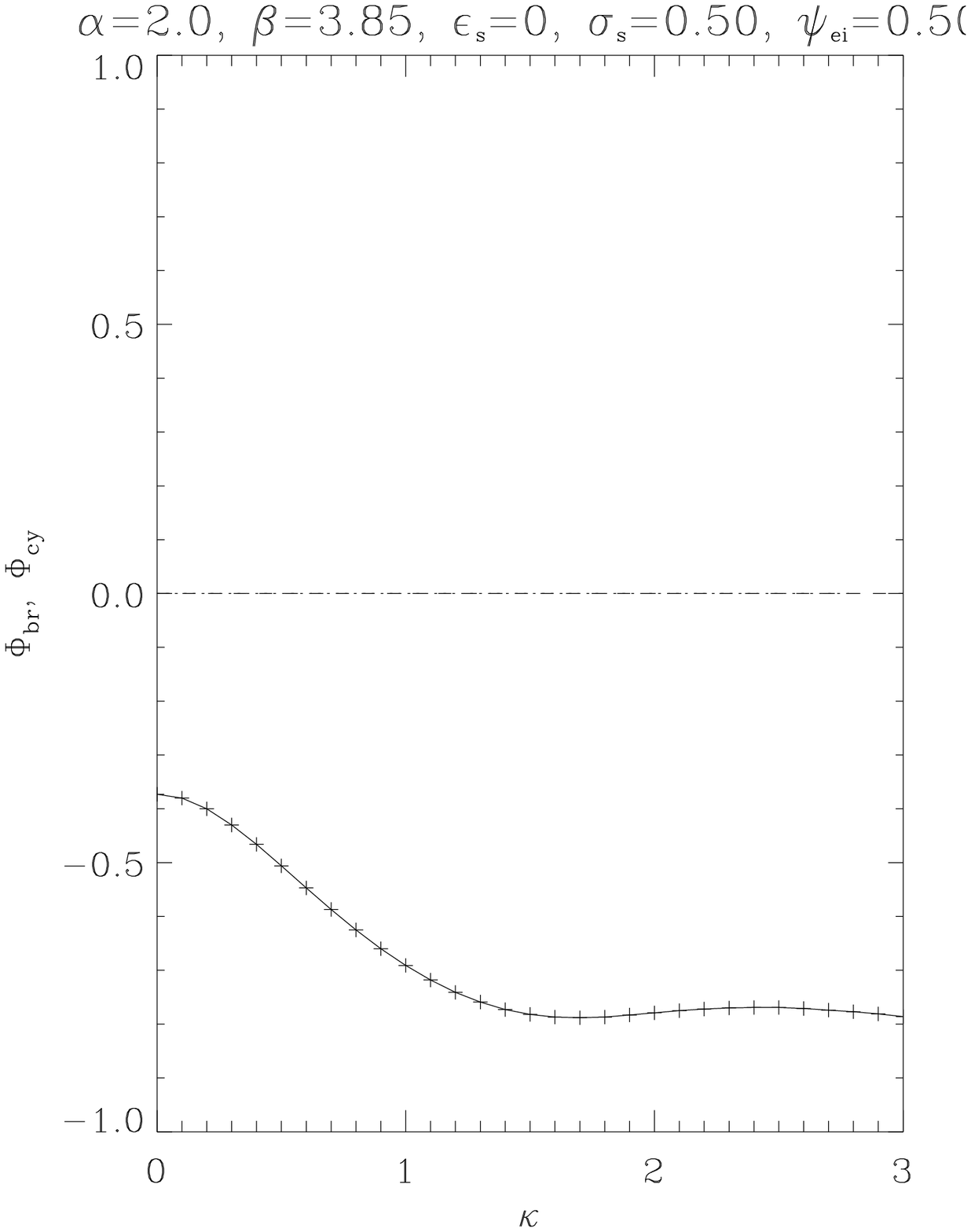}
&
\epsfxsize=5cm
\epsfbox{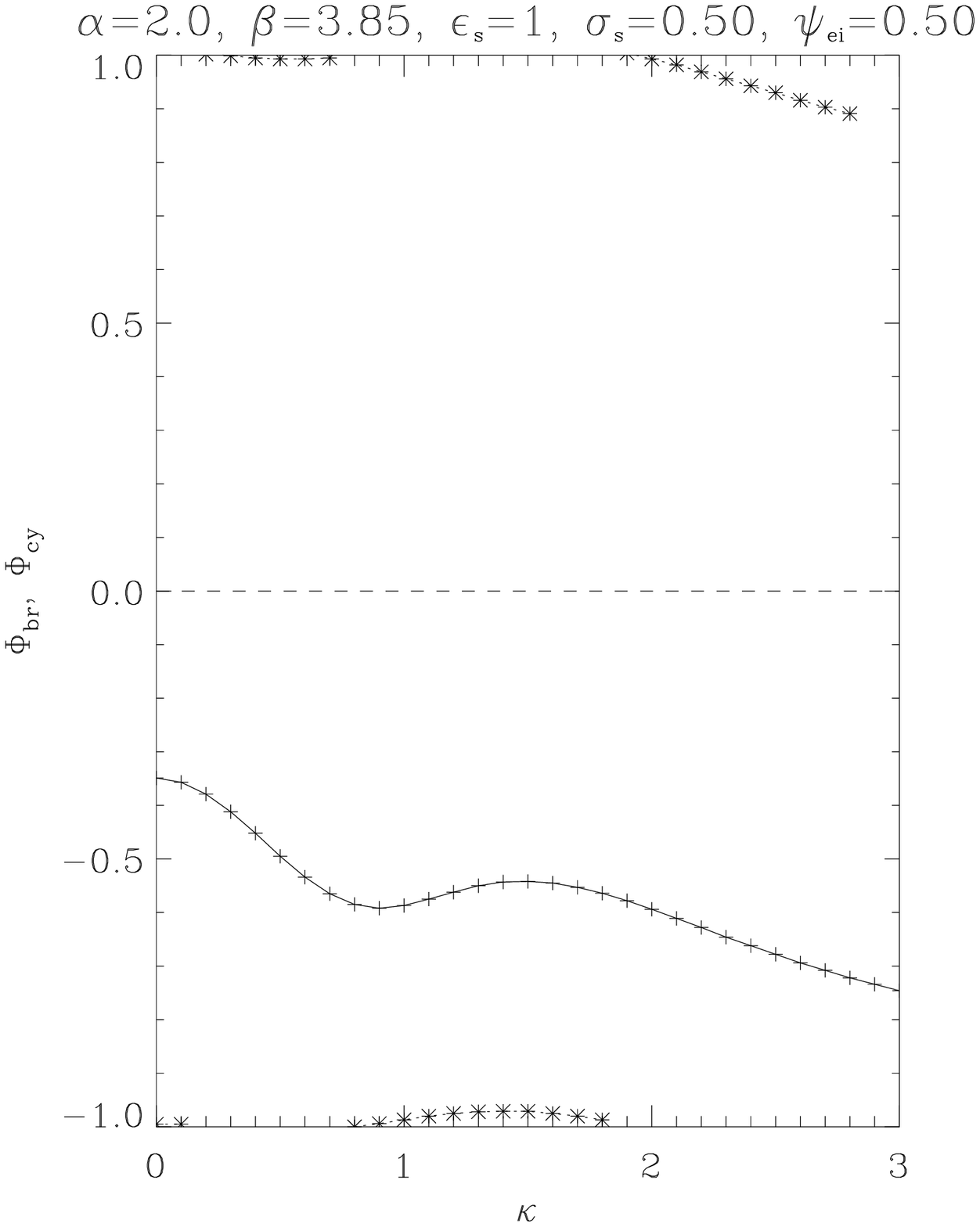}
&
\epsfxsize=5cm
\epsfbox{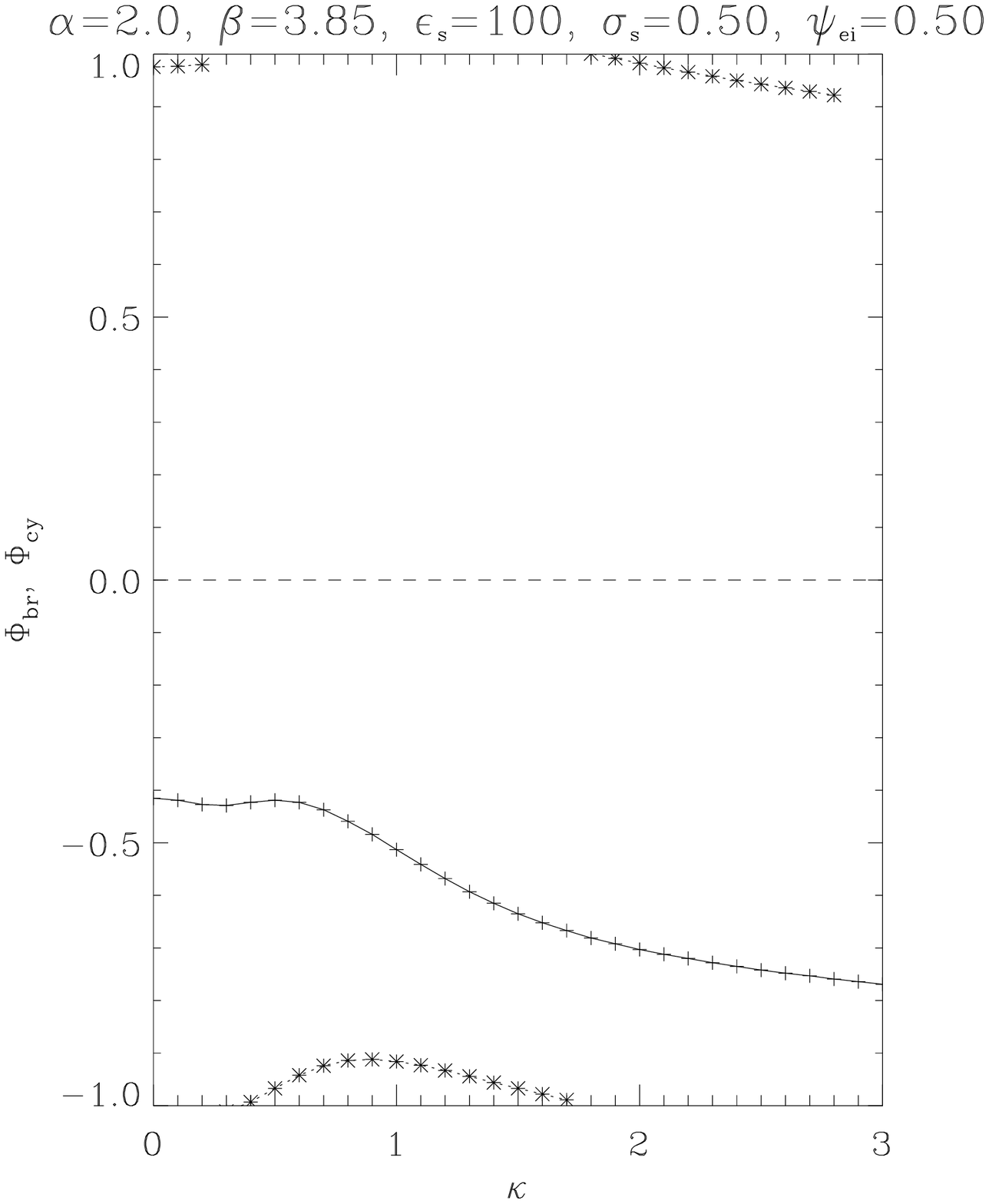}
\end{tabular}
\end{center}
\scriptsize
\caption{
Integrated luminosity response amplitudes and phases for $n=1$ mode
in the presence of transverse perturbations,
for the parameters
$(\sigma\subs{},\psiei)=(0.5,0.5)$
and $\epsilon\subs{}=0,1,100$ from left to right.
The curves corresponding to bremsstrahlung cooling 
are marked with $+$;
the curves corresponding to cyclotron cooling are marked with 
$\times$.
}
\label{'p05s05L1.ps'}
\end{figure}

\begin{figure}
\begin{center}
\begin{tabular}{ccc}
\epsfxsize=5cm
\epsfbox{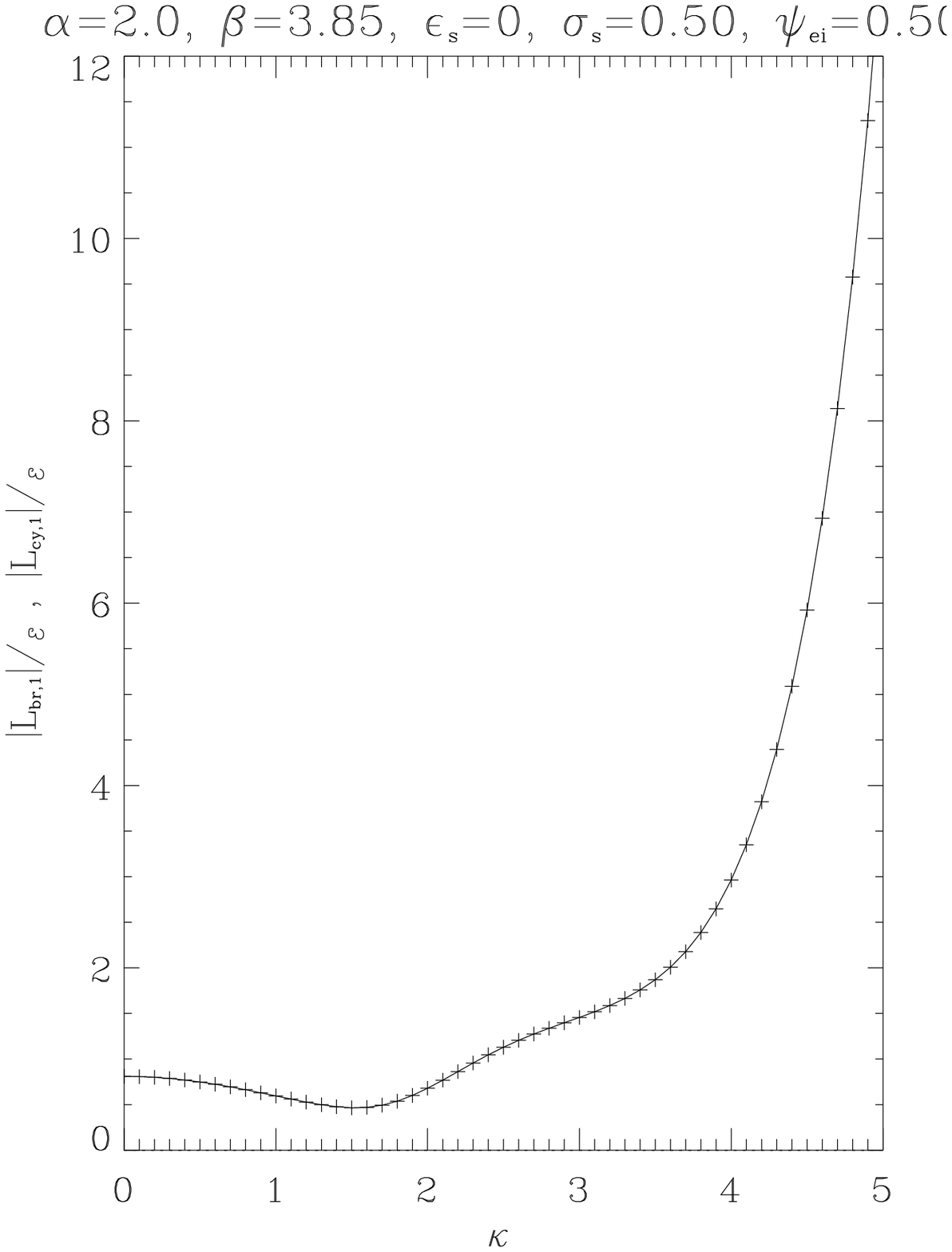}
&
\epsfxsize=5cm
\epsfbox{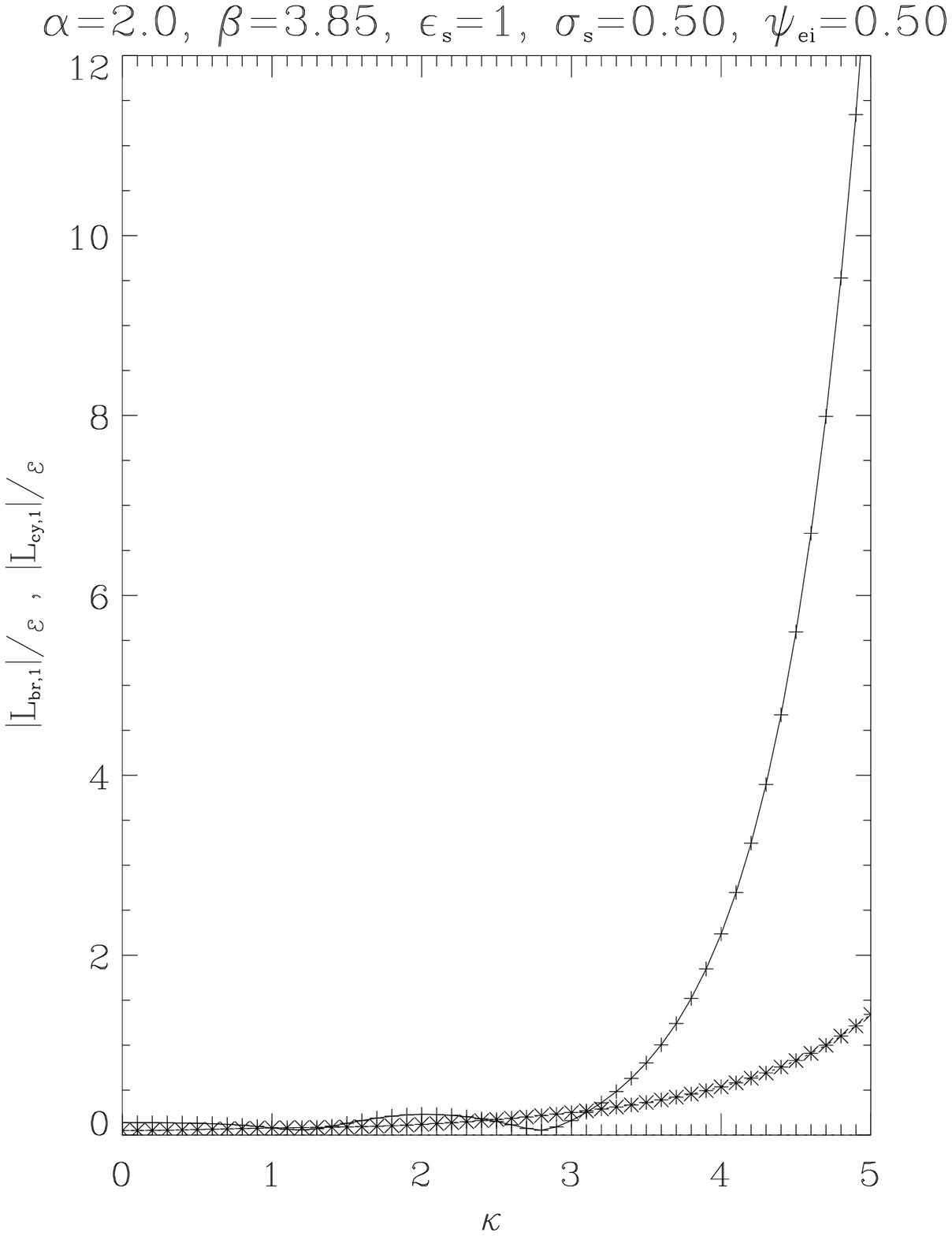}
&
\epsfxsize=5cm
\epsfbox{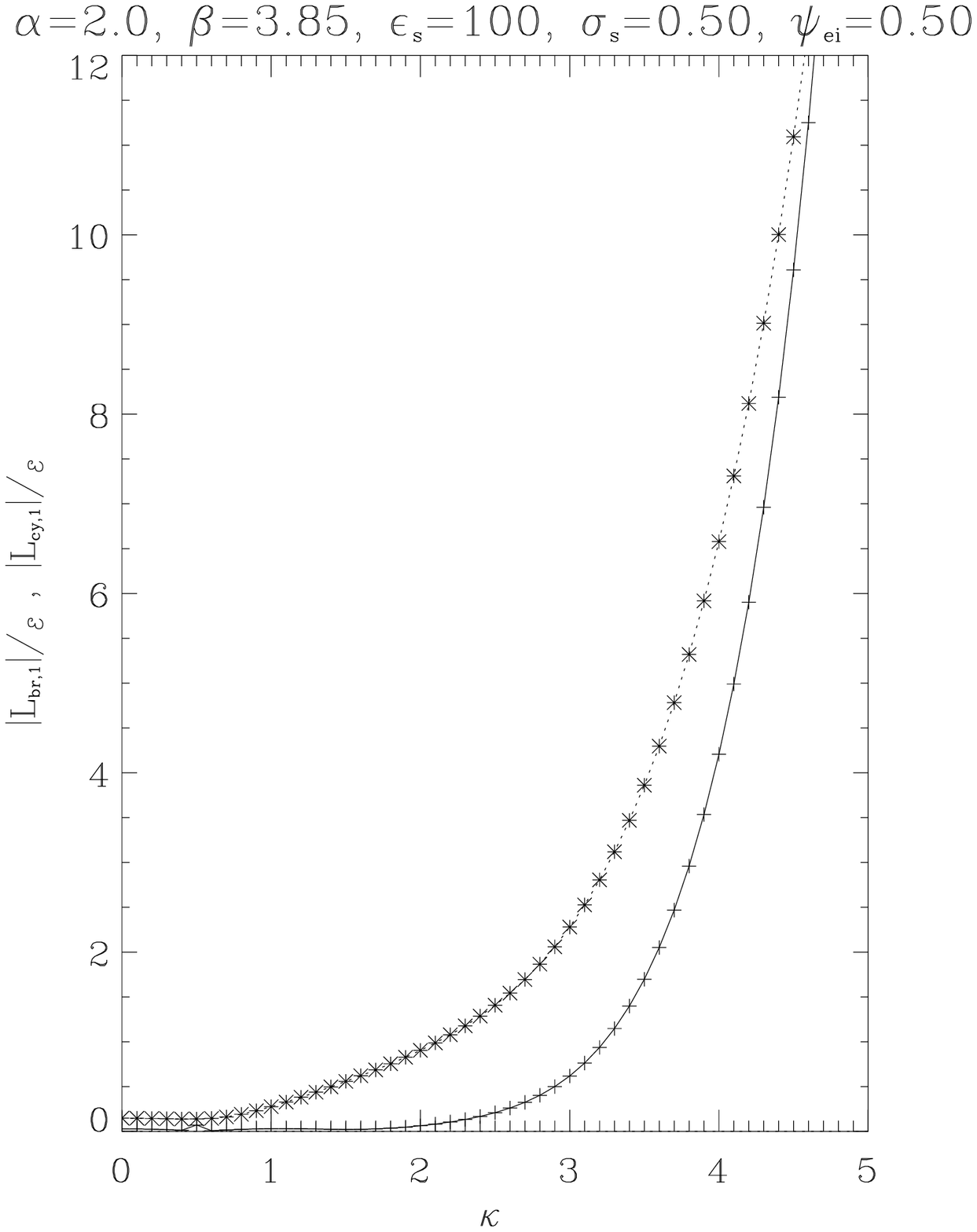}
\\
\epsfxsize=5cm
\epsfbox{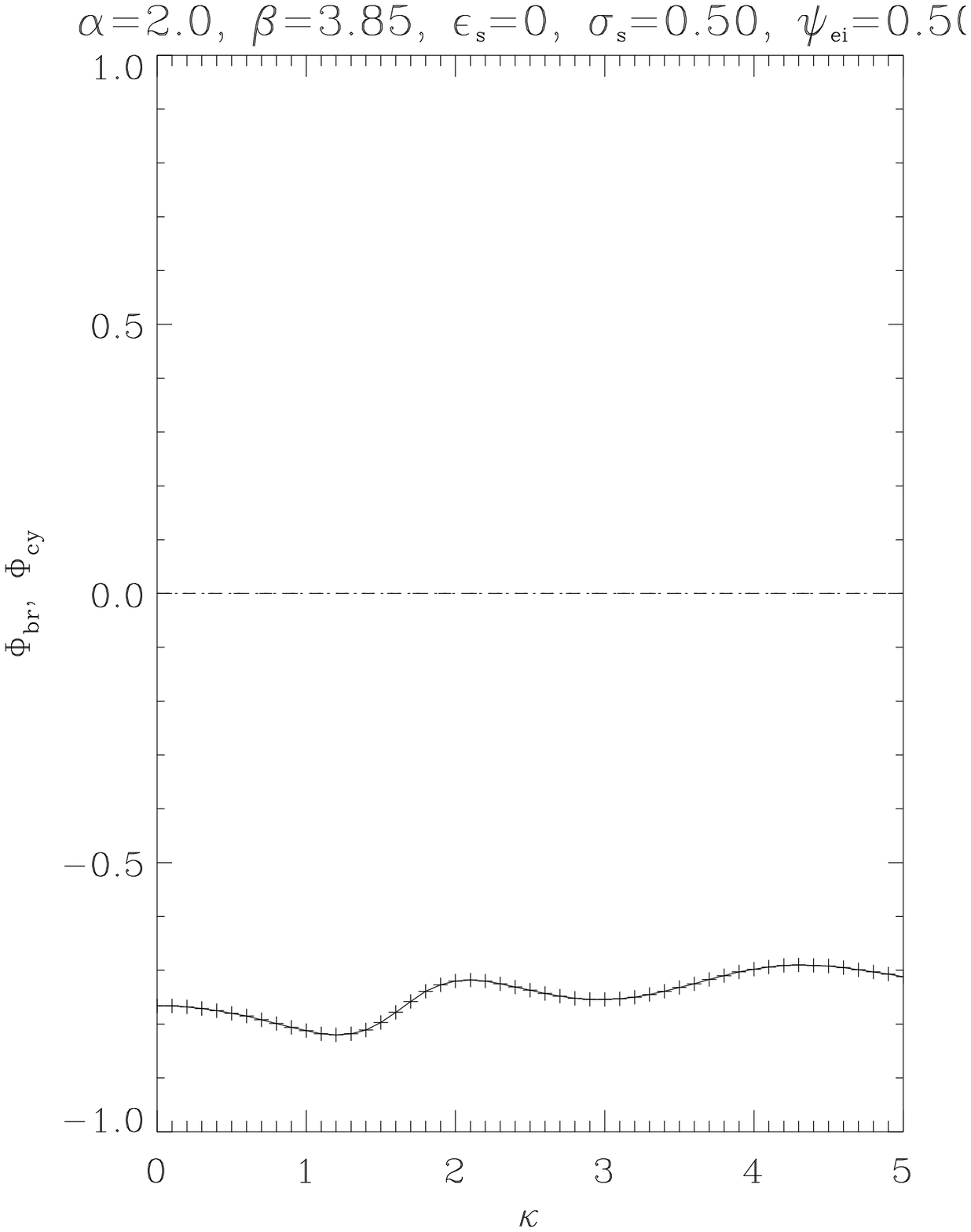}
&
\epsfxsize=5cm
\epsfbox{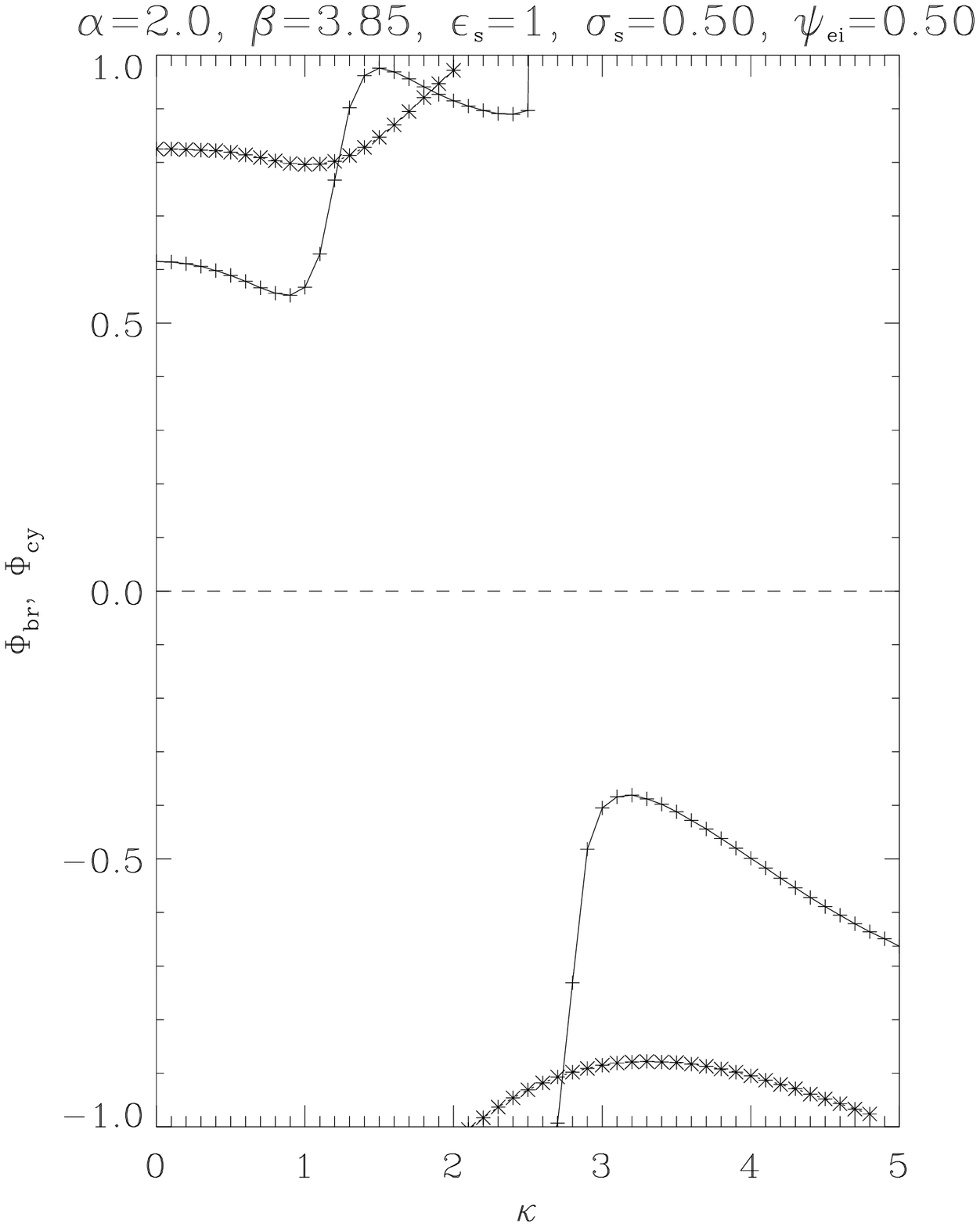}
&
\epsfxsize=5cm
\epsfbox{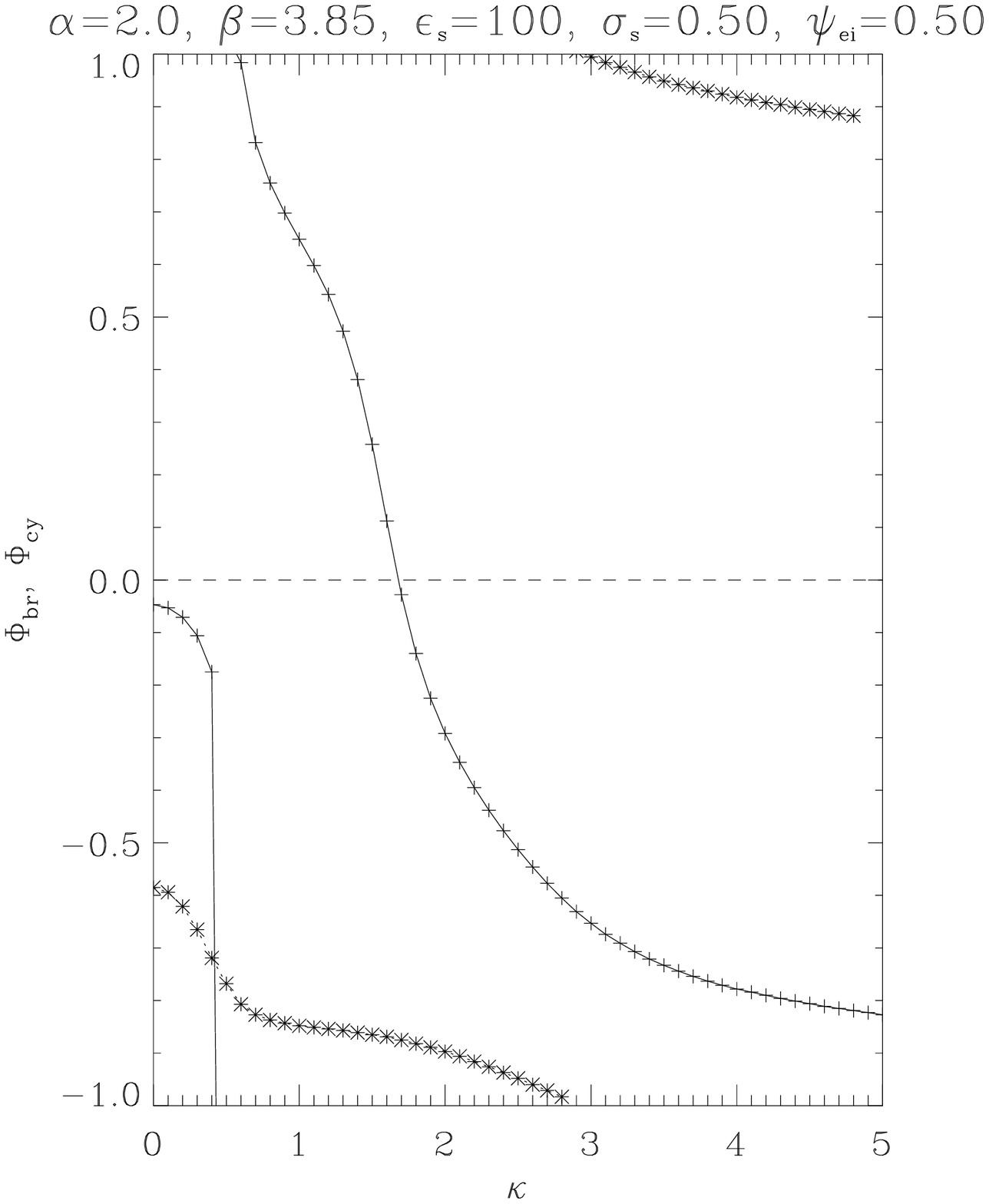}
\end{tabular}
\end{center}
\scriptsize
\caption{
Same as Figure~\ref{'p05s05L1.ps'} but for the $n=2$ mode.
}
\label{'p05s05L2.ps'}
\end{figure}

Figures\ \ref{'p05s05L1.ps'}-\ref{'p05s05L2.ps'}
show the luminosity responses as functions of $\kappa$ for $n=1$ and $2$.
The integrated luminosity amplitudes
$|L_{\rm br,1}|$ and $|L_{\rm cy,1}|$
have minima in $\kappa$
where there are abrupt changes of
the phases $\Phi_{\rm br}$ and $\Phi_{\rm cy}$ respectively.
The number of $\kappa$-minima is determined by $n$.
For $n=1$ and $(\sigma\subs{},\psiei)=(0.5,0.5)$
the $|L_{\rm br,1}|$ minima occur at about
$\kappa\sim1.6,1.0,0.6$ for $\epsilon\subs{}=0,1,100$
and the $|L_{\rm cy,1}|$ minima coincide at approximately
the same $\kappa$-values.
For $n=2$
there are at most two minima of $|L_{\rm br,1}|$.
For small $\epsilon\subs{}$
the minimum corresponding to a larger $\kappa$-value
becomes an inflection,
\eg for $\epsilon\subs{}=0$
the actual amplitude minimum of $|L_{\rm br,1}|$ is at
$\kappa\approx1.8$
and the inflection point is at $\kappa\approx3.4$.
 
Both amplitudes, $|L_{\rm br,1}|$ and $|L_{\rm cy,1}|$
are slowly varying in $\kappa$ when the value of $\kappa$ is below
$\kappa_*$,
where $\kappa_*(n,\epsilon\subs{})
\approx{\frac12}(2n-1)\kappa_\epsilon(\epsilon\subs{})$
and $\kappa_\epsilon$ depends on the system parameters.
For $(\sigma\subs{},\psiei)=(0.5,0.5)$,
$\kappa_\epsilon\sim1.0$ for $\epsilon\subs{}=0$.
It reduces gradually as the cooling efficiency increases,
and $\kappa_\epsilon\sim0.3$ at $\epsilon\subs{}=100$.
For wavenumbers $\kappa>\kappa_*$,
both $|L_{\rm br,1}|$ and $|L_{\rm cy,1}|$
increase rapidly with $\kappa$.
This is a consequence of the increasing amplitudes
of the hydrodynamic variables' $\lambda$-functions 
as $\kappa$ increases.
Because bremsstrahlung cooling is most efficient
in regions near the lower boundary,
$|L_{\rm br,1}|$ tends to rise more steeply with $\kappa$
than $|L_{\rm cy,1}|$ does.
 
The integrated luminosity phases
$\Phi_{\rm br}$ and $\Phi_{\rm cy}$
wind with $\kappa$.
On top of these winding trends, there are phase jumps
where the respective $|L_{\rm br,1}|$, $|L_{\rm cy,1}|$ amplitudes reach minima.
Away from the minima, both of the phases generally decrease
when $\kappa$ increases.
Generally the winding of both phases in $\kappa$ is more rapid
when $\epsilon\subs{}$ is large,
however for sufficiently large $\epsilon\subs{}$
there are modes where $\Phi_{\rm br}$
overtakes $\Phi_{\rm cy}$ in its variation with $\kappa$,
\eg $n=2$ with $(\sigma\subs{},\psiei,\epsilon\subs{})=(0.5,0.5,100)$.
The prevailing winding of the bremsstrahlung phase
$\Phi_{\rm br}$ is usually more sensitive to $\kappa$
than the $\Phi_{\rm cy}$ is,
\ie $-d\Phi_{\rm br}/d\kappa$
tends to be greater than $-d\Phi_{\rm cy}/d\kappa$.
 
In summary, the presence of transverse perturbations
may significantly alter the instability of a mode
and modify the luminosity response.
 
\section{Conclusions}
\label{'conclusions'}

We have presented a general formulation
   for the linear analysis of two-temperature radiative shocks
   with multiple cooling processes.
The formulation recovers the restrictive cases in the previous studies
   such as the one-temperature flows with a single cooling function
   (Chevalier \& Imamura 1982),
   the one-temperature flows with multiple cooling processes
   (Saxton et al.\ 1998)
   and the two-temperature flows with a single cooling function
   (Imamura et al\ 1996).
We have applied the formulation to mCVs and
   investigated the hydrodynamic and emission properties
   of the time-dependent post-shock accretion flows in these systems.
Our finding are summarised as follows.

The amplitude profiles of $\lambda$-eigenfunctions
  show local minima and maxima,
  which we identify as nodes and antinodes.
The nodes and antinodes are prominent only
  in the eigenfunction profiles of density and electron-pressure.
The eigenfunctions for the fundamental and first overtone
  are more similar to each other than any of the higher overtones.
The eigenfunctions for higher-order modes have more nodes.
 
The phase profiles of the eigenfunctions
  describing particular hydrodynamic variables
  circulate about the complex plane as $\xi$ varies
  from the shock down to the lower boundary.
This circulation can be positive or negative overall,
  or there may be reversals of the winding sense between distinct zones.
The abrupt jumps in the phase profiles
  always coincide the nodes in the $\lambda$-eigenfunction profiles.
 
The luminosity responses of cyclotron and bremsstrahlung
  are determined by the $\lambda$-eigenfunctions.
There is no obvious general relationship between
the amplitude or phase luminosity responses
  and the stability properties of the flow.
There are situations in which a mode is unstable
  but the amplitude of oscillations are larger
  in the cyclotron luminosity than in the bremsstrahlung luminosity.
 
For the same stationary condition ($\tau=0$ at $\xi=0$),
all out choices of perturbed boundary conditions
do not show significant differences in the stability properties.
We therefore conclude that the stability properties of
the flow with stationary-wall boundary
is mainly determined by the energy-transport processes.

The presence of a transverse perturbation modifies
  the eigenfunction profiles of all the hydrodynamic variables.
The profiles of electron-pressure and total-pressure eigenfunctions
  are less affected in comparison with the other eigenfunction profiles.
In some range of transverse wavenumber $\kappa$,
  the density and longitudinal-velocity eigenfunction profiles
  develop extra node features.
When $\kappa$ is large enough ($\LS 3$ for $n=1,2$),
  the amplitudes of all eigenfunctions
  become large near the lower boundary.
The amplitudes, however, decrease as the height $\xi$ increases.
For some values of the transverse wavenumber ($1\LS \kappa \LS 3$),
  a mode which is stable in the absence of the transverse perturbation
  can become unstable.
However, when $\kappa$ is very large, the mode is stablised.
The phase difference between
  the oscillations in the bremsstrahlung and cyclotron luminosity
  are also modified in the presence of transverse perturbations.

\appendix
\section{Reduction to restricted systems}
\label{'appendix.reductions'}

The matrix equation
describing the perturbation is
\begin{equation}
{d\over{d\tau_0}}
{
\left[
\begin{array}{c}
\lambda_\zeta\\
\lambda_\tau\\
\lambda_y\\
\lambda_\pi\\
\lambda\sube{}
\end{array}
\right]
}
=
{1\over\tL}
{
\left[
\begin{array}{ccccc}
1&-1&0&{1/{\tau_0}}&0\\
{-{\gamma\pi_0}/{\tau_0}}&1&0&-{1/{\tau_0}}&0\\
0&0&-{{(\gamma\pi_0-\tau_0)}/{\tau_0}}&0&0\\
\gamma&-\gamma&0&{1/{\pi_0}}&0\\
\gamma&-\gamma&0&{\gamma/{\tau_0}}&
 -{{(\gamma\pi_0-\tau_0)}/{\tau_0\pi\sube{}}}
\end{array}
\right]
}
{
\left[
\begin{array}{c}
F_1\\
F_2\\
F_3\\
F_4\\
F_5\\
\end{array}
\right] \ . 
}
\label{'eq.2d2t.matrix.3'}
\end{equation}
For systems with a purely longitudinal perturbation,
$\kappa=0$
and the third row and third column of the matrix can be eliminated.
The $\lambda_y$ terms vanish from the remaining $F$ functions, and
the equation for $\lambda_y$ decouples from the rest of the perturbed variables
is simply $(\ln\lambda_y)'=\delta/\tau_0 -(\ln\tau_0)'$
or otherwise
$\lambda_y'=\lambda_y(\delta-\tau_0')/\tau_0$.
The two-temperature system with purely longitudinal perturbations
is described by this reduced matrix equation,
with corresponding $F$ functions that omit all terms of $\lambda_y$:
\begin{equation}
{d\over{d\tau_0}}
{
\left[
\begin{array}{c}
\lambda_\zeta\\
\lambda_\tau\\
\lambda_\pi\\
\lambda\sube{}
\end{array}
\right]
}
=
{1\over\tL}
{ 
\left[
\begin{array}{cccc}
1&-1&{1/{\tau_0}}&0\\
{-{\gamma\pi_0}/{\tau_0}}&1&-{1/{\tau_0}}&0\\
\gamma&-\gamma&{1/{\pi_0}}&0\\
\gamma&-\gamma&{\gamma/{\tau_0}}&
 -{{(\gamma\pi_0-\tau_0)}/{\tau_0\pi\sube{}}}
\end{array}
\right]
}
{
\left[
\begin{array}{c}
F_1\\
F_2\\
F_4\\
F_5\\
\end{array}
\right] \ .
}
\label{'eq.2d2t.matrix.2.k0'}
\end{equation}

In the single-temperature limit,
$\psiei$ becomes large
and the electron and ion pressures both equal half of the total pressure,
\ie we have 
$2\pi\sube{}\rightarrow\pi_0=1-\tau_0$,
and
$\lambda\sube{}\rightarrow\lambda_\pi$
throughout the entire post-shock flow.
Moreover, $F_5\rightarrow{\frac12}F_4$.
Then we can eliminate the fifth row
of (\ref{'eq.2d2t.matrix.3'}),
yielding
\begin{equation}
{d\over{d\tau_0}}
{
\left[
\begin{array}{c}
\lambda_\zeta\\
\lambda_\tau\\
\lambda_y\\
\lambda_\pi\\
\end{array}
\right]
}
=
{1\over\tL}
{
\left[
\begin{array}{ccccc}
1&-1&0&{1/{\tau_0}}\\
{-{\gamma\pi_0}/{\tau_0}}&1&0&-{1/{\tau_0}}\\
0&0&-{{(\gamma\pi_0-\tau_0)}/{\tau_0}}&0\\
\gamma&-\gamma&0&{1/{\pi_0}}\\
\end{array}
\right]
}
{
\left[
\begin{array}{c}
F_1\\
F_2\\
F_3\\
F_4\\
\end{array}
\right] \ .
}
\label{'eq.2d2t.matrix.1t'}
\end{equation}
Reducing the system to a one-temperature form
with purely longitudinal perturbations
leaves only three non-trivial perturbed variables
and hence a $3\times3$ coefficient matrix
\begin{equation}
{d\over{d\tau_0}}
{
\left[
\begin{array}{c}
\lambda_\zeta\\
\lambda_\tau\\
\lambda_\pi
\end{array}
\right]
}
=
{1\over\tL}
{
\left[
\begin{array}{ccc}
1&-1&{1/{\tau_0}}\\
{-{\gamma\pi_0}/{\tau_0}}&1&-{1/{\tau_0}}\\
\gamma&-\gamma&{1/{\pi_0}}\\
\end{array}
\right]
}
{
\left[
\begin{array}{c}
F_1\\
F_2\\
F_4\\
\end{array}
\right] \ ,
}
\label{'eq.2d2t.matrix.1t.k0'}
\end{equation}
which is equivalent to that in
Saxton \etal\shortcite{saxton97}
and Saxton \etal\shortcite{saxton98}.


\begin{figure}
\section{Eigenfunction profiles}
\begin{center}
\epsfxsize=15cm
\epsfbox{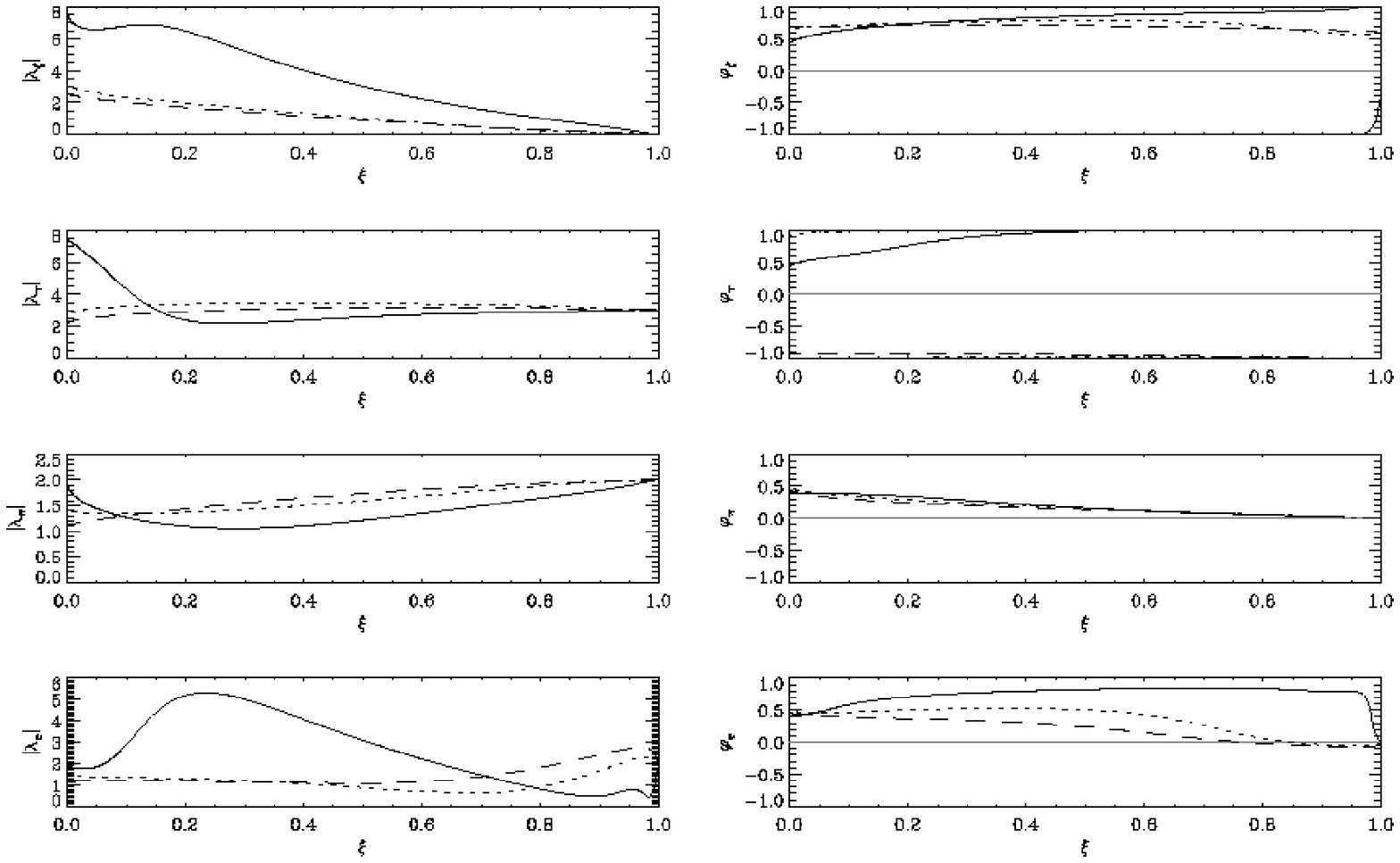}
\end{center}
\scriptsize
\caption{
Eigenfunctions for $n=1$ mode
with $(\sigma\subs{},\psiei)=(0.2,0.1)$
and $\epsilon\subs{}=0,1,100$ for dashed, dotted and solid lines respectively.
Amplitudes are shown in the left column, phase profiles on the right.
}
\label{'a1ef1'}
\begin{center}
\epsfxsize=15cm
\epsfbox{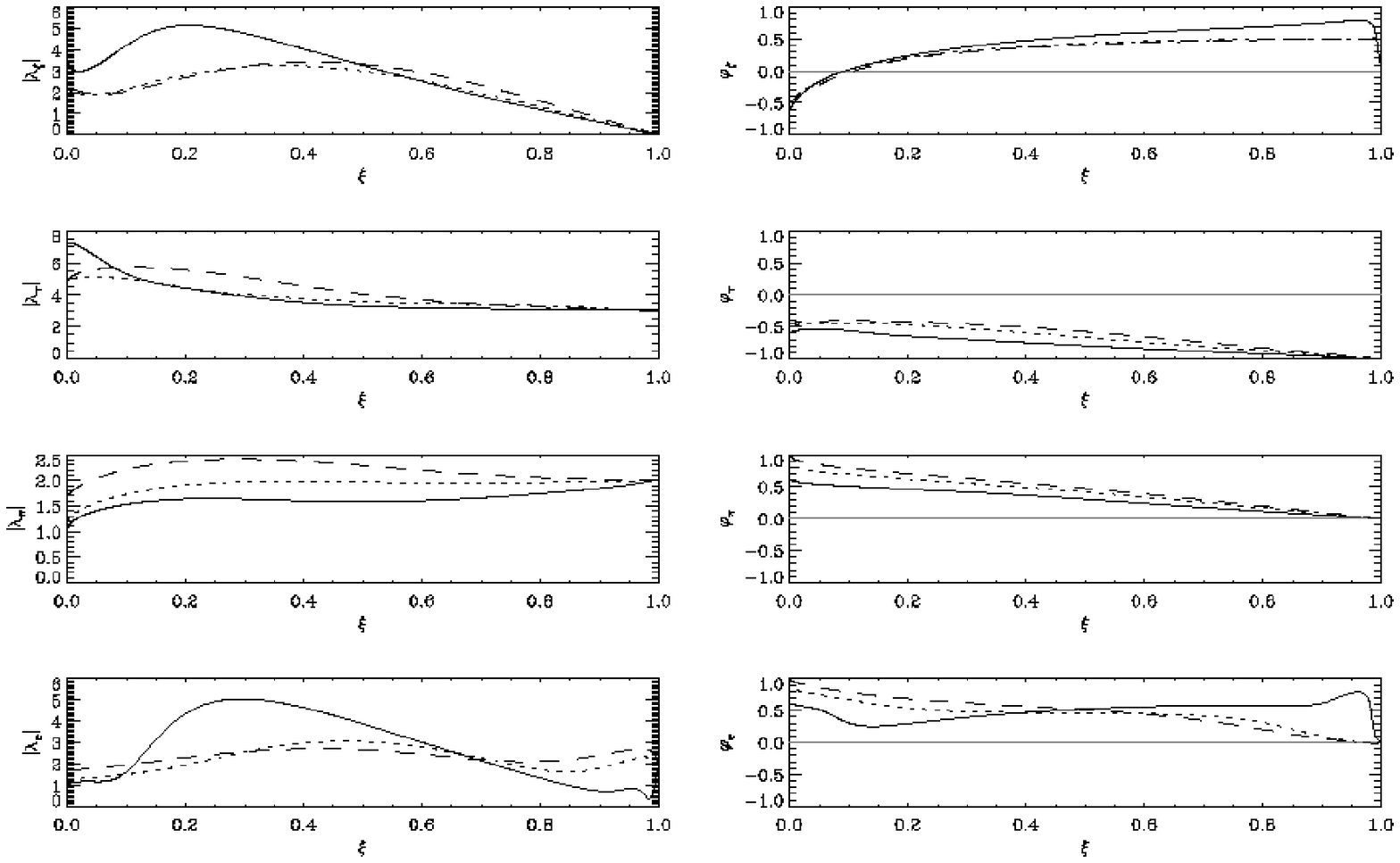}
\end{center}
\scriptsize
\caption{
Same as Figure\ \ref{'a1ef1'} but with
$n=2$ and $(\sigma\subs{},\psiei)=(0.2,0.1)$.
}
\label{'a1ef2'}
\end{figure}

\begin{figure}
\begin{center}
\epsfxsize=15cm
\epsfbox{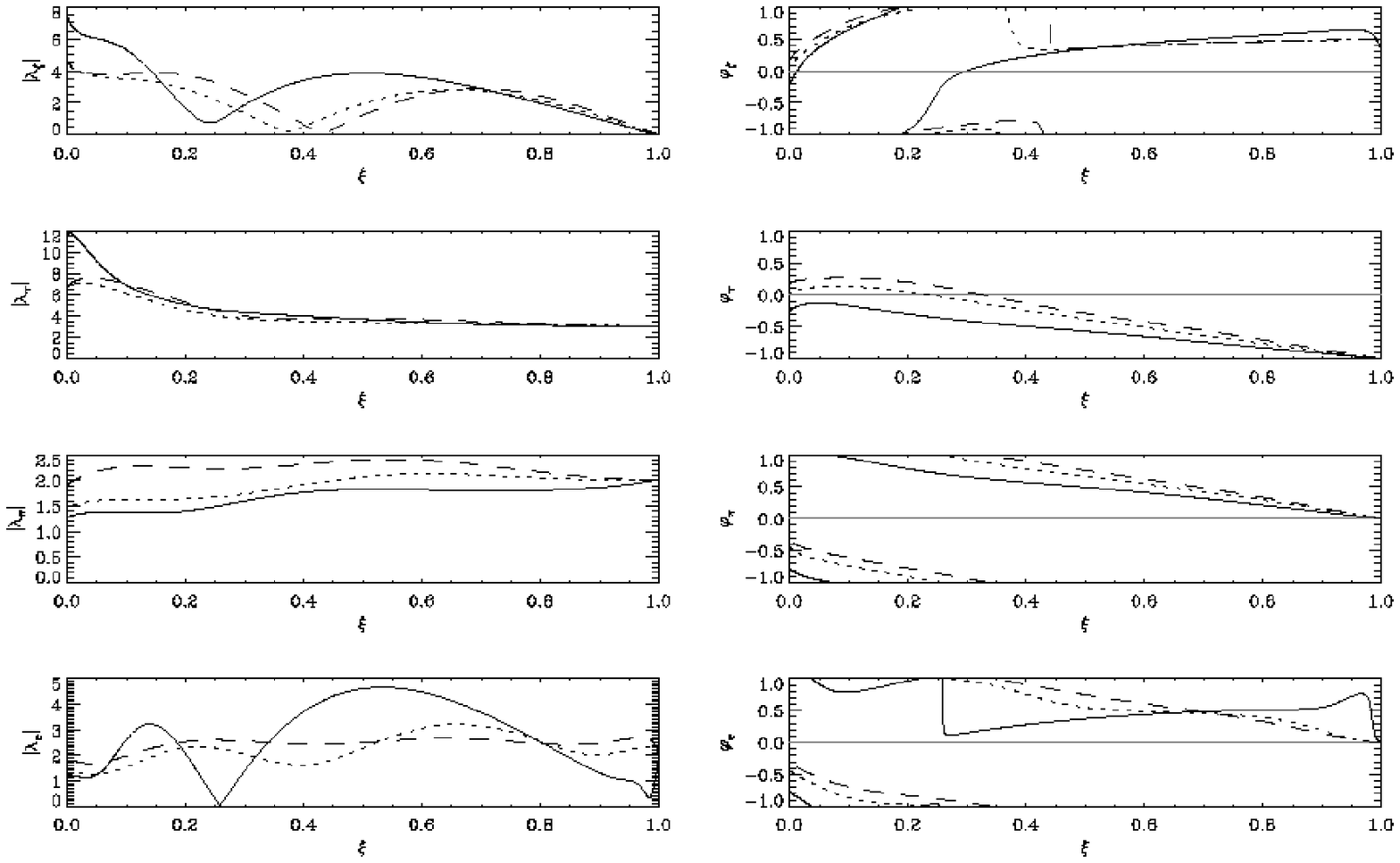}
\end{center}
\scriptsize
\caption{
Same as Figure\ \ref{'a1ef1'} but with
$n=3$ and $(\sigma\subs{},\psiei)=(0.2,0.1)$.
}
\label{'a1ef3'}
\begin{center}
\epsfxsize=15cm
\epsfbox{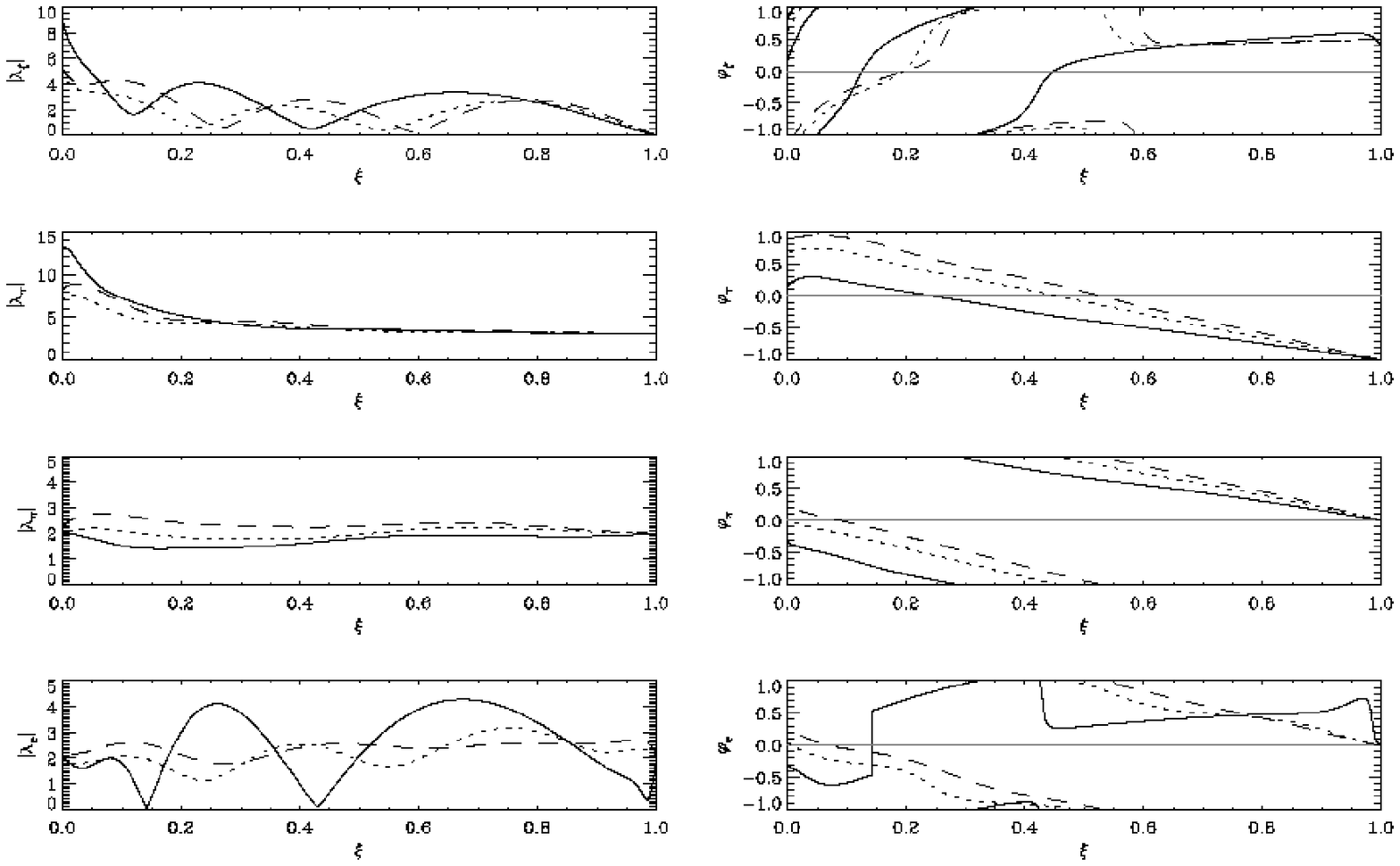}
\end{center}
\scriptsize
\caption{
Same as Figure\ \ref{'a1ef1'} but with
$n=4$ and $(\sigma\subs{},\psiei)=(0.2,0.1)$.
}
\label{'a1ef4'}
\end{figure}



\begin{figure}
\begin{center}
\epsfxsize=15cm
\epsfbox{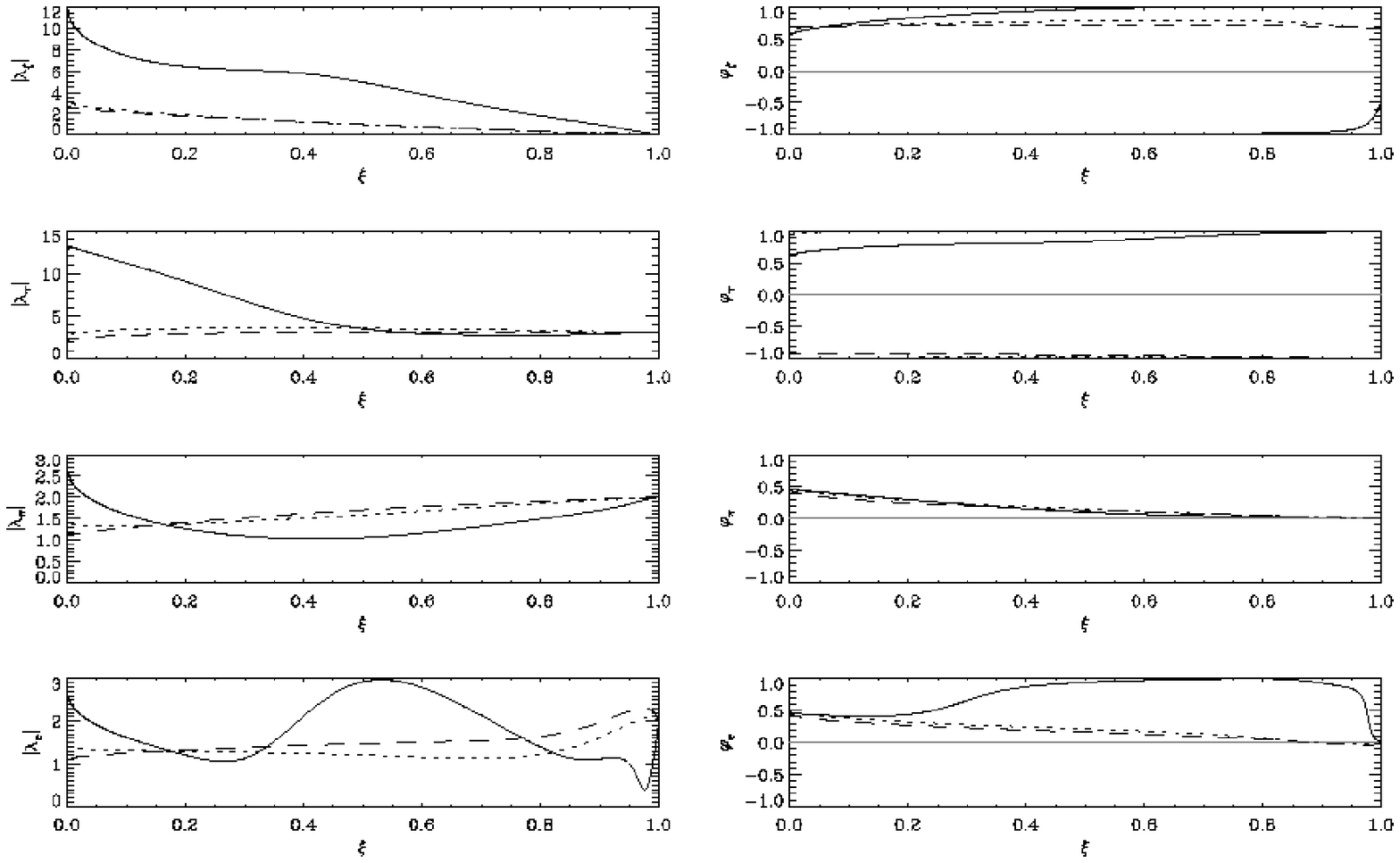}
\end{center}
\scriptsize
\caption{
Same as Figure\ \ref{'a1ef1'} but with
$n=1$ and $(\sigma\subs{},\psiei)=(0.5,0.5)$.
}
\label{'b2ef1'}
\begin{center}
\epsfxsize=15cm
\epsfbox{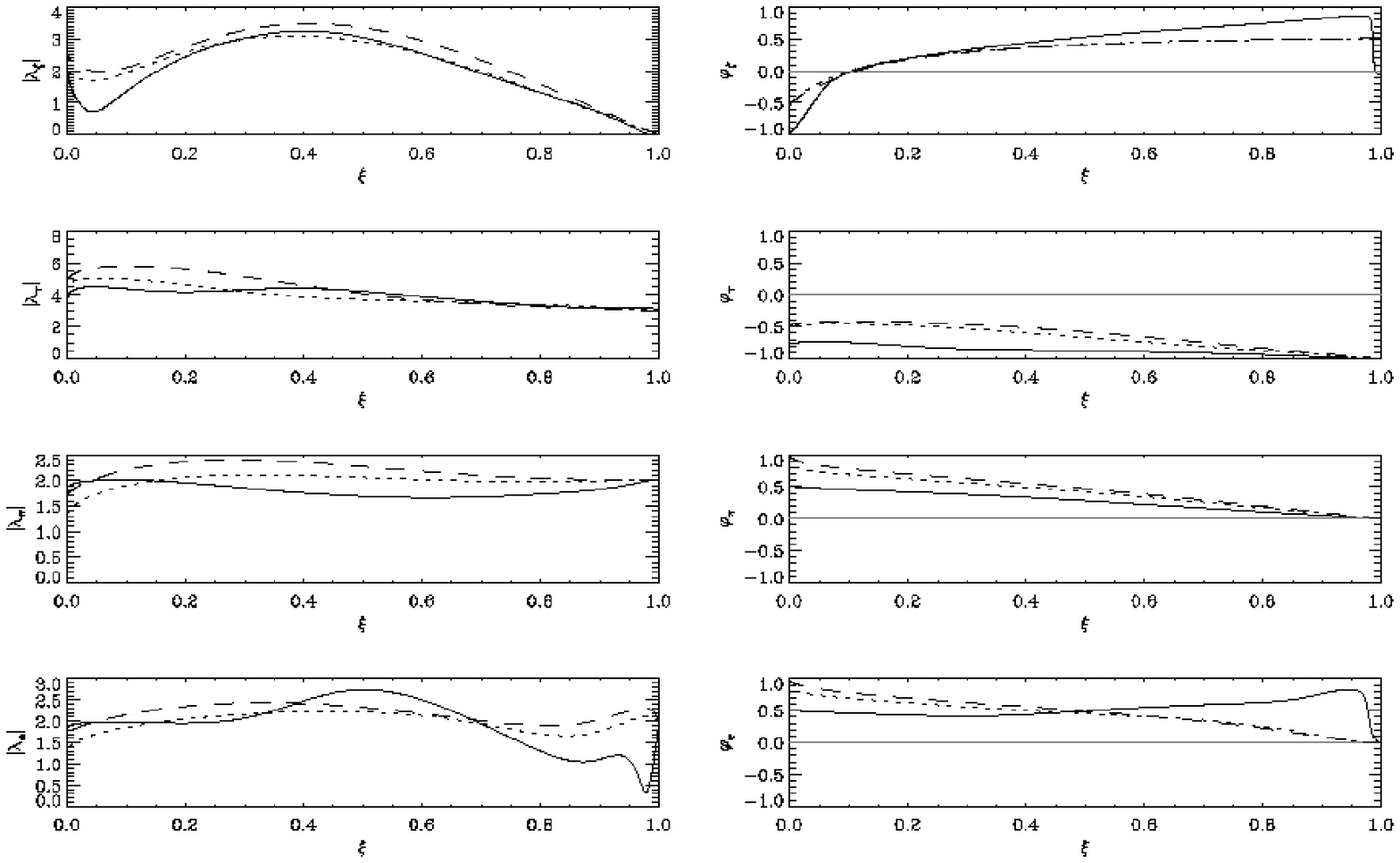}
\end{center}
\scriptsize
\caption{
Same as Figure\ \ref{'a1ef1'} but with
$n=2$ and $(\sigma\subs{},\psiei)=(0.5,0.5)$.
}
\label{'b2ef2'}
\end{figure}

\begin{figure}
\begin{center}
\epsfxsize=15cm
\epsfbox{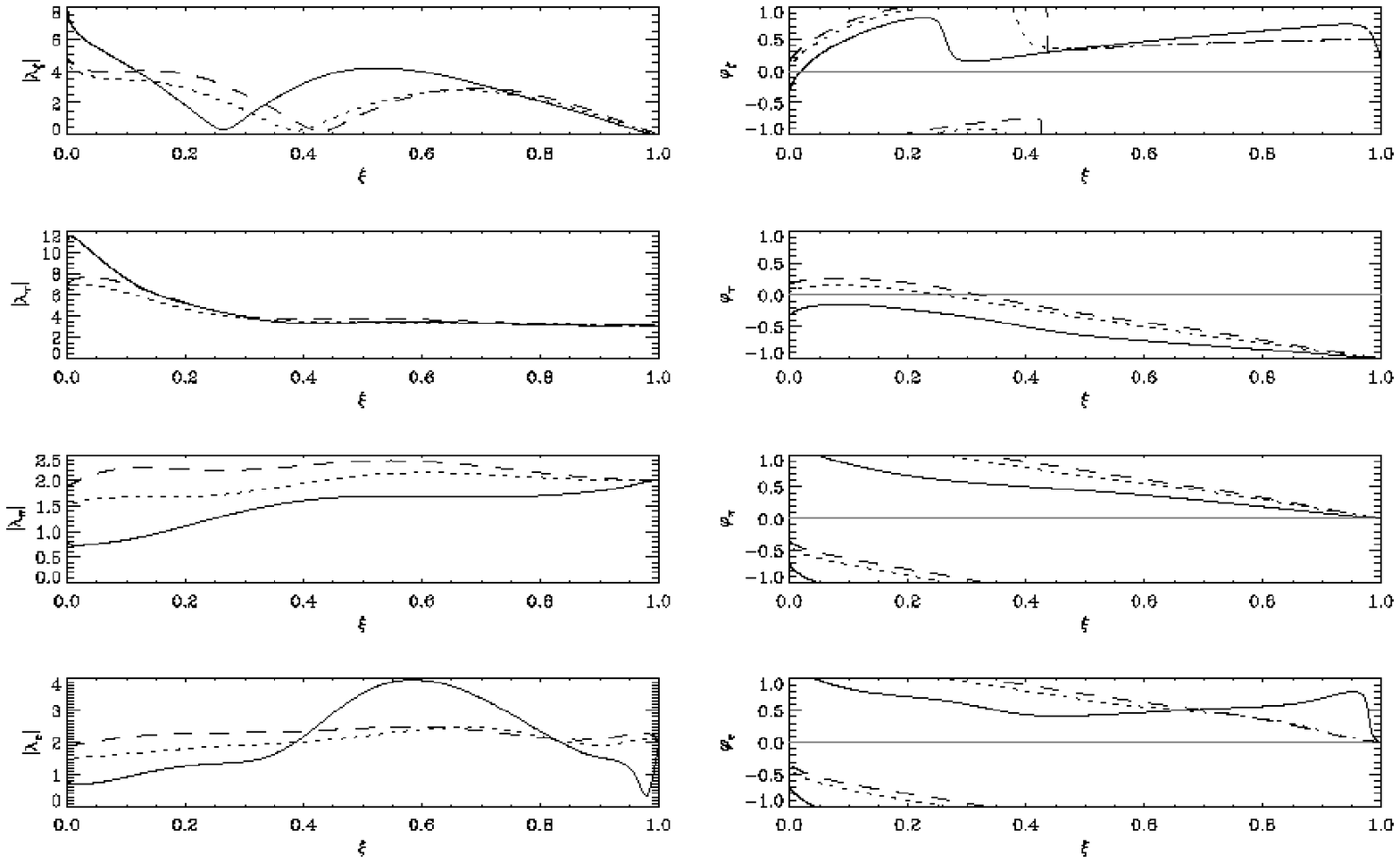}
\end{center}
\scriptsize
\caption{
Same as Figure\ \ref{'a1ef1'} but with
$n=3$ and $(\sigma\subs{},\psiei)=(0.5,0.5)$.
}
\label{'b2ef3'}
\begin{center}
\epsfxsize=15cm
\epsfbox{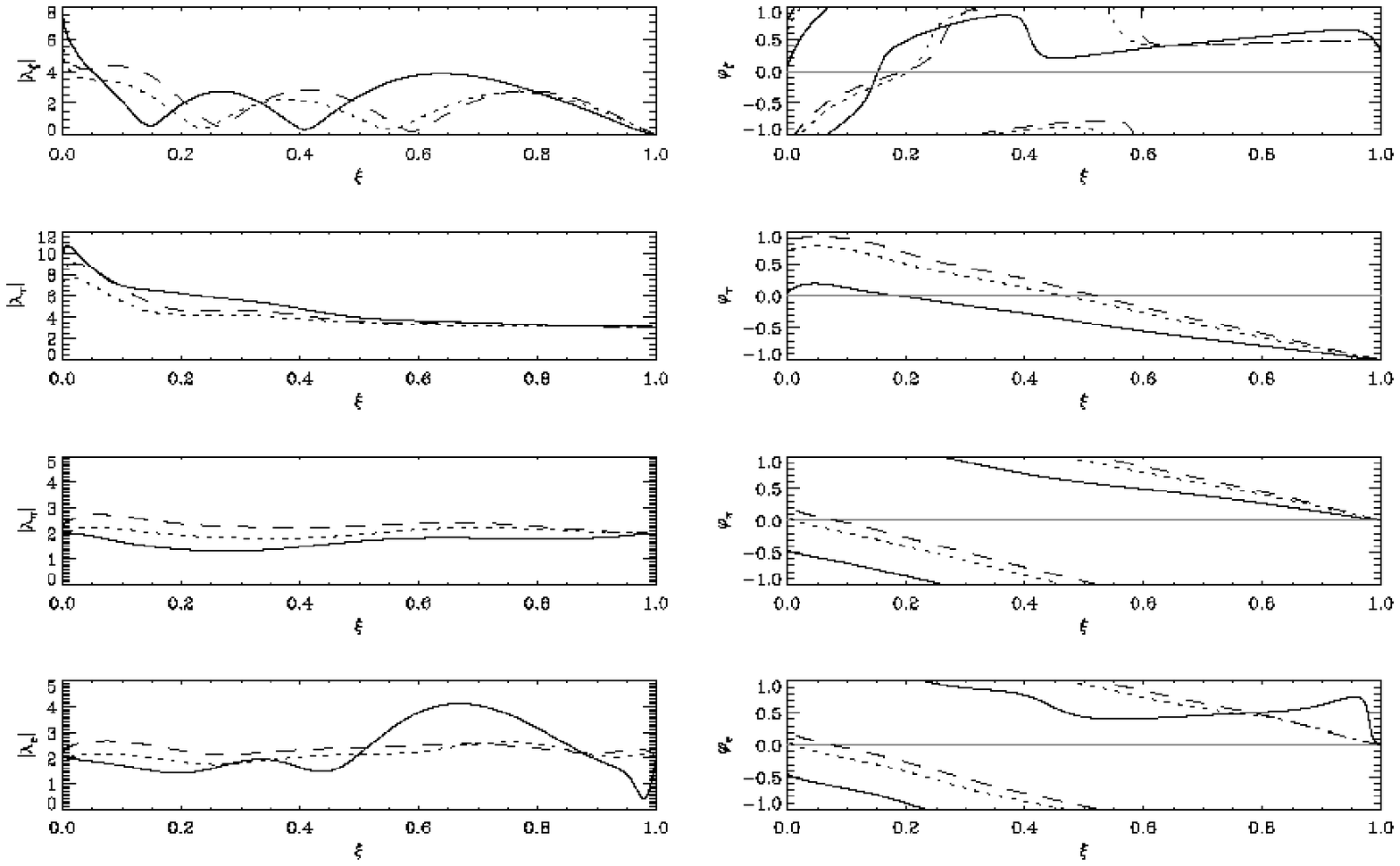}
\end{center}
\scriptsize
\caption{
Same as Figure\ \ref{'a1ef1'} but with
$n=4$ and $(\sigma\subs{},\psiei)=(0.5,0.5)$.
}
\label{'b2ef4'}
\end{figure}

\begin{figure}
\begin{center}
\epsfxsize=15cm
\epsfbox{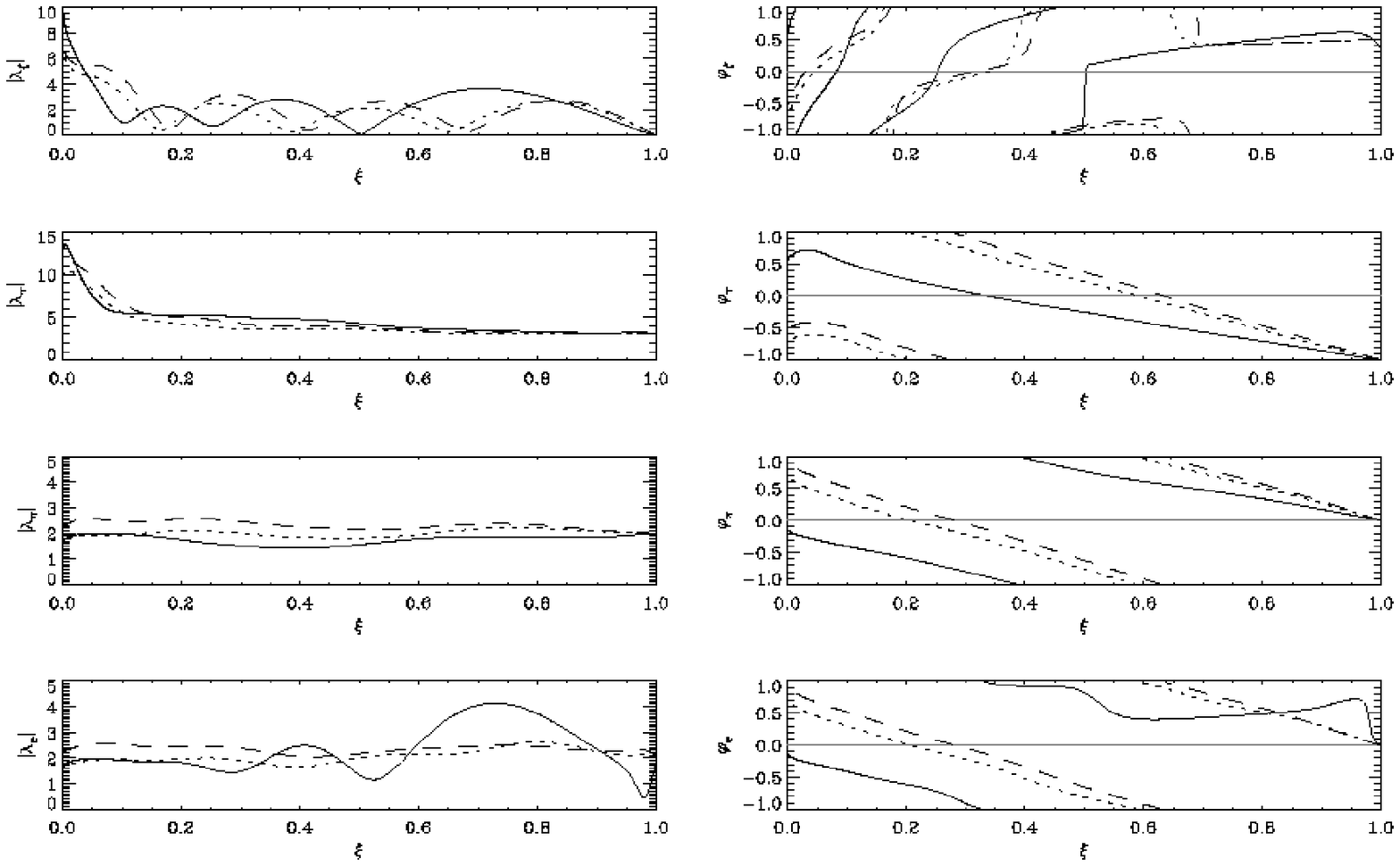}
\end{center}
\scriptsize
\caption{
Same as Figure\ \ref{'a1ef1'} but with
$n=5$ and $(\sigma\subs{},\psiei)=(0.5,0.5)$.
}
\label{'b2ef5'}
\begin{center}
\epsfxsize=15cm
\epsfbox{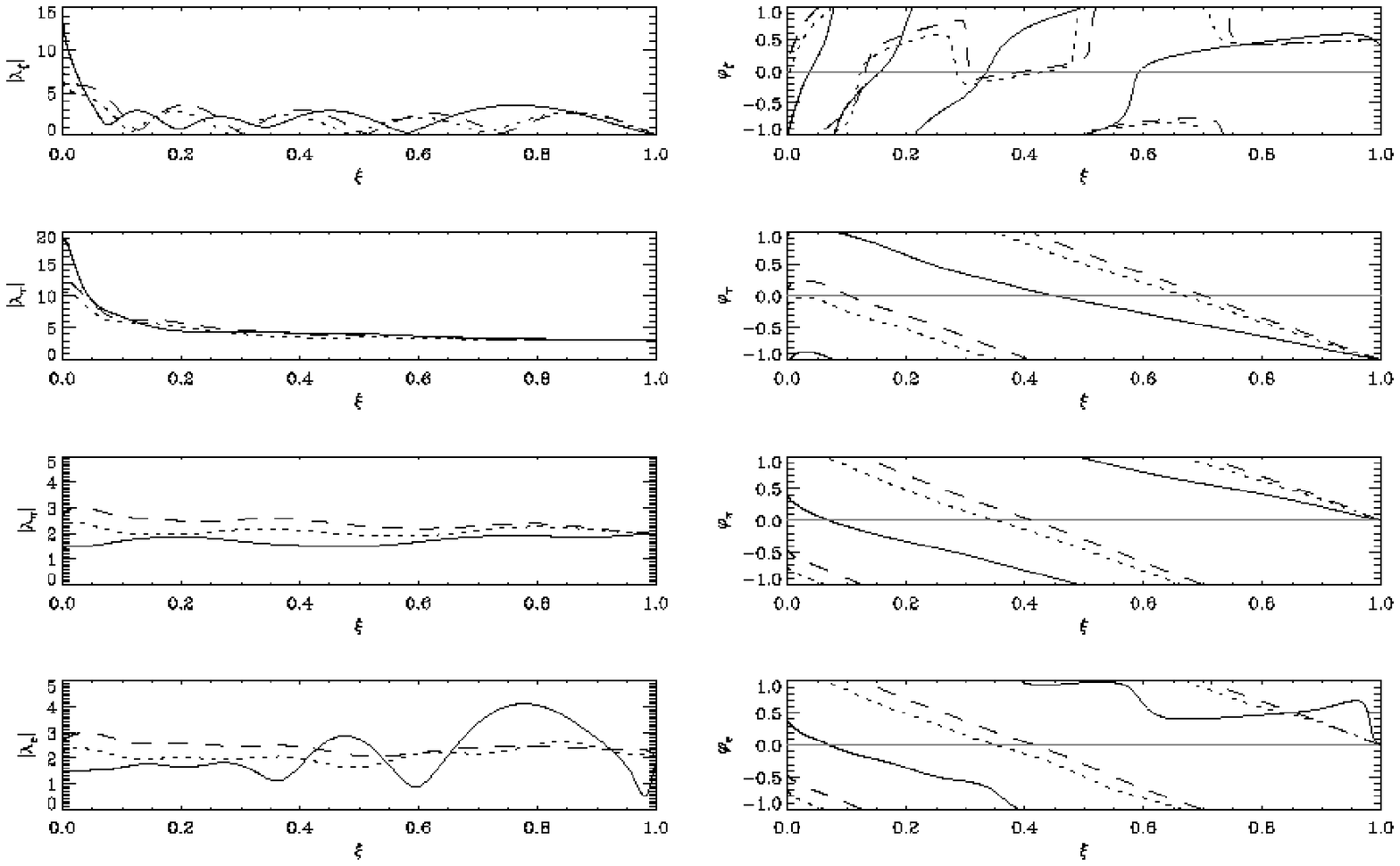}
\end{center}
\scriptsize
\caption{
Same as Figure\ \ref{'a1ef1'} but with
$n=6$ and $(\sigma\subs{},\psiei)=(0.5,0.5)$.
}
\label{'b2ef6'}
\end{figure}


\begin{thebibliography}{99}

\bibitem[\protect\citename{Aizu }1973]{aizu}
Aizu, K., 1973. Prog. Theor. Phys., 49, 1184  
\bibitem[\protect\citename{Bertschinger }1986]{bertschinger}
Bertschinger, E., 1986, ApJ, 304, 154
\bibitem[\protect\citename{Chanmugam et al.\ }1985]{cls}
Chanmugam G., Langer, S. H., Shaviv, G., 1985, ApJ, 299, L87 
\bibitem[\protect\citename{Chevalier \& Imamura }1982]{chevalier}
Chevalier, R. A., Imamura, J. N., 1982, ApJ, 261, 543 
\bibitem[\protect\citename{Cropper }1990]{cropper}
Cropper, M., 1990, Sp. Sci. Rev., 54, 195
\bibitem[\protect\citename{Cropper, Wu, Ramsay \& Kocabiyik }1999]{cropper99}
Cropper, M., Wu, K., Ramsay, G. Kocabiyik, A., 1999, MNRAS, 306, 684
\bibitem[\protect\citename{Dgani \& Soker}1994]{dgani}
Dgani, R., Soker, N., 1994, ApJ, 434, 262
\bibitem[\protect\citename{Gaetz, Edgar \& Chevalier }1988]{gaetz}
Gaetz, T. J., Edgar, R. J., Chevalier, R. A., 1988, 
    ApJ, 329, 927 
\bibitem[\protect\citename{Falle }1975]{falle75}
Falle, S. A. E. G, 1975, MNRAS, 172, 55
\bibitem[\protect\citename{Falle }1981]{falle81}
Falle, S. A. E. G, 1981, MNRAS, 195, 1011
\bibitem[\protect\citename{Houck \& Chevalier }1992]{houck}
Houck, J. C., Chevalier, R. A., 1992, ApJ, 395, 592 
\bibitem[\protect\citename{Hujeirat \& Papaloizou }1998]{hujeirat}
Hujeirat, A., \& Papaloizou, J. C. B., 1998, A\&A, 340, 593
\bibitem[\protect\citename{Imamura }1985]{imamura85}
Imamura, J. N., 1985, ApJ, 296, 128   
\bibitem{rashed}
Imamura, J. N., Rashed, H., Wolff, M. T., 1991, ApJ, 376, 665  
\bibitem[\protect\citename{Imamura \& Wolff }1990]{wolff90}
Imamura, J. N., Wolff, M. T., 1990, ApJ, 355, 216   
\bibitem[\protect\citename{Imamura, Wolff \& Durisen }1984]{durisen}
Imamura, J. N., Wolff, M. T., Durisen, R. H., 1984, 
    ApJ, 276, 667 
\bibitem[\protect\citename{Imamura et al.\ }1996]{aboasha}
Imamura, J. N., Aboasha, A., Wolff, M. T., Wood, K. S., 
    1996, ApJ, 458, 327 
\bibitem[\protect\citename{Innes, Giddings \& Falle }1987a]{innes87a}
Innes, D. E., Giddings, J. R., Falle, S. A. E. G., 1987a,  
    MNRAS, 224, 179  
\bibitem[\protect\citename{Innes, Giddings \& Falle }1987b]{innes87b}
Innes, D. E., Giddings, J. R., Falle, S. A. E. G., 1987b,  
    MNRAS, 226, 67   
\bibitem[\protect\citename{King \& Lasota }1979]{king}
King, A. R., Lasota, J. P., 1979, MNRAS, 188, 653 
\bibitem[\protect\citename{Lamb \& Masters }1979]{lamb}
Lamb, D. Q., Masters, A. R., 1979, ApJ, 234, L117  
\bibitem[\protect\citename{Langer, Chanmugam \& Shaviv }1981]{langer81}
Langer, S. H., Chanmugam, G., Shaviv, G., 1981, ApJ, 245, L23 
\bibitem[\protect\citename{Langer, Chanmugam \& Shaviv }1982]{langer82}
Langer, S. H., Chanmugam, G., Shaviv, G., 1982, ApJ, 258, 289 
\bibitem[\protect\citename{Melrose }1986]{melrose}
Melrose, D. B., 1986, Instabilities in Laboratory and Space Plasmas,
Cambridge University Press, Cambridge
\bibitem[\protect\citename{Rybicki \& Lightman }1979]{rybicki}
Rybicki, G. B., Lightman, A. P., 1979, Radiative Processes 
    in Astrophysics, John Wiley \& Sons, New York
\bibitem[\protect\citename{Saxton }1999]{saxtonthesis}
Saxton, C. J., 1999, PhD Thesis, University of Sydney, Australia
\bibitem[\protect\citename{Saxton }2001]{saxton2001}
Saxton, C. J., 2001, Publ. Astr. Soc. Australia, submitted.
\bibitem[\protect\citename{Saxton \& Wu }1999]{saxton99}
Saxton, C. J., Wu, K., 1999, MNRAS, 310, 677
\bibitem[\protect\citename{Saxton, Wu \& Pongracic }1997]{saxton97}
Saxton, C. J., Wu, K., Pongracic, H., 1997, 
    Publ. Astr. Soc. Australia, 14, 164 
\bibitem[\protect\citename{Saxton, Wu, Pongracic \& Shaviv}1998]{saxton98}
Saxton, C. J., Wu, K., Pongracic, H., Shaviv, G., 1998, 
    MNRAS, 299, 862
\bibitem{strickland} 
Strickland, R., Blodin, J. M., 1995, ApJ, 449, 727 
\bibitem[\protect\citename{T\'{o}th \& Draine}1993]{toth}
T\'{o}th, G., Draine, B. T., 1993, ApJ, 413, 176  
\bibitem[\protect\citename{Wolff, Gardiner \& Wood }1989]{wolff89}
Wolff, M. T., Gardner, J., Wood, K. S., 1989, ApJ, 346, 833 
\bibitem[\protect\citename{Wu }1994]{wu}
Wu, K., 1994, Proc. Astr. Soc. Australia, 11, 61  
\bibitem[\protect\citename{Wu }2000]{wu2000}
Wu, K., 2000, Space Science Rev., 93, 611
\bibitem[\protect\citename{Wu, Chanmugam \& Shaviv }1992]{wcs92}
Wu, K., Chanmugam, G., Shaviv, G., 1992, ApJ, 397, 232 
\bibitem[\protect\citename{Wu, Chanmugam \& Shaviv }1994]{wcs94}
Wu, K., Chanmugam, G., Shaviv, G., 1994, ApJ, 426, 664
\bibitem[\protect\citename{Wu et al.\ }1996]{wpcs}
Wu, K., Pongracic, H., Chanmugam, G., Shaviv, G., 1996, 
    Publ. Astron. Soc. Aust., 13, 93

\end{thebibliography}
\end{document}